\documentclass[preprint]{elsarticle}

\usepackage{amssymb}

\usepackage{graphicx}
\usepackage{wrapfig}
\usepackage{amsmath, amssymb, amsthm}
\usepackage{subfig}
\usepackage{verbatim}
\usepackage{xspace}
\usepackage[usenames,dvipsnames]{color}

\newcommand{\para}[1]{\noindent{\bf\em #1}\xspace}
\newcommand{\paraskip}[1]{\medskip \noindent{\bf\em #1}\xspace}
\newcommand{\pf}{\smallskip\noindent{\em Proof.\xspace}\xspace}

\newcommand{\edge}[2]{#1 #2\xspace}

\newcommand{\base}{\operatorname{base}}

\newcommand{\slab}{\operatorname{slab}}
\newcommand{\slabs}{\operatorname{slabs}}

\newcommand{\R}{{\bf R}\xspace}

\newtheorem{theorem}{Theorem}[section]
\newtheorem{lemma}[theorem]{Lemma}
\newtheorem{corollary}[theorem]{Corollary}

\usepackage[margin=1in]{geometry}

\usepackage{lipsum}
\makeatletter
\def\ps@pprintTitle{%
 \let\@oddhead\@empty
 \let\@evenhead\@empty
 \def\@oddfoot{}%
 \let\@evenfoot\@oddfoot}
 \makeatother

\journal{Computational Geometry}

\begin{document}

\begin{frontmatter}



\title{Faster Reductions from Straight Skeletons to Motorcycle Graphs}


\author{John Bowers\fnref{Research supported by an NSF graduate fellowship under Grant No. S121000000211.}}

\address{Department of Computer Science, University of Massachusetts, Amherst, MA 01003, USA.}

\begin{abstract}
We give an algorithm that reduces the straight skeleton to the motorcycle graph in $O(n\log n)$ time for (weakly) simple polygons and $O(n(\log n)\log m)$ time for a planar straight line graph with $m$ connected components. The current fastest algorithms for computing motorcycle graphs are an $O(n^{4/3 + \epsilon})$ time algorithm for non-degenerate cases and $O(n^{17/11 + \epsilon})$ for degenerate cases. Together with our algorithm this results in an algorithm computing the straight skeleton of a non-degenerate (weakly) simple polygon with $r$ reflex vertices in $O(n\log n + r^{4/3 + \epsilon})$ time and of a non-degenerate planar straight line graph with $m$ connected components in $O(n(\log n)\log m + r^{4/3 + \epsilon})$ time. For degenerate cases the algorithm takes $O(n\log n + r^{17/11 + \epsilon})$ and $O(n(\log n)\log m + r^{17/11 + \epsilon})$ time respectively.
\end{abstract}

\begin{keyword}
Computational geometry\sep Straight skeletons\sep Motorcycle graphs\sep Roof construction
\end{keyword}

\end{frontmatter}


\section{Introduction}
\label{sec:introduction}

\begin{figure}
\vspace{-10pt}
\centering
\includegraphics[width=0.99\textwidth]{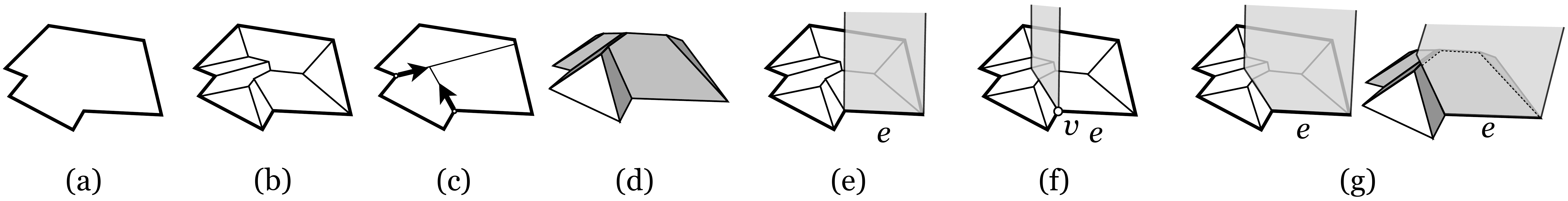}
\vspace{-6pt}
\caption{\small{(a) A polygon. (b) Its straight skeleton. (c) The induced motorcycle graph. (d) The straight skeleton roof. (e) An edge slab. (f) The motorcycle slab for $v$ with respect to $e$. (g) Shows a view of $\slab(e)$, which is the union of the edge and motorcycle slabs for $e$ from $z=+\infty$ (left) and in perspective (right).}}
\vspace{-14pt}
\label{fig:sskel0}
\end{figure}

The straight skeleton of a simple polygon (Fig.~\ref{fig:sskel0}b) is a tree-like structure that subdivides its interior into regions. It was first defined by Aichholzer et al.\ in \cite{aaag-ntsp-95} by tracing the vertices of the polygon during a wavefront process in which the sides of the polygon are moved inwards in parallel at constant speed. It was later generalized to planar straight line graphs (PSLGs) \cite{aa-ssgpf-96}. The trace of the vertices during the wavefront process forms the straight-skeleton. It has a wide array of applications including polygon interpolation 
\cite{Barequet:2003:SBC:644108.644129}, procedural modeling of urban environments \cite{Vanegas:2012:PGP:2322116.2322132}, biomedical imaging \cite{Cloppet:2000:ABN:331097.331118}, and polygon decomposition \cite{Tanase:2003:PDB:777792.777802}, to name a few. For convex polygons, the straight skeleton is identical to the medial axis and is linear time computable, but for general simple polygons and PSLGs the computational complexity is still an open problem. In the polygon case, the fastest algorithms for computing it first compute a structure called the induced motorcycle graph, which was introduced by Eppstein and Erickson \cite{EppEri-SCG-98}, and then compute the straight skeleton as a post-processing step. 
 In the case of PSLGs, however, no sub-quadratic reduction of the straight skeleton to the motorcycle graph is known. This results of this paper are summarized by:

\begin{theorem}[Main Results] \label{theorem:main}The straight skeleton problem can be reduced to the motorcycle graph problem in $O(n\log n)$ time and $O(n)$ space for simple polygons or $O(n(\log n)\log m)$ time and $O(n\log m)$ space for a PSLG with $m$ connected components.\end{theorem}

The current fastest algorithms for computing the induced motorcycle graph of a polygon or PSLG with $r$ reflex vertices are the $O(r^{4/3 + \epsilon})$ algorithm from  \cite{Vigneron:2013vw} for non-degenerate input and the $O(r^{17/11 + \epsilon})$ time algorithm from \cite{EppEri-SCG-98} for degenerate input. Together with our results this gives an algorithm computing the straight skeleton of a non-degenerate (weakly) simple polygon with $r$ reflex vertices in $O(n\log n + r^{4/3 + \epsilon})$ time and of a non-degenerate planar straight line graph with $m$ connected components in $O(n(\log n)\log m + r^{4/3 + \epsilon})$ time. For degenerate cases the algorithm takes $O(n\log n + r^{17/11 + \epsilon})$ and $O(n(\log n)\log m + r^{17/11 + \epsilon})$ time respectively.

\paraskip{Overview.} In addition to the wavefront process, the straight skeleton has alternatively been defined as a terrain, sometimes called a {\em roof}, which can be characterized as a lower envelope of certain infinite strips in $\R^3$ \cite{aaag-ntsp-95, Cheng:2007go, HuberH12}. Our general approach for polygons is a divide and conquer algorithm which computes this roof. Given a polygon (or a sub-chain of the polygon) our algorithm subdivides the polygon into two sub-chains, recursively computes an intermediate structure we call a {\em partial roof} for each sub-chain, and then merges the result. The partial roof captures certain properties of the final roof which allow us to reconstruct the roof in the final merge operation. A key insight of this paper is that although the final roof is a terrain, which in particular implies no self-intersections, this restriction is unnecessary for the intermediate partial roofs. Indeed, our partial roofs are {\em intrinsically} topological disks, but their particular {\em realizations} in $\R^3$ may be more topologically complicated.  This distinction may be made more clear by a familiar analog: a Klein bottle is intrinsically a 2D manifold, meaning locally it always looks like a small patch of the Euclidean plane, but any realization in $\R^3$ exhibits self-intersections and so their are points on the realization that are more complicated. However, if one is intrinsically walking along the surface (say, as a flatlander), one would never encounter such a self-intersection. The intersection is an extrinsic property of the particular realization chosen, but is not intrinsic to the underlying surface. Our result is extended to PSLGs by using a modification of the vertical subdivision procedure from \cite{chengESA2014}. This allows us to subdivide the PSLG into polygons and then employ our divide and conquer approach to compute the part of the straight skeleton roof which lies above each polygon.

\paraskip{Related Work.} Our algorithm for polygons has been circulating since Nov. 2013 and was posted to the ArXiV in May 2014 \cite{bowersArxiv2014}. Subsequently, two related papers have appeared. The first gives a reduction from the straight skeleton problem to the motorcycle graph problem taking $O(n(\log n)\log r)$ time for a polygon with $r$ reflex vertices with or without holes \cite{chengESA2014}. We adapt a technique from that paper--the vertical decomposition algorithm--to extend our polygon algorithm to PSLGs. The second, which appeared in EuroCG \cite{eurocg2014}, studies the problem of computing the straight skeleton for the special case of monotone polygons without holes. The main idea is similar to ours: subdivide the polygon into its two monotone chains, compute an intermediate terrain for the each chain, and merge the result. Their result makes use of the fact that the two chains are monotone, which allows them to efficiently compute a terrain for each chain and merge the result. This is different than our method, since we cannot assume our input polygons are monotone and thus drop the restriction that the intermediate results be terrains. The first sub-quadratic straight skeleton algorithm is due to Eppstein and Erickson \cite{EppEri-SCG-98} and takes $O(n^{1+\epsilon} + n^{8/11 + \epsilon}r^{9/11+\epsilon})$ time. Prior to the present work and that of \cite{chengArxiv2014}, this was the fastest for PSLGs and the fastest {\em deterministic} algorithm for polygons (with or without holes). They introduced {\em motorcycle graphs} as an abstraction of the main difficulty, but did not give an algorithm for straight skeletons that uses motorcycle graph as input. The first such algorithm was described by Cheng and Vigneron \cite{Cheng:2007go}. They give an algorithm computing a motorcycle graph in $O(n^{3/2}\log^2 n)$ time and a post-processing step computing the straight skeleton of a polygon with $h$ holes from its motorcycle graph in {\em expected} $O(n\sqrt{h}\log^2 n)$ time. The first step was recently improved to $O(n^{4/3 + \epsilon})$ time by Vigneron and Yan \cite{Vigneron:2013vw} for non-degenerate inputs. The best known lower bounds for straight skeletons are $\Omega(n\log n)$ for PSLGs \cite{EppEri-SCG-98} and polygons with holes \cite{Hub11}, and $\Omega(n)$ for simple polygons. A parallel thread of research focuses on algorithms which perform better in practice than their theoretical upper bounds. Huber and Held \cite{Huber:2011kr}, describe an $O(n^2\log n)$ time algorithm for computing the straight skeleton of planar straight-line graphs that uses the motorcycle graph which behaves like $O(n\log n)$ in practice--though worst case examples can be constructed. Similarly, Palfrader et al., \cite{phh-2012} investigate the algorithm from \cite{aa-ssgpf-96} and show that it behaves like $O(n\log n)$ in practice, though examples requiring $O(n^2\log n)$ are known. It remains open to close the gap between theoretical upper and lower bounds and experimental observation.

\section{Preliminary Terms}
\label{sec:preliminaries}

\paraskip{Straight Skeletons.} Historically, the straight skeleton of a polygon $P$ has been defined by a wavefront process: move the edges of $P$ towards its interior at unit speed while keeping each edge parallel to its original position. Each edge grows or shrinks to maintain incidence with its neighboring edges. An edge may shrink to zero-length, in which case it is replaced by a vertex in the wavefront, or may hit some other edge of the wavefront, in which case the wavefront polygon is split into two, and the wavefront continues independently in each. The trace of the vertices during this process is the {\em straight skeleton}, denoted $SS(P)$. For a more thorough treatment of the wavefront definition of the straight skeleton, see \cite{aaag-ntsp-95}. The wavefront model is extended to PSLGs in \cite{aa-ssgpf-96}.

\paraskip{Motorcycle graphs.} Place ``motorcycles'' at points $p_1,\dots,p_n$ in the plane with velocity vectors $v_1,\dots,v_n$. A motorcycle $M_i$ begins at $p_i$ and moves along the ray $p_i + t v_i$, leaving a track behind it. It crashes if it encounters another motorcycle's track. The {\em motorcycle graph} is given by vertices for the initial positions $p_1, \dots, p_n$ and the crash sites $c_1,\dots,c_n$ for each motorcycle, and an edge for each track. The {\em motorcycle graph induced by a polygon $P$ (or PSLG $G$)}, denoted $MG(P)$, is given by creating a motorcycle for each reflex vertex $v$ of the polygon, with speed equal to $1/\sin{(\theta/2)}$, where $\theta$ is the interior angle at $v$ in $P$. In a PSLG a degree 1 vertex induces two motorcycles, each making an angle of $3\pi/2$ on either side with the incident edge. We show how to handle this more generically below. The speed of each motorcycle is the same as the speed a vertex moves in the wavefront algorithms for straight skeleton computation. In addition to the tracks, the polygon/PSLG edges are treated as obstacles and a motorcycle crashes if it encounters either an edge or a track. See Fig.~\ref{fig:sskel0}c. \looseness=-1

\paraskip{The roof model of the straight skeleton.} An alternative view of the straight skeleton to the wavefront model is the {\em roof model} \cite{aaag-ntsp-95}. In the roof model the straight skeleton is a polygonal ``roof'' of faces in ${\bf R}^3$ each lying in the upper half space $z\geq 0$ with the boundary edges embedded in the $xy$-plane. The roof model is given by lifting each vertex $v$ of the straight skeleton by augmenting its position with a $z$-coordinate equal to the time $t$ at which the wavefront reaches $v$. We call this the {\em straight skeleton roof}, denoted $R(P)$. The non-boundary edges of the roof is the (lifted) straight-skeleton, denoted $SS(P)$. See Fig.~\ref{fig:sskel0}d. Each face of the roof lies in a plane through its base edge making a dihedral angle of $\pi/4$ with the $xy$-plane. \looseness=-1

\paraskip{Edge and motorcycle slabs.} An alternative characterization of $R(P)$ is given in \cite{Cheng:2007go}. There $R(P)$ is defined as the lower envelope of a set of partially infinite strips in ${\bf R}^3$ called slabs defined with respect to the edges of the polygon $P$ and the edges of the motorcycle graph $MG(P)$. For each edge $e$ of $P$ they define an {\em edge slab} and for each reflex vertex $v$ of $P$ they define two {\em motorcycle slabs}, one for each edge incident $v$. Before defining the slabs, let us attach a coordinate frame to each edge of $P$. 
Define three unit 3-vectors along $e$: an {\em edge vector} $\vec{E}_e$, a {\em slope vector} $\vec{S}_e$, and a {\em normal vector} $\vec{N}_e$. Given an edge $e$ of $P$, $\vec{E}_e$ is the unit vector pointing along $e$ in counter-clockwise direction around $P$; $\vec{S}_e$ is the unit vector orthogonal to $\vec{E}_e$ lying above the interior of $P$ and making an angle of $\pi/4$ with the $xy$-plane; and $\vec{N}_e = \vec{E}_e\times \vec{S}_e$. The {\em edge slab} of an edge $e$ is defined by $\{p + t \vec{S}_e\,|\,p\in e, t\geq 0\}$. Let $u$ be a reflex vertex of $P$ and $M_u$ be its motorcycle in $MG(P)$, $c_u$ be the crash site of $M_u$, and $t_u$ be the crash time. Lift $c_u$ into ${\bf R}^3$ to obtain $\bar{c}_u$ by augmenting $t_u$ as its $z$-coordinate. Let $e$ be an edge of $P$ incident $u$. Then the {\em motorcycle slab for $u$ with respect to $e$} are the points $\{p + t\vec{S}_e\,|\,p\in(u, \bar{c}_u), t\geq 0\}$ where $p$ is on the line segment $(u, \bar{c}_u)$. We call $(u, \bar{c}_u)$ the {\em lifted motorcycle track}. See Fig.~\ref{fig:sskel0}e, f. Each point on a slab can be written as a linear combination of the slab's slope and edge vectors. In other words if $p$ is a point on a slab $s$ with base edge $e$, then $p$ can be written as $a \vec{E}_e + b \vec{S}_e$ for some $a, b\in \R$. We call $(a, b)$ the {\em local coordinates} of $p$ in $s$. As a shorthand we treat $\vec{S}_e$ as the {\em (local) vertical axis} for a slab and $\vec{E}_e$ as the {\em (local) horizontal axis}. 

\paraskip{The structure $\slabs(P)$.} Each edge $e$ has one edge slab and for both of its endpoints it has a motorcycle slab if the endpoint is reflex. All slabs for $e$ are contained in the plane through $e$ with normal $\vec{N}_e$. As in \cite{Huber:2011kr}  we simplify the notation by referring to the union of the edge slab and any motorcycle slabs for an edge $e$ as {\em the slab for $e$}, denoted $\slab(e)$. See Fig.~\ref{fig:sskel0}g. We denote the set of slabs for all edges of the polygon by $\slabs(P)$ (i.e. $\slabs(P) = \{\slab(e)\,|\,e\in P\}$). The lower envelope of $\slabs(P)$ is given by keeping the part of each slab which is lower (in terms of $z$-coordinate) than all other slabs. In \cite{Cheng:2007go} it is shown that (1) $R(P)$ is equivalent to the part of the lower envelope of $\slabs(P)$ which projects orthogonally onto the interior of $P$ and (2) the face with base edge $e$ can be defined as the lower envelope in the direction of $\vec{S}_e$ in the plane supporting $\slab(e)$ of the line segments given by intersecting all other slabs with $\slab(e)$. We call (2) the {\em local (2D) definition} for a face of the straight skeleton roof and use these two characterizations in the remainder of the paper. Each face of the straight skeleton roof is monotone with respect to the base edge of its supporting slab, and its boundary is the union of two monotone chains. Furthermore, the lower monotone chain, which includes only the base edge, is convex. \looseness=-1

\paraskip{Planar straight line graphs.} The slab based roof definition was extended to PSLGs in \cite{HuberH12}. Vertices may now have degree different from 2. It is convenient to apply the following operation so that the boundary components of each face are combinatorially simple polygons (all vertices of degree 2): for each face compute a walk of the edges of each connected boundary component of the face. Each time a vertex or edge is visited by such a walk it is duplicated so that the walk is combinatorially simple. For any degree 1 vertex encountered on the walk, add a small zero-length edge, and for the purposes of the induced motorcycle graph, consider that it makes an angle of $\pi/2$ with its two incident edges. This gives rise to the same induced motorcycle graph as before, and each reflex vertex is incident to a single motorcycle edge on the interior of the face. The slab for such a zero-length edge is made up of the union of its two motorcycle slabs. 

\paraskip{Vertical Slab.} Given a line, ray, or line segment $l$ in the $xy$-plane, we define its {\em vertical slab} $H(l)$ to be the set of points $(x, y, z)$ in ${\bf R}^3$ where $(x, y)\in l$ and $z\geq 0$. This concept is used in our algorithm for PSLGs. 

\smallskip
\paraskip{Assumptions.} We assume real-RAM computation. For now we also assume that the input polygon or PSLG is {\em non-degenerate}, meaning no two motorcycles crash simultaneously and in {\em general position}, meaning  that the intersection of any two slabs is either empty or a line segment and no four slabs meet at a point. In Sec.~\ref{sec:generalposition} we show how to remove these assumptions while maintaining the same time bounds.

\section{Partial roofs for subchains of simple polygons}\label{sec:partialroofs}
\begin{figure}
\centering
\subfloat[]{\includegraphics[width=0.2\textwidth]{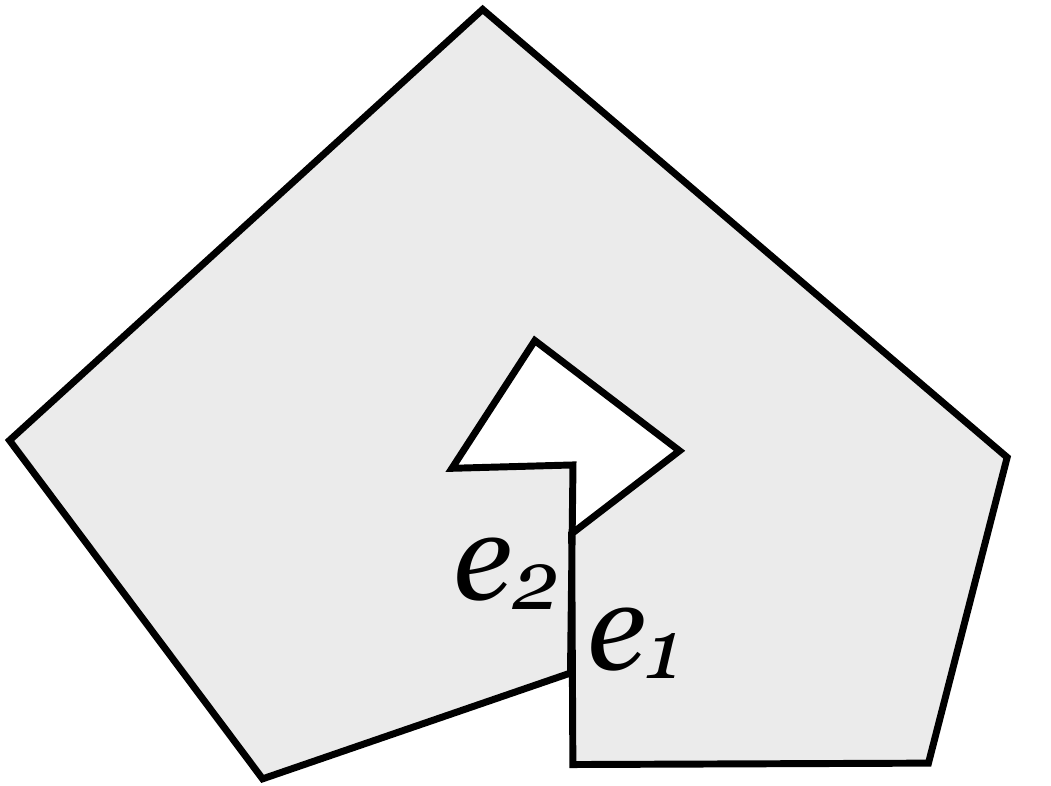}}
\qquad
\subfloat[]{\includegraphics[width=0.25\textwidth]{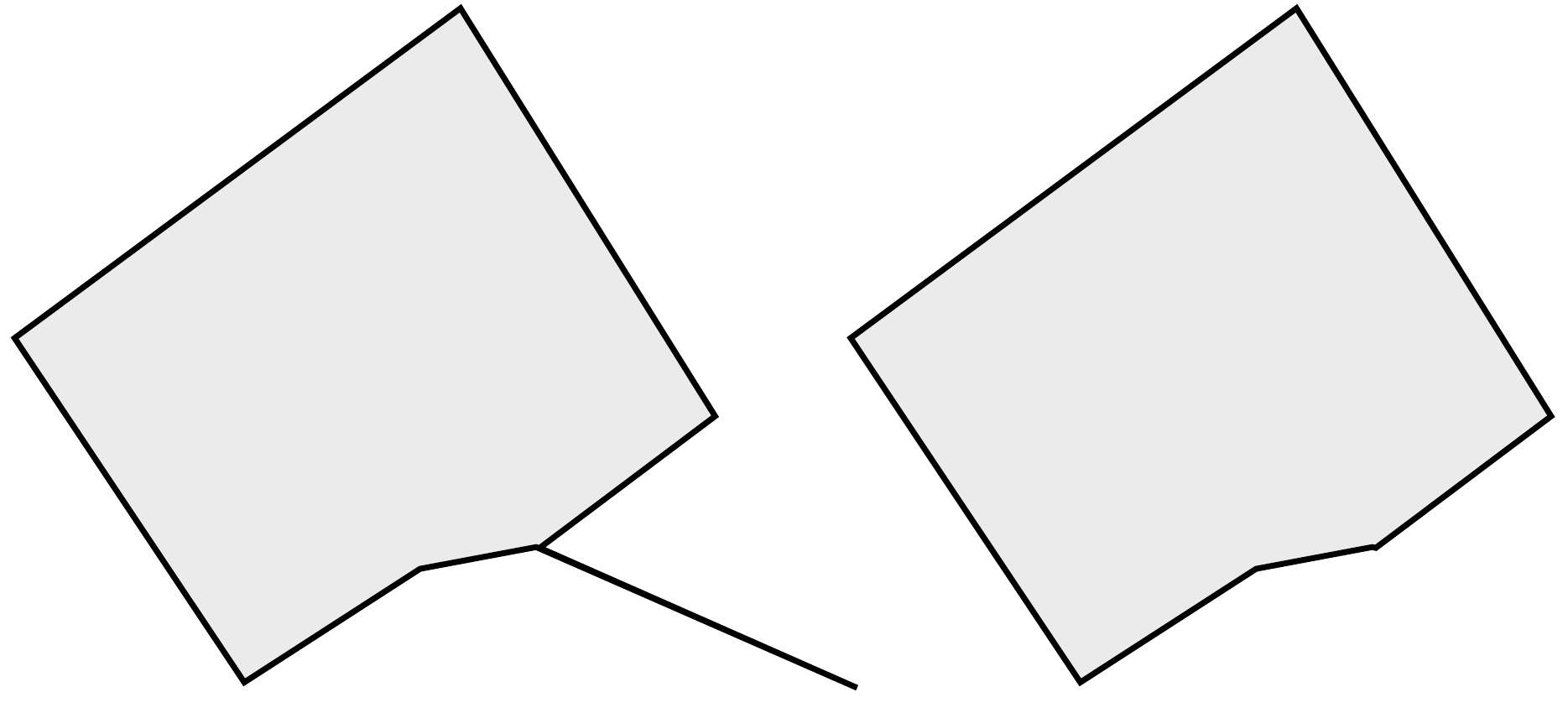}}
\qquad
\subfloat[]{\includegraphics[width=0.4\textwidth]{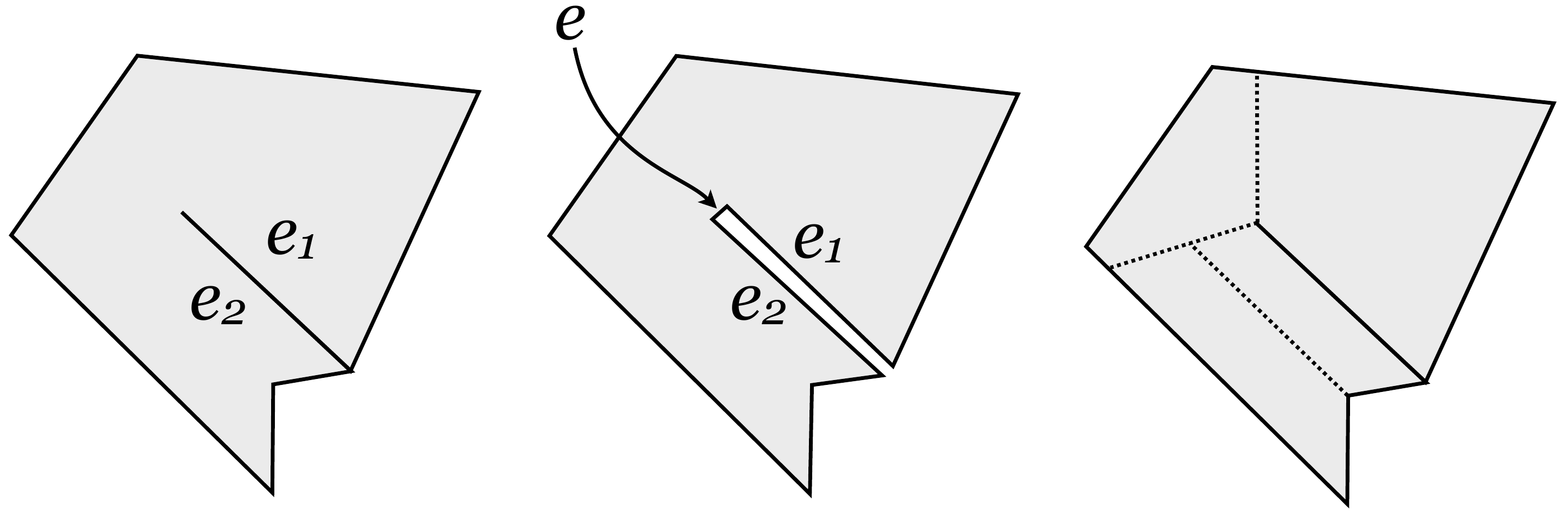}}
\vspace{-4pt}
\caption{\small{Weakly simple polygons. (a) The two edges $e_1$ and $e_2$ overlap, but we consider them disjoint. (b) The left polygon has two edges which have an interior angle of $0$. We handle this by restricting $P$ only to the parts incident to the interior (right). (c) Since we treat the polygon as a walk, the apparent single edge on the interior is represented by two edges $e_1$ and $e_2$ making a sharp turn of $2\pi$ (left). We handle this by adding a little zero length edge $e$ between the two (shown here with positive length for visualization). The induced motorcycle graph is shown on the right.}}
\label{fig:simple}
\end{figure}

\vspace{-6pt}
We now show how to compute the straight skeleton roof $R(P)$ for a (weakly) simple polygon $P$. By ``weakly simple'' we mean that the interior of $P$ is topologically a disk and each part of the boundary of $P$ is incident to the interior. Figure~\ref{fig:simple} shows several examples of weakly simple polygons. This allows, for instance, two boundary edges to coincide as long as the interior of $P$ is on opposite sides of the two edges. See Fig.~\ref{fig:simple}a. We consider such edges and their corresponding slabs disjoint, meaning in particular that when we test for the intersection of two edges of $P$, the result is either a vertex of $P$, if the two edges are consecutive along $P$, or is {\em empty}. We assume that $P$ is given by a walk along the boundary of its interior, meaning in particular that we have no vertices of degree 1. If the polygon has sharp turns, meaning a vertex of degree $2\pi$, then we replace the vertex with a zero-length edge which, for the purposes of the induced motorcycle graph, makes right angles with its incident faces. Such an edge produces two motorcycles in the induced motorcycle graph, and the slab for such an edge is the union of its two motorcycle slabs. See Fig.~\ref{fig:simple}c. This allows us to treat such cases generically, rather than as special cases.

\paraskip{Overview.} A straightforward divide and conquer approach for computing $R(P)$ is to subdivide $P$ into equal length chains $C_1$ and $C_2$, recursively compute the lower envelopes of their defining slab sets and merge the result. However, the combinatorial complexity of the lower envelope of the slabs in a chain may be $\Omega(n^2\alpha(n))$\footnote{Where $\alpha(n)$ denotes the inverse Ackermann function.} \cite{ed1989} and finding all intersections between two lower envelopes is non-trivial. But, not all intersections between the lower envelopes of $\slabs(C_1)$ and $\slabs(C_2)$ appear in the final roof $R(P)$. For instance, an edge of $C_1$ is associated with only one face of $R(P)$ but its slab may appear as multiple faces in the lower envelope of $\slabs(C_1)$. This motivates our definition of a {\em partial roof}: the main idea is to define an intrinsic surface which has edges along all intersections that will eventually be part of the final roof, {\em but may not have edges for all intersections}. We then merge partial roofs of subchains by computing a path of local intersection between the two, reminiscent of Shamos and Hoey's Voronoi diagram algorithm \cite{Shamos:1975:CP:1382429.1382488}.

\paraskip{Extending $\R^3$.} We extend $\R^3$ with points at infinity, each of which is given by an equivalence class of vectors with the same unit vector. This allows us to conveniently represent slabs and parts of slabs as polygons. For example, a slab bounded by a single edge and two rays becomes a triangle with one vertex at the point at infinity equivalent to the slab's slope vector\footnote{In geometric group theory this is known as extending $\R^3$ by the {\em visual boundary}.}. We call a polygon with a point at infinity {\em unbounded}.\looseness=-1

\paraskip{Partial roof: definition.} A {\em partial roof} $R$ for a $k$-length subchain $C$ of a polygon $P$ is a piecewise linear surface, topologically a disk, with $k$ faces (one for each slab of $\slabs(C)$) and satisfies four properties (defined below): face monotonicity, face containment, edge containment, and the boundary property. The geometry of each face is defined by a simple (possibly unbounded) polygon on its supporting slab and two faces may be glued together along an edge which lies on the intersection of the two supporting slabs. We denote the boundary by $\partial R$.\looseness=-1

\begin{wrapfigure}{l}{0.6\textwidth}
\vspace{-26pt}
\centering
\subfloat[]{
	\begin{minipage}[c][1\width]{0.13\textwidth}
		\centering
		\includegraphics[width=\textwidth]{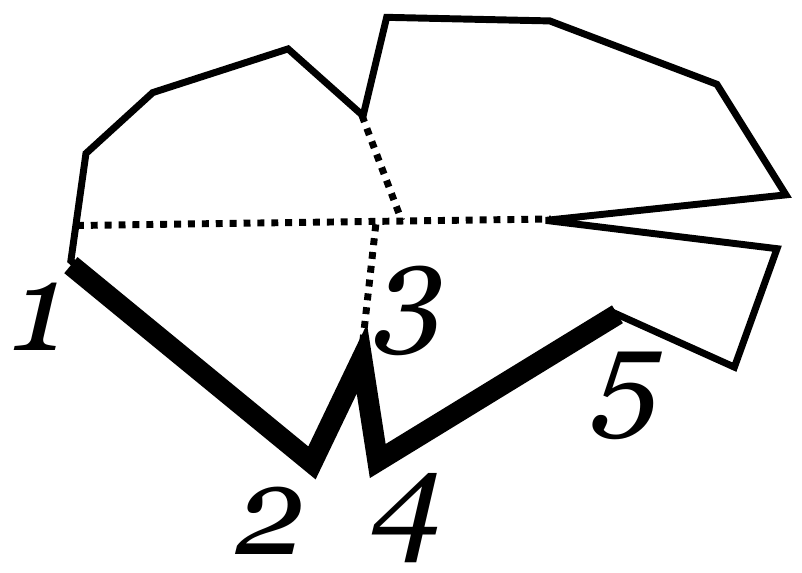}
	\end{minipage}
}
\subfloat[]{
	\begin{minipage}[c][1\width]{0.15\textwidth}
		\centering
		\includegraphics[width=\textwidth]{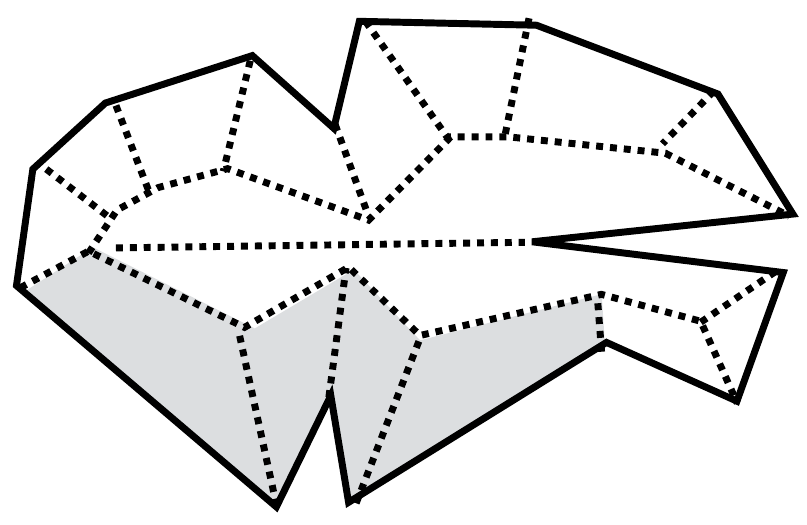}
	\end{minipage}
}
\subfloat[]{

	\begin{minipage}[c][1\width]{0.15\textwidth}
		\centering
		\includegraphics[width=\textwidth]{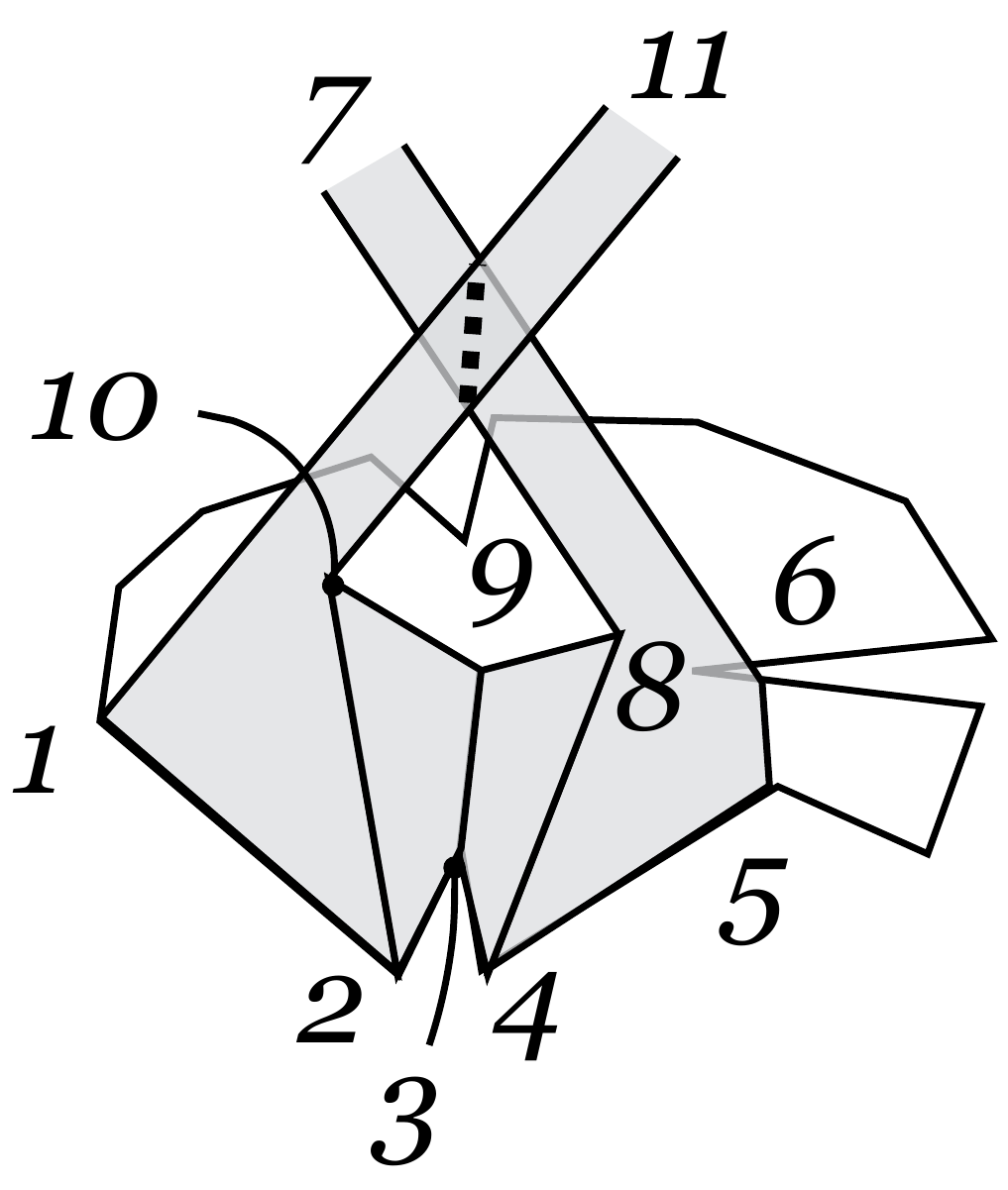}
	\end{minipage}
}
\subfloat[]{

	\begin{minipage}[c][1\width]{0.13\textwidth}
		\centering
		\includegraphics[width=\textwidth]{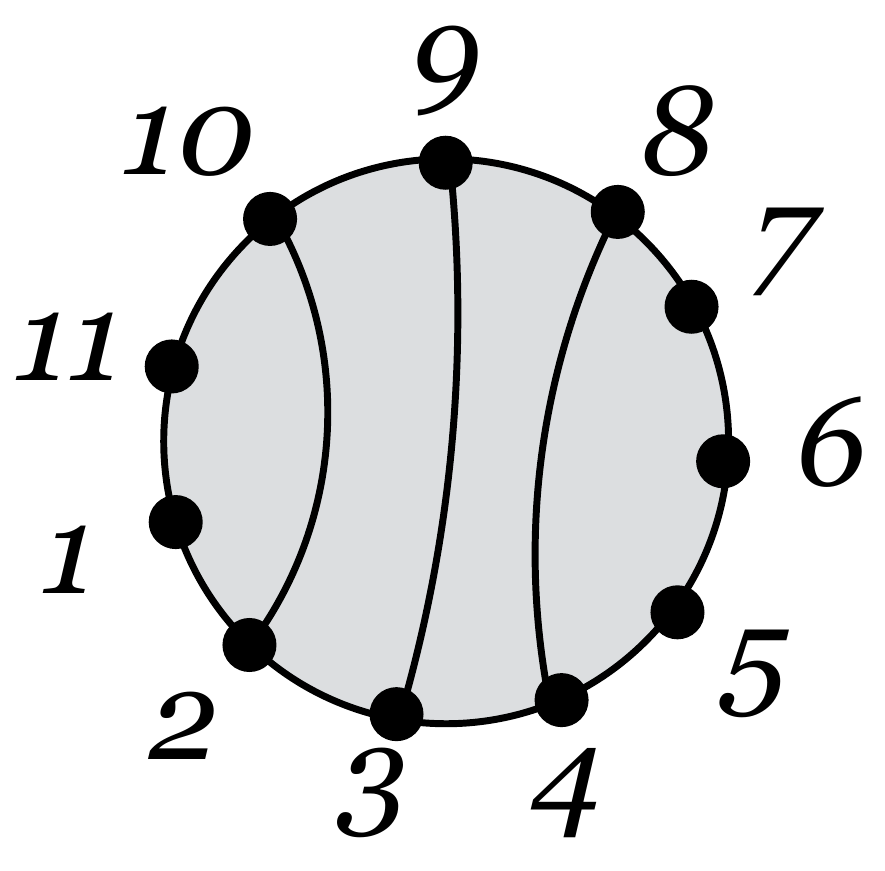}
	\end{minipage}
}
\vspace{-6pt}
\caption{\small{(a) A polygon, induced motorcycle graph (dotted), and a 4 edge subchain (bold). (b) The straight skeleton with the faces incident to the subchain shaded. (c) A partial roof for the subchain from (a). The vertices $7$ and $11$ correspond to points at infinity. (d) The combinatorially representation of the underlying surface as a disk. {\em Note:} the dotted line in (c) shows an intersection between the realizations of the left and right-most faces that is not part of the underlying surface.}}
\vspace{-14pt}
\label{fig:sskel}
\end{wrapfigure}

\paraskip{Intrinsic and extrinsic properties.} An important distinction is between the underlying {\em intrinsic surface}, which is given by the local geometry of each face and how the faces are glued together along edges, and the {\em extrinsic realization} of the surface, which is what the surface ``looks like'' in 3D. It is helpful to think of a partial roof as defined by cutting out each face from its supporting slab independently of the other faces and then gluing the faces back together along certain edges. In doing this we temporarily forget where each face sits in $\R^3$: each face simply has its own local geometry and the neighboring faces it is glued to. A realization is given by mapping back the vertices, edges, and faces into $\R^3$ in such a way that the local geometry of each face and its incidence with faces it is glued to is respected. Since the geometry of each face is defined on the surface of a slab residing in $\R^3$, we define the canonical realization of the surface, which we refer to as {\em the (canonical) realization} and denote by vertical bars $|\cdot|$, to be the one mapping each vertex, edge, and face back onto the original position on the slab that it was cut out from. It is important to note, however, that other realizations exist. (Think of ``folding'' the surface along its edges. This changes how the surface is situated in $\R^3$ and possibly introduces self-intersections, but does not change the underlying surface.) Note that a surface may exhibit self-intersections in a particular realization that are not present in the underlying intrinsic surface\footnote{A familiar example is that of a Klein bottle, which always locally looks (to any observer sitting on the surface) like a patch of the Euclidean plane, even though in any realization of the surface in $\R^3$ there is a self-intersection.}. A partial roof is intrinsically a disk, even though it may exhibit self-intersections in $\R^3$ which make the realization (if we forget the underlying intrinsic surface) something more topologically complicated. See Fig.~\ref{fig:sskel}c, d.

\paraskip{Properties.} The {\em face monotonicity property} is that each face is a (simple, possibly unbounded) polygon that is monotone with respect to the base edge of its supporting slab. The {\em face containment property} is that the realization of each face geometrically contains the final face for that slab in the final straight skeleton roof. The {\em edge containment property} is that if there is an edge $e$ of the final straight skeleton roof between slabs $s_1$ and $s_2$, and $s_1, s_2\in \slabs(C)$, then there exists an edge between the faces of $R$ supported by $s_1$ and $s_2$ whose realization geometrically contains $|e|$. 

\begin{figure}
\vspace{-10pt}
\centering
\includegraphics[width=0.75\textwidth]{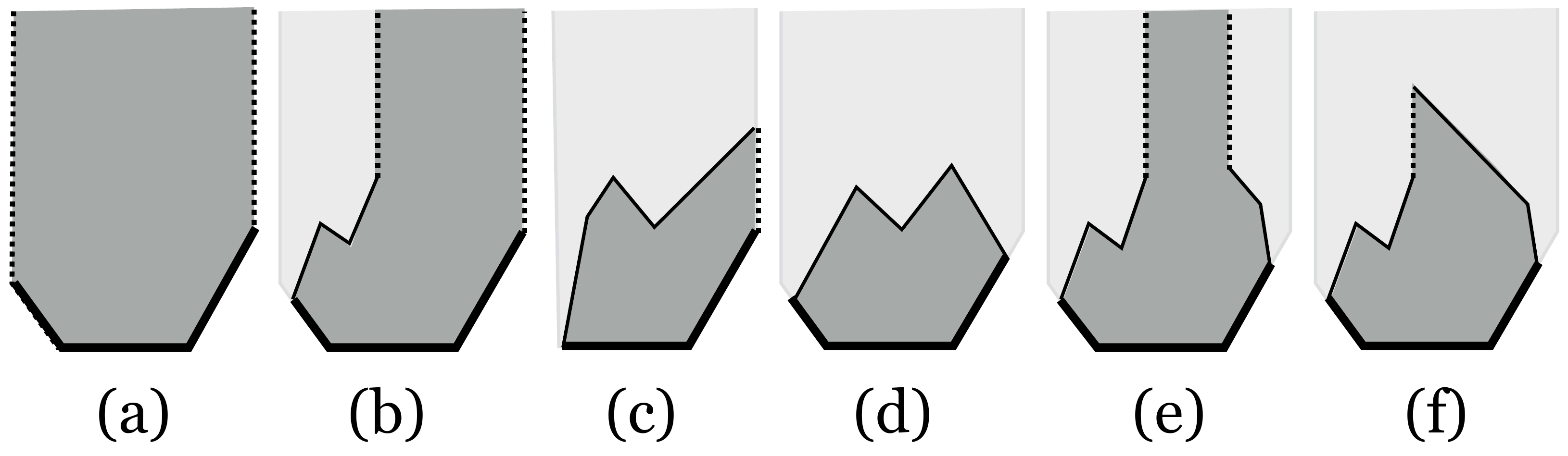}
\vspace{-6pt}
\caption{\small{Examples of possible faces (dark gray) defined on their supporting slabs (light gray). The slabs are oriented so that the base edge is the horizontal edge at the bottom. Thick lines denote the slab border chain; thin lines denote the interior chains; and dotted lines denote the slope edges.}}
\vspace{-14pt}
\label{fig:facetypes}
\end{figure}

\paraskip{Faces and the boundary property.} Each face is a (possibly unbounded) polygon defined on the surface of its defining slab by at most two chains we call its {\em interior chains} which are monotone with respect to the slab's base edge. There are six distinct types of faces that are illustrated in Fig.~\ref{fig:facetypes}. Each interior chain starts at a {\em base point} and ends at a {\em terminal point}. The base point starts either on the left (right) motorcycle edge, if it exists, or left (right) endpoint of the base edge otherwise. Each interior chain is a monotone chain moving rightwards (leftwards). If there are two interior chains, then one starts on the left motorcycle edge or base edge endpoint and moves rightwards, and the other starts on the right motorcycle edge or base edge endpoint and moves leftwards. The two chains may only overlap vertically at their terminal points (i.e. the union of the two chains is monotone except possibly at the terminal points). The face is defined by shooting a ray from the terminal point of each interior chain upwards along the slab's slope vector. This subdivides the slab into (possibly unbounded) polygons, and the face is the polygon containing the base edge. In Figure~\ref{fig:facetypes} (a) illustrates a face defined by no interior chains, in this case the face is simply the entire slab. (b)-(d) illustrate faces defined by one interior chain, and (e) and (f) illustrate faces defined by two interior chains.  In each case the thick black lines denote the parts of the face lying along the base and motorcycle edges, the thing black lines denote the interior chains, and the dotted lines denote the edges lying along the rays extended from each terminal point. In (b) the terminal point of the chain is somewhere on the interior of the face, in (c) the terminal point lies on the boundary of the slab, in (d) the terminal point lies on the opposite motorcycle edge. Note that in (c) the chain begins at the left endpoint of the base edge, since there is no left motorcycle edge. (e) and (f) illustrate faces defined by two interior chains. In (f) the terminal point of the right chain lies directly above the terminal point from the left chain.

Each face is made up of at most four chains: the (at most) two interior chains, the chain of edges lying along the base and motorcycle edges of the slab, which we call its {\em slab border chain} (drawn as thick lines in Fig.~\ref{fig:facetypes}), and the (at most) two {\em slope edges} of the {\em slope chain} (drawn as the dotted lines in Fig.~\ref{fig:facetypes}). The {\em boundary property} is (1) that the boundary edges of each partial roof are formed by two distinct chains: a {\em defining chain} containing the base edges of each face, and a {\em fringe chain} containing the remaining edges and (2) for a given face each edge on its interior chain is internal in the partial roof, each slope edge lies on fringe of the partial roof, and each motorcycle edge of the slab border chain are on the fringe of the partial roof if and only if the defining chain of the partial roof ends at the base edge endpoint incident to the motorcycle edge. 


\paraskip{Discussion.} The face containment property ensures that every face is large enough that it contains the final face for its slab, so that the merge operation can ``cut down'' each face until it eventually becomes equal to the final face. The edge containment property (we will see below) ensures that when two slabs intersect {\em and we need to know about the intersection}, the intersection is represented by an edge of the partial roof. This is crucial, and is the reason we can forget about the other intersections. The basic idea is this: if an intersection between two faces exists geometrically on the realization of a partial roof but is not an actual edge of the underlying intrinsic surface, then by the edge containment property {\em no edge} along that intersection exists in the final roof. Thus, if we are merging two partial roofs along a path that hits one of these non-edge intersections, at that point we no longer care exactly how the partial roofs are represented--we are now on parts of faces that will eventually be cut away and are not part of the final roof. The face monotonicity property is used to bound the complexity of the merge operation. Finally, the boundary property allows us to bound the combinatorial complexity of the partial roof. 

\begin{lemma}[Linear complexity of partial roofs]\label{lem:partialroofcomplexity}
The combinatorial complexity of a partial roof for an $k$-length subchain of a simple polygon is $O(k)$. 
\end{lemma}

\pf The internal edges of $R$ form a forest (otherwise there would be a cycle of internal edges, contradicting that each face is incident to $\partial R$ along its base edge). We are going to ensure that any vertex of the forest lying on the boundary is a leaf node, which guarantees that all non-leaf nodes of the forest have degree at least 3. This is violated when a face $f$ is incident to the boundary at a vertex $v$ but the two edges of $f$ incident to $v$, $e_1$ and $e_2$ are internal to $R$. Conceptually split the tree at each such a vertex: replace any such vertex $v$ with a zero-length dummy edge between two vertices $v'$ and $v''$ such that $e_1$ is incident to $v'$ and $e_2$ is incident to $v''$. Having done this for all such vertices in every incident face, any vertex of the forest incident to $\partial R$ is now a leaf node and all leaf nodes of the forest lie on $\partial R$ (since each face of $R$ is simple). We now show that the number of leaves in this forest is $O(k)$, which bounds the number of internal edges and vertices of $R$ by $O(k)$. The result then follows from the fact that each face has at most $O(1)$ edges on $\partial R$ (by the boundary property) and $R$ has exactly $k$ faces. Since each face has at most $O(1)$ edges incident to $\partial R$, aside from the dummy vertices added above, there are $O(k)$ vertices on $\partial R$. We now bound the number of added dummy vertices. Assume that a face has two vertices $v_1$ and $v_2$ that are replaced by dummy vertices. First: neither $v_1$ nor $v_2$ are incident to the defining chain, since this would imply that they are endpoints of the base edge, and the base edge is incident to the defining chain, a contradiction. Now, since both $v_1$ and $v_2$ lie on the fringe and the face $f$, there must be a chain of interior edges incident to $f$ between $v_1$ and $v_2$. But this implies that there exists a face not incident to the base chain, a contradiction. Thus any face has at most one vertex that is replaced with a zero-length dummy edge. After replacing all such vertices, we are still left with $O(k)$ vertices on the boundary, completing the proof. \qed
%
%

We now show that the only object which meets the definition of a partial roof for the entire polygon is the final roof $R(P)$:

\begin{lemma}[A partial roof of the entire polygon is the straight skeleton]\label{lem:partialrooftoroof}Let $P$ be a simple polygon, $R(P)$ be its straight skeleton roof, and $R$ be a partial roof for $P$. Then $R(P)=R$.\end{lemma}
\pf Let $e_1$ be an edge of $P$. Then $e_1$ is the base edge of a face $f_1$ in $R$ and a face $f_1'$ in $R(P)$. 
We first claim that for each edge of $f_1'$ there is a corresponding edge of $f_1$ which is equal to it (in ${\bf R}^3$). Let $e'$ be any edge of $f_1'$ which is not the base edge. Then there is a second face $f_2'$ of $R(P)$ incident to $e'$. Denote its base edge by $e_2$ and let $f_2$ denote the corresponding face in $R$. By the edge containment property there must exist an edge $e$ in $R$ which is incident to both $f_1$ and $f_2$ such that $|e|$ contains $|e'|$. Further, if $|e|$ strictly contains $|e'|$, then $f_1$ is not simple because the edges incident $e'$ also have corresponding edges in $f_1$ that contain them and one must be crossed by $|e|$, a contradiction. Thus $|e| = |e'|$. It follows that the faces $f$ and $f'$ are identical.\looseness=-1 \qed

\section{Merging partial roofs for a simple polygon}\label{sec:mergeop}

\begin{figure}
\vspace{-30pt}
\centering
\subfloat[]{\includegraphics[width=0.21\textwidth]{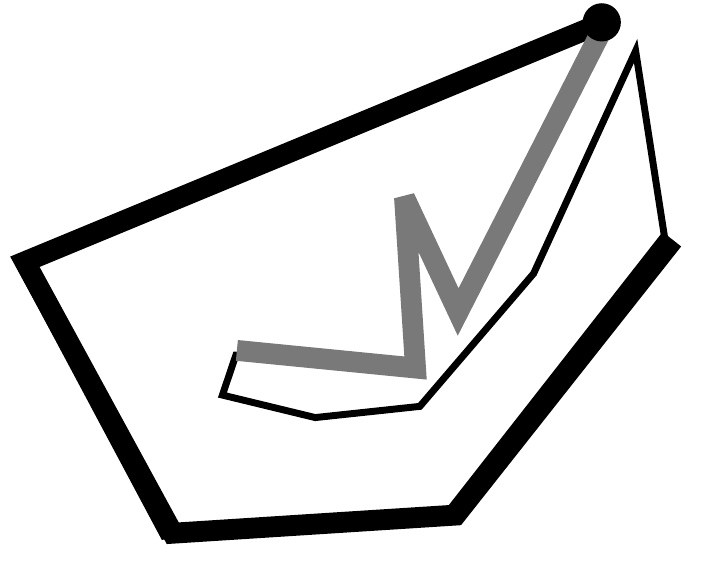}}
\subfloat[]{\includegraphics[width=0.35\textwidth]{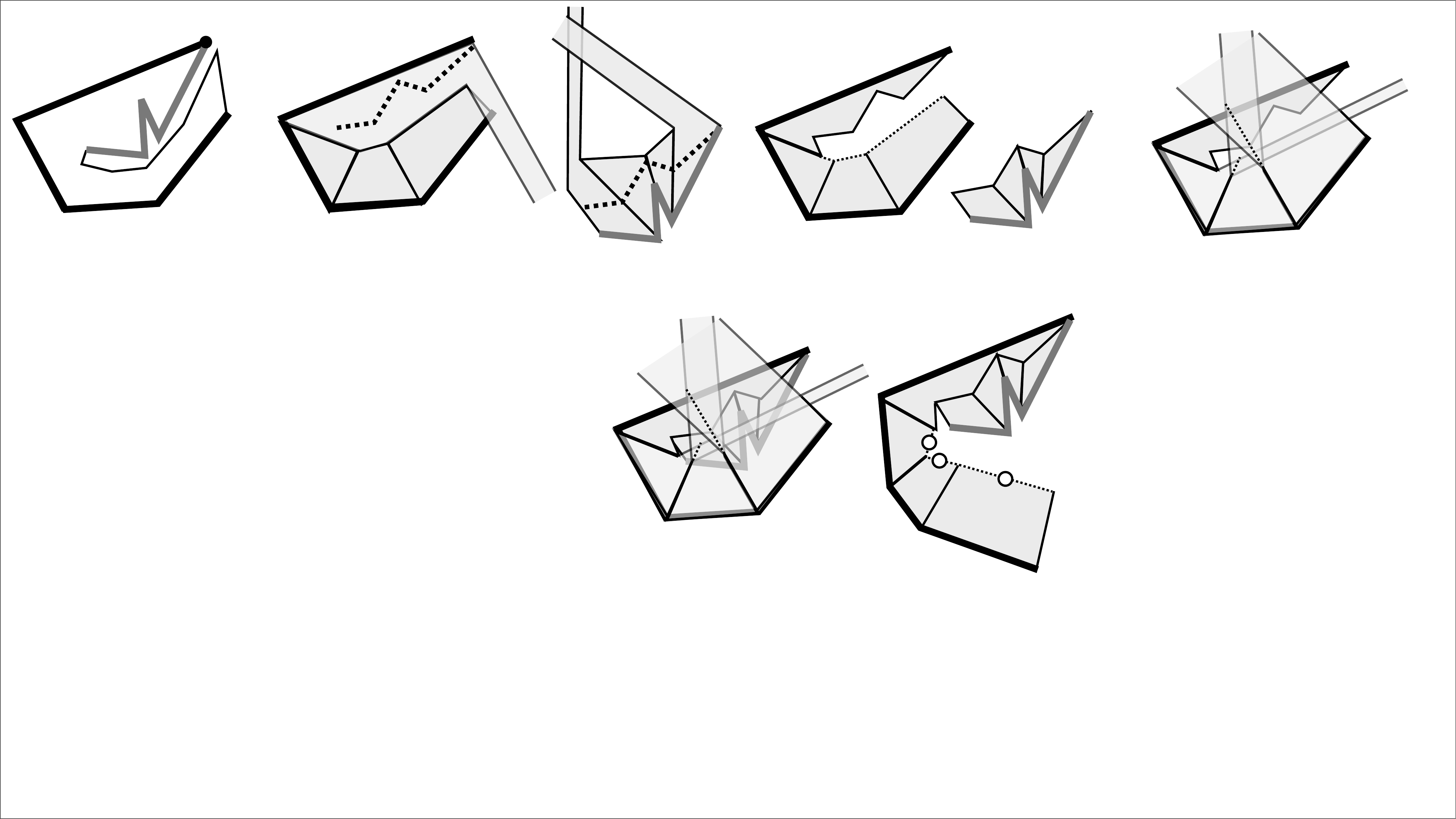}}
\subfloat[]{\includegraphics[width=0.2\textwidth]{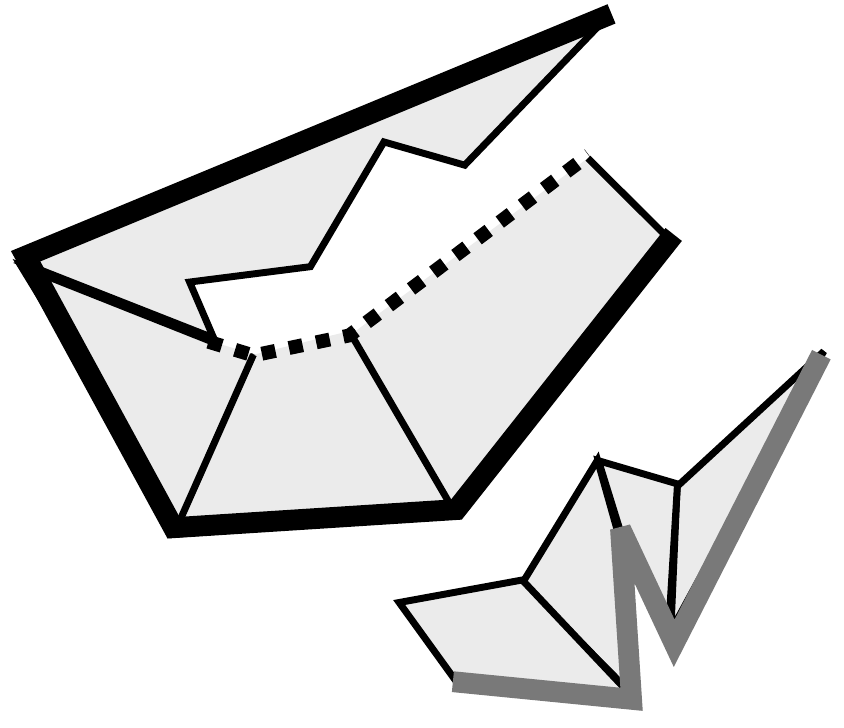}} \\
\subfloat[]{\includegraphics[width=0.25\textwidth]{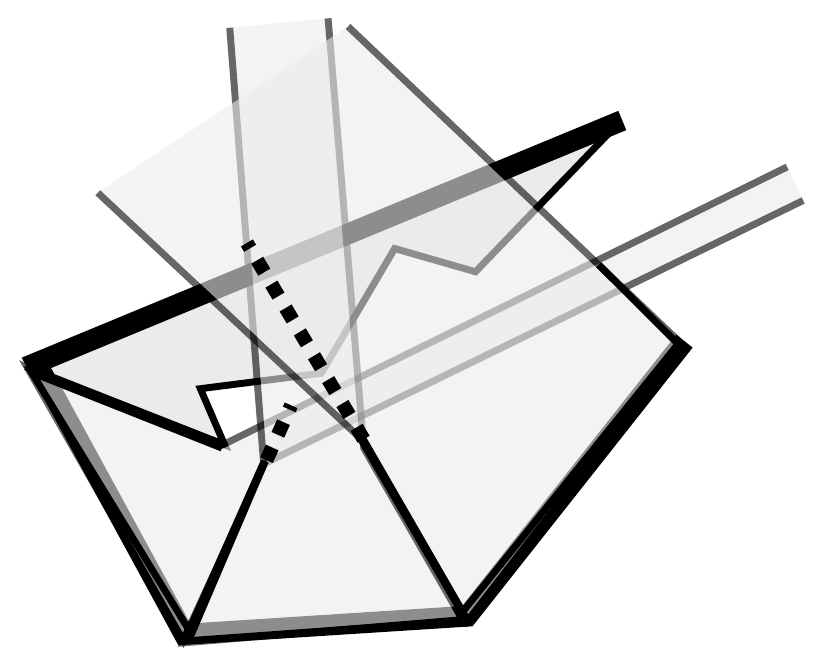}}
\subfloat[]{\includegraphics[width=0.24\textwidth]{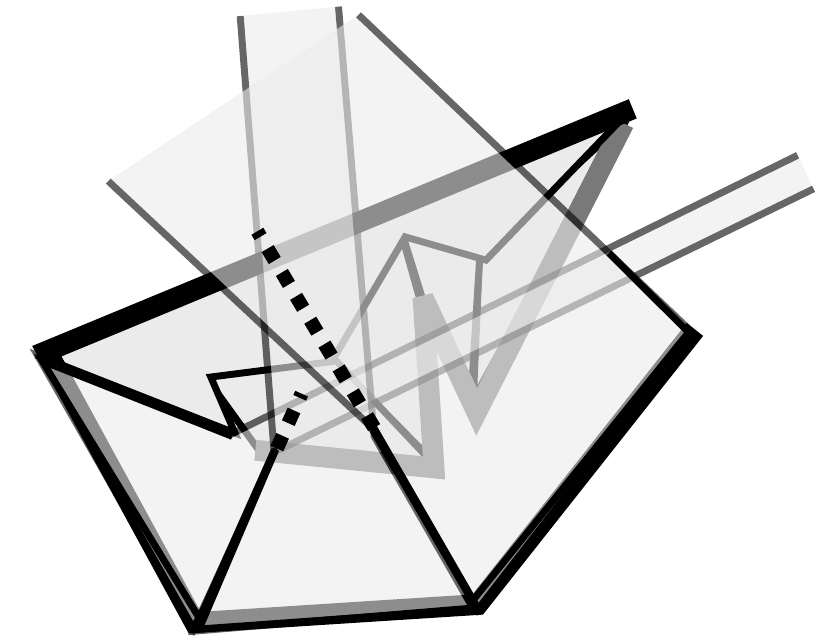}}
\subfloat[]{\includegraphics[width=0.15\textwidth]{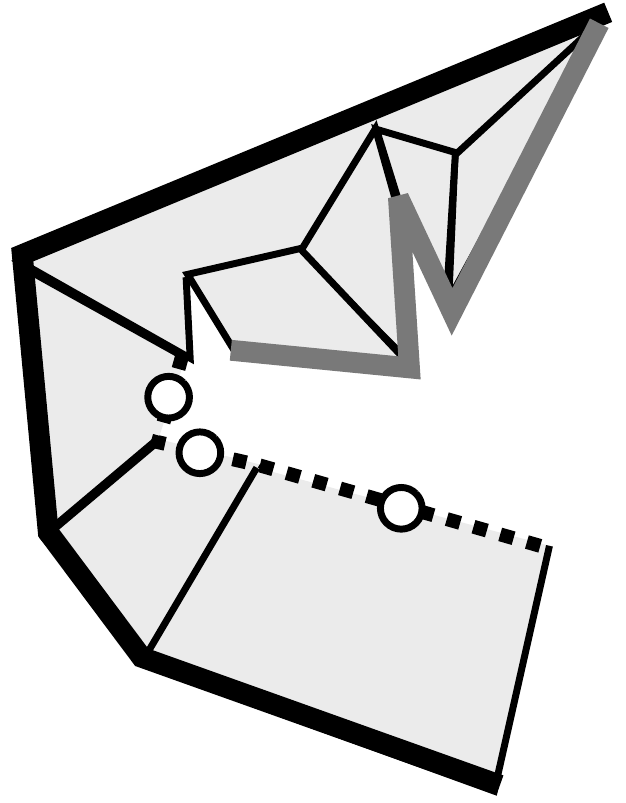}}
\vspace{-4pt}
\caption{\small{The steps of the merge operation. (a) A polygon with two subchains denoted by thick black and gray lines. (b) Partial roofs for the subchains and the splicing path between them (dotted line). Note that in reality the partial roofs overlap each other in $\R^3$, but are depicted here separately for visualization purposes. (c) Cutting along the splicing path. The dotted boundary edges on the roof of the black chain require fringe simplification since they were previously interior edges but now lie on the fringe. (d) The fringe simplification step removes these edges from each face's interior chains, replacing them with unbounded edges. The dotted lines represent intersections between faces that are not part of the underlying intrinsic surface. (e) Gluing the two together along the splicing path. (f) A view of the topology of the underlying intrinsic surface. The white vertices represent vertices at infinity. Note that (e) shows that the three bottom faces of (f) intersect each other, but this intersection is not part of the intrinsic surface.}}
\label{fig:mergeop}
\vspace{-10pt}
\end{figure}

\paraskip{Procedure.} The merge operation takes as input two partial roofs $R_1$ and $R_2$ for co-incident subchains of a simple polygon and produces a partial roof $R$ for the the combined subchain. The basic idea is to start at the {\em gluing vertex} common to both chains and compute a walk of each surface which {\em locally} lies on the intersection of $|R_1|$ and $|R_2|$. The purpose of this walk is to detect all intersections between $|R_1|$ and $|R_2|$ which must exist as edges in $R$ to satisfy the edge containment property. {\em This may detect other intersections between $|R_1|$ and $|R_2|$ but will not necessarily detect all intersections}. We then cut each surface along the path discarding some of the subdivided faces and glue the two surfaces together along the path to form $R$.

More specifically, we (1) compute the {\em splicing path} on each surface by starting at the vertex $\hat{v}$ (which is incident to exactly one face $f_1$ of $R_1$ and one face $f_2$ of $R_2$). We then compute the intersection of $|f_1|$ and $|f_2|$ and walk along this intersection until we hit an edge of either face (say $f_2$). If the edge is not a boundary edge of the surface, we traverse across it to the next edge, say $f_3$, and continue along the intersection of $|f_1|$ and $|f_3|$. It should be noted that this computation is {\em local}, meaning that it ignores self-intersections which may intersect the path in the realization in $\R^3$ that are not represented in the underlying intrinsic surface. The walk stops if we hit a boundary edge of either surface, or we detect that the next edge of the walk is an edge that is provably not required to satisfy the edge containment invariant: i.e. the edge is on an intersection of two slabs that we can prove does not appear in the final straight skeleton roof (more on this below). Once we have computed the splicing path, then (2) we cut each face it traverses along the path which subdivides the face into two. One of the subdivided faces will be incident to the defining chain, and we discard the other. This makes the path a chain of boundary edges, and (3) we glue the two surfaces together along the two corresponding boundary chains. Finally (4) we perform a ``clean-up'' operation to ensure that the boundary property is maintained. Figure~\ref{fig:mergeop} illustrates a single merge. We now give more details on steps (1), (2), and (4). Step (3) is a common operation on piecewise linear surfaces.

\begin{wrapfigure}{r}{0.5\textwidth}
\vspace{-24pt}
\centering
\includegraphics[width=0.4\textwidth]{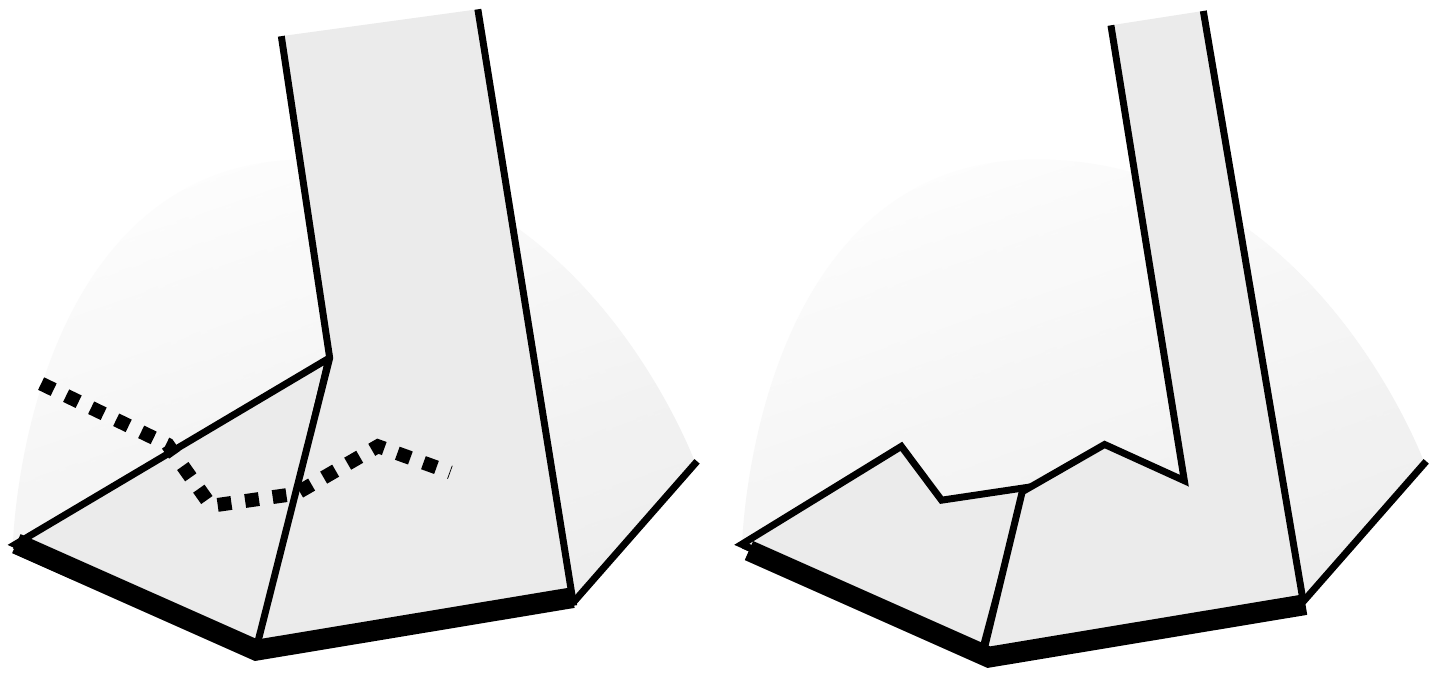}
\vspace{-8pt}
\caption{\small{The subdivision of two faces along the splicing path (dotted). Left: before the subdivision. Right: afterwards. The side containing the base edge (bold) is retained.}}
\vspace{-8pt}
\label{fig:subdividestep}
\end{wrapfigure}

\para{Subdividing the faces.} For most faces the splicing path traverses the entire face, and subdividing the face along the splicing path is well defined. The only special case is the last face encountered. If the splicing path does not simultaneously encounter a boundary edge in both partial roofs, then in one of the roofs, say $R_2$, the last face $f$ encountered by the path is not completely cut into two. Let $x$ be the endpoint of the splicing path in $f$. Intrinsically on f, start at $x$ and trace the ray emanating from $x$ along the slope vector of $\slab(f)$. This either hits an edge of $f$ or escapes to infinity. In the first case, split the hit edge at the hit point by adding a vertex $y$ and subdivide $f$ by $p$ and an edge from $x$ to $y$. Otherwise, $f$ must have an infinite vertex, say $v_\infty$. Split $f$ by cutting along $p$, and then adding an edge from $x$ to $v_\infty$. See Fig.~\ref{fig:subdividestep}. \looseness=-1

\paraskip{Stopping the walk.} The basic idea behind the stopping conditions is that we only need the splicing path to find  edges along the intersection of the two roofs that are necessary to satisfy the edge containment property. In Lemma~\ref{lem:mergeproducespartialroof} we prove by induction that all such edges constitute the first edges along the splicing path and correspond to a simple path of internal edges on the final straight skeleton roof. Because of this, if at any point in the computation of the splicing path we detect that the an edge (or the next possible edge) of the splicing path is provably not needed to satisfy the edge containment invariant, we can stop the splicing path walk. We use properties of the final straight skeleton roof $R(P)$ to detect when we arrive at an edge that provably cannot be an edge of $R(P)$. 

In the following we use properties of a simple path $p$ of interior edges of the final roof $R(P)$ that starts at a boundary vertex $\hat{v}$ to determine the stopping conditions for the splicing path walk. First: since interior edges of the straight skeleton form a tree, it is not possible that two edges of $p$ be incident to the same face such that there is an intermediate edge between the two that is not incident to the face. Thus we stop the splicing path walk if the path re-enters a face it has already traversed. 
 Second: the edges of $p$ incident to a face $f$ that are not motorcycle edges form the upper monotone chain of $f$. Thus, if we detect that adding the next potential edge of the splicing path walk makes the splicing path non-monotone with respect to the base edge of the face it is currently in, we stop. 
 Third: let $f_1$ and $f_2$ be the faces incident to the left and right (resp.) of the last edge of $p$. The base edges of $f_1$ and $f_2$ split $P$ into two subchains. Let $C$ denote the subchain containing $\hat{v}$, $C_1$ denote the part of $C$ from $\hat{v}$ to $\base(f_1)$ and $C_2$ denote the part from $\hat{v}$ to $\base(f_2)$. Since $R(P)$ is a disk, all faces to the left of $p$ have their base edges on $C_1$. Similarly all faces to the right of $p$ have their base edges on $C_2$. Furthermore the interior edges of a straight skeleton roof are valleys if and only if they lie along lifted motorcycle tracks (cf. \cite{Cheng:2007go}). Thus we require (1) that the non-discarded parts of all faces of $R_1$ traversed by splicing path lie to the same side of the path (similarly for $R_2$ on the opposite side) and (2) that a splicing path edge becomes a valley in $R$ if and only if the edge lies along a lifted motorcycle track. If we detect that the next edge added along the splicing path will violate this condition, we stop.

\paraskip{Fringe simplification.} 
Discarding split faces in Step (2) may result in edges from the interior chain of a face becoming boundary edges on $\partial R$, thus violating the boundary property. Let $e_1, \dots, e_m$ denote the edges of a face $f$ that lie on $\partial R$ but are not part of the slab border chain. These edges form a connected chain along the boundary (we prove this in Lemma~\ref{lem:mergeproducespartialroof}). We use this property to perform the following clean-up which ensures that the boundary property is maintained. Let $e_1,\dots,e_m$ be the chain of edges of a face $f$ on the boundary of $R$ that are not slab border edges (i.e. lie along the base edge and motorcycle edges) and $u$ and $v$ be the endpoints of the chain. Each interior vertex of the chain has degree 2. We replace the chain by adding a new vertex $w$ at the point at infinity equivalent to the slope vector of the slab supporting $f$ and swapping out the chain with two edges $\edge{u}{w}$ and $\edge{w}{v}$. Note that this step may introduce self-intersections to the realization of the surface, but all operations are performed on the underlying intrinsic surface, and are simply contractions of boundary chains to shorter boundary chains. Thus, though the realization may not be a disk, the underlying surface remains one. See Fig.~\ref{fig:mergeop}e, f.\looseness=-1

We now investigate several properties used to prove the correctness of the merge operation. 

\begin{lemma}\label{lem:splicingpathdefiningchain}The splicing path does not intersect the defining chain of either $R_1$ or $R_2$.\end{lemma}
\pf Assume that it does, then for some edge of the polygon, there exists a slab that intersects the edge. But each slab is incident to the $xy$-plane only along an edge of $P$, so $P$ is not (weakly) simple, a contradiction.\qed

\begin{lemma}\label{lem:disk}The (intrinsic) surface produced by a merge operation is topologically a disk.\end{lemma}
\pf By definition the splicing path is a simple path on the interior of the (intrinsic) surface $R_1$ (resp. $R_2$) which is incident to the boundary at the gluing vertex (and possibly along a motorcycle edge). Lemma~\ref{lem:splicingpathdefiningchain} implies that if the splicing path traverses all the way to a boundary edge of $R_1$ (which splits $R_1$ into two disks) only one of them will contain the base edges. Thus all faces in the other are discarded. (Similarly for $R_2$.) We now show that discarding the remaining faces maintains that $R_1$ (resp. $R_2$) is a disk. Assume not. Then either the remaining faces form at least two separate connected components, or they form at least two topological disks that are incident only at a vertex. In the first case, since each face is incident to a base edge, the base edges from one component are disconnected from the base edges of the other, so the splicing path must cut the defining chain contradicting Lemma~\ref{lem:splicingpathdefiningchain}. In the second case, if the vertex is on the defining chain, we contradict Lemma~\ref{lem:splicingpathdefiningchain}; if it is not on the defining chain, then again we have that the base edges are dicsonnected. Finally, by the stopping conditions every discarded face lies on the same side of the splicing path, which is opposite to the side that is glued to the other surface. Removing these faces, then, removes a topological disk from $R_1$ which is incident to the boundary and does not touch the defining path. Thus $R_1$ (resp. $R_2$) is a disk after the faces are discarded and the gluing path remains an intact series of edges along the boundary, so $R$ must be (intrinsically) a disk. \looseness=-1 \qed

\medskip
\paraskip{Correctness.}
%
%
%
%
We now prove correctness by showing that $R$ satisfies the properties of a partial roof (Sec.~\ref{sec:partialroofs}):

\begin{lemma}\label{lem:mergeproducespartialroof}
The merge operation correctly computes a partial roof. 
\end{lemma}

\pf  Let $R$ denote the output surface, $R_1$ and $R_2$ denote the input partial roofs, $R(P)$ denote the final straight skeleton roof, and $C$, $C_1$, and $C_2$ denote the defining chains for $R$, $R_1$ and $R_2$. By Lemma~\ref{lem:disk} $R$ is topologically a disk. The face monotonicity property follows directly from the second stopping condition.

\paraskip{Face containment.} Suppose the face containment property does not hold in $R$. Then there exists some slab $s$ such that (without loss of generality) $s\in \slabs(R_1)$ and the face $f$ corresponding to $s$ in $R$ violates the face containment property. Let $f'$ denote the corresponding face of $R(P)$ and $f''$ denote the corresponding face of $R_1$. In particular, this means that $|f|$ does not contain $|f'|$. However, since $R_1$ is a partial roof $|f''|$ contains $|f'|$. Thus the splicing path must have cut through $f$. But the splicing path can only cut along intersections between the slab $s$ and other slabs in $\slabs(R)$. This means that there is an intersection between $|f'|$ and a slab in $\slabs(R)$ which contradicts that $f'$ is a face of $R(P)$. \looseness=-1 

%

\paraskip{Edge containment.} This property has two pieces: an existence claim and a geometric containment claim. The basic idea of the proof of existence is to use induction along the splicing path to show that it contains all of the edges required to satisfy the property. Once we have that, the geometric containment follows the same line of reasoning as face containment above. 
 Let $e'$ be an edge of $R(P)$ incident to faces supported by slabs $s_1$ and $s_2$ such that $s_1$ and $s_2$ are slabs in $\slabs(C)$. Without loss of generality, there are two cases, either $s_1, s_2\in \slabs(C_1)$ or $s_1\in\slabs(C_1)$ and $s_2\in\slabs(C_2)$. 

\paraskip{Case 1:} Since $R_1$ is a partial roof, there is an edge $e$ in $R_1$ incident to the faces $f_1$ and $f_2$ of $R_1$ that are supported by $s_1$ and $s_2$ such that $|e|$ geometrically contains $e'$. For contradiction, suppose that no such edge exists in $R$. Then the splicing path must cut $f_1$ and $f_2$ below $e$ (otherwise $e$ cannot have been discarded) or in such a way that $|e|$ only partially covers $|e'|$. In either case the after cutting $f_1$ and $f_2$ along the splicing path, they no longer maintain the face containment property, a contradiction.\looseness=-1

\paraskip{Case 2:} We claim that the faces $f_1$ and $f_2$ in $R$ supported by $s_1$ and $s_2$ are incident along some edge $e$ and $|e|$ contains $|e'|$. Since the edges of the straight skeleton form a tree, there exists a unique path $p'$ along the interior edges of the straight skeleton roof $R(P)$ from $\hat{v}$ to $e'$. We claim that $p'$ corresponds to the first part of the splicing path $p$. Let $k$ be the length of $p'$. The proof is by induction for $i$ from 1 to $k$.

\paraskip{Base step:} By definition, the first edge of both $p'$ and $p$ is along the intersection of the slabs of the base edges incident to $\hat{v}$. Geometric containment follows the argument as above.\looseness=-1

\begin{wrapfigure}{r}{0.41\textwidth}
\vspace{-24pt}
\centering
\includegraphics[width=0.40\textwidth]{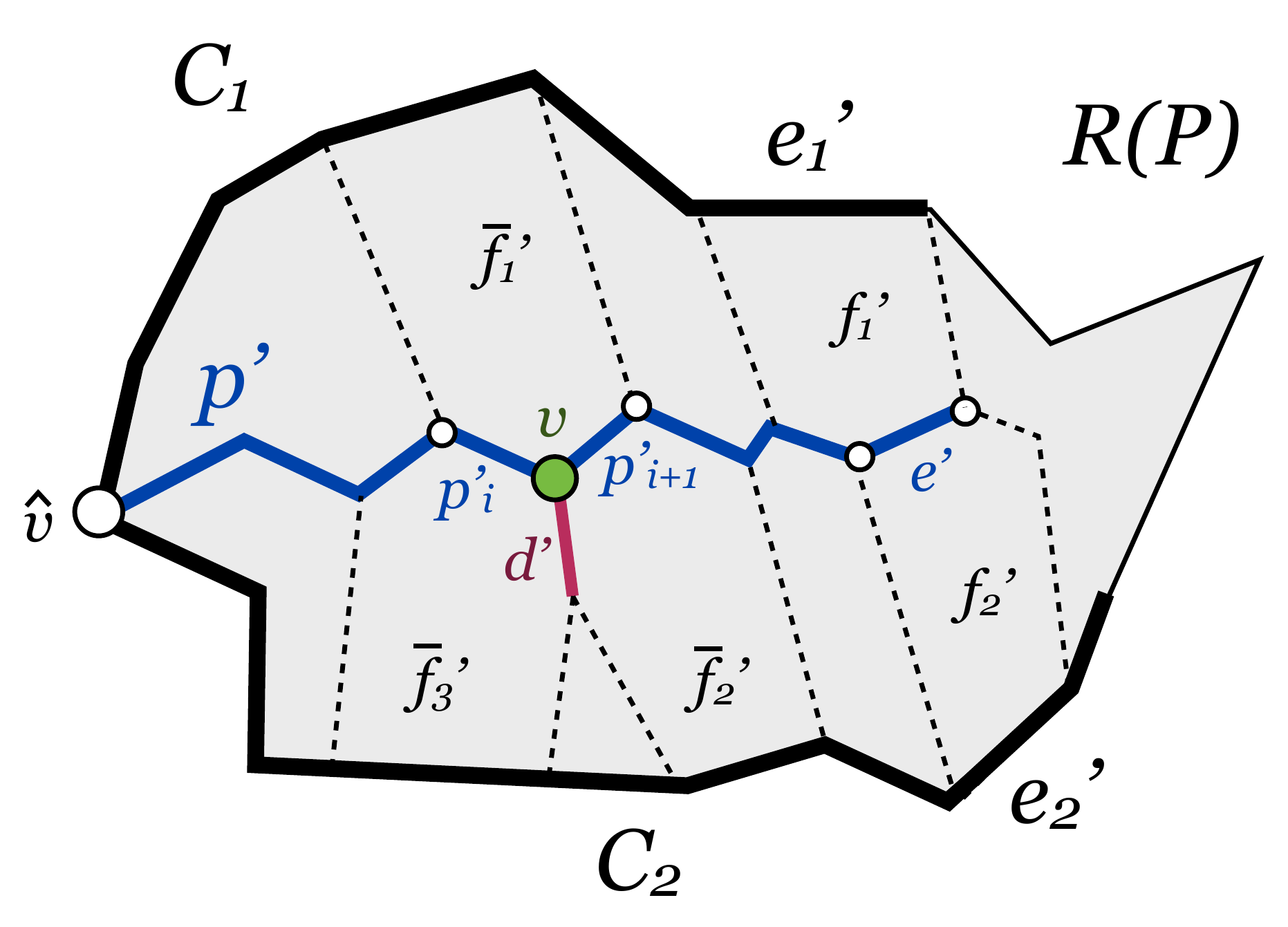}
\vspace{-4pt}
\caption{\small{The setup in $R(P)$ for the inductive step of the edge containment property.}}
\label{fig:prop7}
\vspace{-12pt}
\end{wrapfigure}

\paraskip{Inductive step:} Now assume the claim is true for the first $i<k$ edges of $p'$. Denote the edges of $p$ and $p'$ in order from $\hat{v}$ by $p_1, p_2, \dots$ and $p'_1, p'_2, \dots$, resp. Let $\bar{f}'_1$ and $\bar{f}'_2$ denote the faces incident to $p'_i$ and $\bar{f}_1$ and $\bar{f}_2$ be the faces of $R$ with the same base edges. Then $\bar{f}_1$ and $\bar{f}_2$ are incident along $p_i$ and $|p_i|$ contains $|p'_i|$. We prove that this holds for $i+1$. We first need to prove that $p$ contains an edge with index $i+1$. Let $v$ be the vertex between $p'_i$ and $p'_{i+1}$. By genericity there is one other internal edge, say $d'$, that is also incident to $v$. Denote the faces incident to $p'_{i+1}$ by $\bar{f}'_1$ and $\bar{f}'_2$ such that the base edges are on $C_1$ and $C_2$ (resp.). Without loss of generality assume that $d'$ is incident to $\bar{f}'_2$. Let $\bar{f}'_3$ be the other face incident to $d'$. Since $\bar{f}'_2$ and $\bar{f}'_3$ lie on the same side of $p'$ (by the stopping condition) the base edge of $\bar{f}'_3$ is on $C_2$. By the edge containment property on $R_2$, there is an edge $d$ in $R_2$ which is incident to two faces with base edges equal to the base edges of $\bar{f}'_2$ and $\bar{f}'_3$. Let $\bar{f}_1$, $\bar{f}_2$, and $\bar{f}_3$ be the faces of $R$ with base edges corresponding to $\bar{f}'_1$, $\bar{f}'_2$, and $\bar{f}'_3$ (resp.). By Case 1 above, $d$ is an edge of $R$ between $\bar{f}_2$ and $\bar{f}_3$. Since $d$ borders $\bar{f}_3$ and $|d|$ contains $|d'|$, then the splicing path between $\bar{f}_1$ and $\bar{f}_3$ must hit $d$ at $f_2$. Since $d$ is not a boundary edge, the splicing path continues along the intersection of $\bar{f}_1$ and $\bar{f}_2$. This edge is $p_{i+1}$. Geometric containment then follows by the same proof by contradiction as above. \looseness=-1

\paraskip{Boundary property.} Since the path enters a face along an interior edge, the new edges along the path are part of the interior chain defining the face. However, the cutting and discarding step may have introduced edges to the boundary of $R$ that were previously part of the interior chains. 
We claim that before the fringe simplification step, such edges form a connected chain. Assume not, then there are at least two distinct subchains of such edges incident to the same face $f$ of $R$ that lie on $\partial R$. Since these chains are disconnected, there must be some interior edge of $R$ between them. But that edge must be incident to a face that is incident to a base edge. Since $R$ is a disk, however, this implies that the defining chain is disconnected, which can only happen if the splicing path intersects the defining chain, contradicting Lemma~\ref{lem:splicingpathdefiningchain}. Thus the fringe simplification step is able to find a single connected chain containing all the violating edges and replaces them with a single slope chain, restoring the boundary property. 
\qed

\medskip
\paraskip{Existence.} Existence of a partial roof follows from the observation that for any edge $e$ of a simple polygon $P$, $\slab(e)$ is itself a partial roof for $e$. That a partial roof exists for any subchain $C$ now follows by induction.

\medskip
\para{Running time.} We store each partial roof as a doubly-connected edge list (\cite{guibas:stolfi:85}) which handles most of the operations we need efficiently out-of-the-box. The only non-trivial part is finding the splicing path. 
 The basic idea is that to compute the walk we need to use ray shooting across each face, which is in general an expensive operation. However, due to the monotonicity of each face and the walk we can quickly compute a trapezoidal decomposition of each face in order to accelerate the ray shooting. This gives us:

\begin{lemma}\label{lem:splicingpathcomplexity}The splicing path can be computed in linear time and space.\end{lemma}

\pf Suppose we merge a partial roof $R_1$ for a subchain of length $k_1$ to a partial roof $R_2$ for a subchain of length $k_2$ and let $k=k_1 + k_2$. 
To prove the lemma, we  show that the splicing path has at most $O(k)$ edges, that each potential next edge can be found in $O(1)$ time, and that the stopping conditions can be checked in $O(1)$ time per potential edge. 

At each iteration, the walk lies on one face of $R_1$ and one face of $R_2$. The basic idea is to use the intersection of the realization of the two faces to compute a direction and shoot a ray (intrinsically) across each face to find the first edge of either face hit, then advance the splicing path in both faces along this ray to the closer hit-point. This adds an edge to the splicing path and in one of the partial roofs we cross an edge into a new face. Since we only continue until we hit a boundary edge or a face we have already traversed, the length of the final splicing path is at most $k$ and requires shooting $O(k)$ rays. 

To compute the ray shooting, we exploit the monotonicity of the splicing path across each face (second stopping condition). Subdivide the the faces of both partial roofs into trapezoids by extending chords from each vertex on the interior of each face perpendicular to its base edge. Each internal vertex of a partial roof has degree 3 and thus is incident to at most 6 trapezoids. This gives us a bounds of $O(k)$ on the number of trapezoids generated. We now perform the same ray-shooting/walking scheme as above except in the trapezoids. The path now traverses trapezoids, but still cannot cross the same trapezoid twice since by the stopping conditions it must remain monotone to the base edge and cannot re-enter a face. Thus it has at most one edge on any trapezoid for a total length of $O(k)$. Shooting a ray in a trapezoid takes $O(1)$ time.

The first stopping condition is handled by marking each face when the splicing path enters it. When the splicing path enters a new face, we simply check whether the face has already been marked. This requires an $O(k)$ time preprocessing step to initialize each marker, and then an $O(1)$ check each time we compute an edge of the splicing path. The second stopping condition requires us to check whether the next potential edge of the splicing path is non-monotone to the previous edge and can be checked in constant time. The final check requires us to check for each edge if we stop the splicing path at that edge not along a motorcycle track, which side of the splicing path the discarded face lies on. If we stop the splicing path before it traverses the entire face, then we complete the subdivision of the face by shooting a ray upwards along the slope vector. From this it follows that the subdivided face that includes the base edge slopes downwards away from the splicing path. The only way this changes is if we exit a face before the splicing path has traversed above the base edge. In other words, the splicing path enters and exits the same motorcycle edge. In this case, the subdivided face containing the base edge lies on the upper side of the splicing path. To check this property we check whether the bottom edge of each trapezoid lies along a base edge or a motorcycle edge. We can then check, if the next edge is going to exit a face, whether that edge flips the base edge onto the wrong side of the splicing path in constant time.

By Lemma~\ref{lem:partialroofcomplexity} each partial roof can be represented by a DCEL using $O(k)$ storage, and the additional storage requirements are only a representation of the splicing path and storing the additional $O(k)$ trapezoids, each of which takes $O(k)$ space in total.
\qed

Then we have as a direct corollary:

\begin{corollary}[Merging partial roofs in linear time]\label{lem:merge} Given two partial roofs with defining chains that are co-incident subchains of a simple polygon $P$, there is an algorithm for computing a partial roof of the concatenated defining chains in $O(k)$ time. \looseness=-2
\end{corollary}

\medskip
\para{Proof of Main Theorem for simple polygons.} Given the merge operation the procedure for computing the straight-skeleton is surprisingly straightforward: subdivide the polygon into equal length subchains, recursively compute a partial roof for each, and merge the results to produce the straight-skeleton roof. Taking Lemmas~\ref{lem:partialroofcomplexity} and \ref{lem:partialrooftoroof} and Corollary \ref{lem:merge} together we have:

\begin{theorem}\label{thm:polygon}The straight skeleton of a simple polygon can be computed from its induced motorcycle graph in $O(n\log n)$ time and $O(n)$ space.\end{theorem}



\section{Extending to planar straight line graphs}\label{sec:extension}

\paraskip{Overview.} We now show how to compute the straight skeleton roof of a planar straight line graph. The straight skeleton on the interior of each face of a PSLG is independent of the other faces. For this reason we focus here on the straight skeleton of the outer face. This is the most general case because its straight skeleton may include both bounded and unbounded faces. Let $R(G)$ denote the straight skeleton roof for a PSLG $G$ and $F$ denote its outer face. We will call the part of $R(G)$ that orthogonally projects onto $F$ in the $xy$-plane the {\em restriction of $R(G)$ to the region above $F$} and denote this $R(F)$. In fact, for any planar region $C$ we will denote by $R(C)$ the restriction of $R(G)$ to the region above $C$. The main idea is to subdivide the interior of $F$ into a set of cells $\{ C_i\}$, each of which captures the essential properties that allow the divide and conquer algorithm for simple polygons to work. Each cell $C_i$ has a defined slab set $\slabs(C_i)$ called a {\em subdivided slab set}, and the restriction of the lower envelope of $\slabs(C_i)$ to the area above $C_i$ is equal to $R(C_i)$. As before we define a notion of a {\em partial roof} to subchains of the boundary of each $C_i$ and then use our divide and conquer approach for polygons to compute $R(C_i)$. 

\paraskip{Essential properties.} We now informally examine the essential properties used by our polygon algorithm that each cell $C_i$ captures. The first essential property is that the edges of the polygon $P$ represent known parts of the roof $R(P)$, namely $P$ is equal to the intersection of $R(P)$ with the $xy$-plane. For each cell $C_i$ we compute a lifting of its boundary, denoted $\partial C_i$, onto the final roof $R(F)$. This lifting can be computed efficiently for all cells without having to compute the entire $R(F)$. The lifted edges $\partial C_i$ take the place of the edges of $P$ in the polygon algorithm. The second essential property is that each slab in $\slabs(P)$ is incident to $P$ along its base edge. For a cell $C_i$, we require that each slab in $\slabs(C_i)$ be incident to $\partial C_i$ along a connected chain of edges.  In particular these first two properties mean that we can employ the same divide and conquer approach: given two subchains of $\partial C_i$ incident to the same vertex, the vertex represents a known starting point along an intersection of two slabs that must appear on any partial roof for the merged chains. The third essential property is that the straight skeleton of a polygon $P$ is a tree (and thus acyclic), which is used in the proof of correctness to prove the edge containment property. For a cell $C_i$ we require that the interior edges of $R(C_i)$ (which are the lifted straight skeleton edges on the interior of $C_i$) be acyclic.
%

\paraskip{The subdivided roof.} Our algorithm computes a subdivision of the final roof $R(F)$. Let $C_1, C_2, \dots$ be a subdivision of $F$ into cells. Denote by $\partial C_i$ the lifting of the boundary of the cell $C_i$ onto $R(F)$. The lifting of the boundary of all cells induces a subdivision on $R(F)$ we call the {\em subdivided roof} and denote by $\hat{R}(F)$.  Figure~\ref{fig:pslg14}a depicts the projection of the final straight skeleton roof $R(G)$ of a PSLG $G$ onto the $xy$-plane. Figure~\ref{fig:pslg58}b shows a subdivided roof $\hat{R}(F)$ for the outerface $F$ of $G$ produced by a particular cellular subdivision (the blue dashed and solid lines are the added edges used to form the subdivision). To perform the subdivision we use a modified version of the vertical subdivision procedure from \cite{chengArxiv2014} to divide $F$ into cells. We then use our divide and conquer approach for polygons to compute $R(C_i)$ for each cell $C_i$. The output of our algorithm is the subdivided roof $\hat{R}(F)$. From this the final roof $F$ can be constructed in a linear time by merging subdivided faces across cell boundary edges. 

\paraskip{The subdivided slab set.} The subdivided roof presents one technical problem. A slab $s\in\slabs(F)$ supports only one face of $R(F)$, but may support more than one face of $\hat{R}(F)$. To handle this we extend the subdivision procedure from \cite{chengArxiv2014} to subdivide each slab $s$ resulting in a set of {\em subdivided slabs} for each cell $C_i$, denoted $\slabs(C_i)$. It is the lower envelope of these subdivided slabs that forms the roof $R(C_i)$. 

\paraskip{Notation and assumptions.} As before we extend $\R^2$ and $\R^3$ with points at infinity which are equivalence classes over vectors that point in the same direction. We let $n$ denote the number of edges in $F$ and $m$ denote the number of boundary components. To simplify the presentation we assume that: (1) no edge of $F$ is parallel to the $x$ or $y$-axes and (2) no angle bisector line of the lines supporting any two edges of $\partial F$ is parallel to either axes\footnote{This can be enforced by the following $O(n\log n)$ preprocessing step which finds a small rotation to apply to $G$ to ensure the property holds. Let $L_1, \dots, L_n$ denote the lines through the origin parallel to the $n$ edges of $G$. Sort these by angle made with the $x$-axis. Let $\epsilon_x$ and $\epsilon_y$ be the smallest non-zero angles between any line and the $x$ and $y$ axes (resp.). Clearly, applying a small rotation by any angle $\alpha < \min\{\epsilon_x, \epsilon_y\}$ is sufficient to ensure that now edges are parallel to the $x$ or $y$ axes, since the rotation will make any edges currently parallel to either axes non-parallel but is too small to make any edge not yet parallel into a parallel edge. Now, for each line $L$, find the line $L'$ such that the slope $M$ of $L$ and the negative slope $M'$ of $L$ are closest but not equal (this can be done easily in $O(n\log n)$ time using a modified 1D closest pair algorithm). Given an $L$, this is the same as finding the line $L'$ such that the bisector lines of $L$ and $L'$ are closer to parallel to the $x$ and $y$ axes than $L$ with any other line $L''$. Let $\theta$ be the smallest angle of rotation that aligns the bisector lines of $L$ and $L'$ with the axes. Let $\theta'$ be the smallest non-zero angle over all such $\theta$. Then any rotation $\alpha < \theta'$ will ensure that the bisector lines of all pairs of lines $(L, L')$ are not the axes. Thus we apply a rotation of $\alpha  = (\min\{\epsilon_x, \epsilon_y, \theta'\})/2$ to $G$.  }. We also assume that each connected boundary component of $F$ is combinatorially a simple polygon. This can be ensured by ``walking around'' the part of each component incident to $F$. Sharp turns in the walk (of angle $2\pi$) are handled as in the polygon case by adding a zero-length line segment which we think of as making right angles with its neighboring edges. See Fig.~\ref{fig:pslg14}c.

\subsection{The subdivision procedure}
 
We use a modification of the vertical subdivision procedure from \cite{chengArxiv2014}.  The basic idea is to use vertical lines (in the $xy$-plane) to partition the interior of $F$ into cells. The lines are defined with respect to a particular set of subdivision points. In \cite{chengArxiv2014} these points are chosen so that the part of the final roof on the interior of each cell is convex, which allows them to use existing techniques to efficiently compute the lower envelope of the supporting planes of slabs for a particular cell rather than the lower envelope of the slabs. However, to achieve this they require $O(r)$ subdivision points where $r$ is the number of reflex vertices. We use strictly fewer subdivision points: exactly one for each connected component of $F$. We also note that though they employ their subdivision procedure only for polygons with holes, the subdivision algorithm extends naturally to PSLGs. {\em Note the the reader:} there is one difference in terminology: in order to treat edge and motorcycle slabs in a unified manner we have defined a single slab for each edge of $F$ to be the union of the edge and motorcycle slabs (see Sec.~\ref{sec:preliminaries}). As was noted previously, this simplification was also employed by \cite{Huber:2011kr}. In \cite{chengArxiv2014}, instead of merging the edge and motorcycle slabs into a single slab, they subdivide each face of the final straight skeleton into the components that lie entirely on the edge and motorcycle slabs using what they call {\em flat edges} that lie along the common boundary of an base edge's edge and motorcycle slabs. Both of these methods are essentially equivalent and are semantic tools to allow us to simplify the discussion.

\begin{figure}
\centering
\subfloat[]{\includegraphics[width=0.45\textwidth]{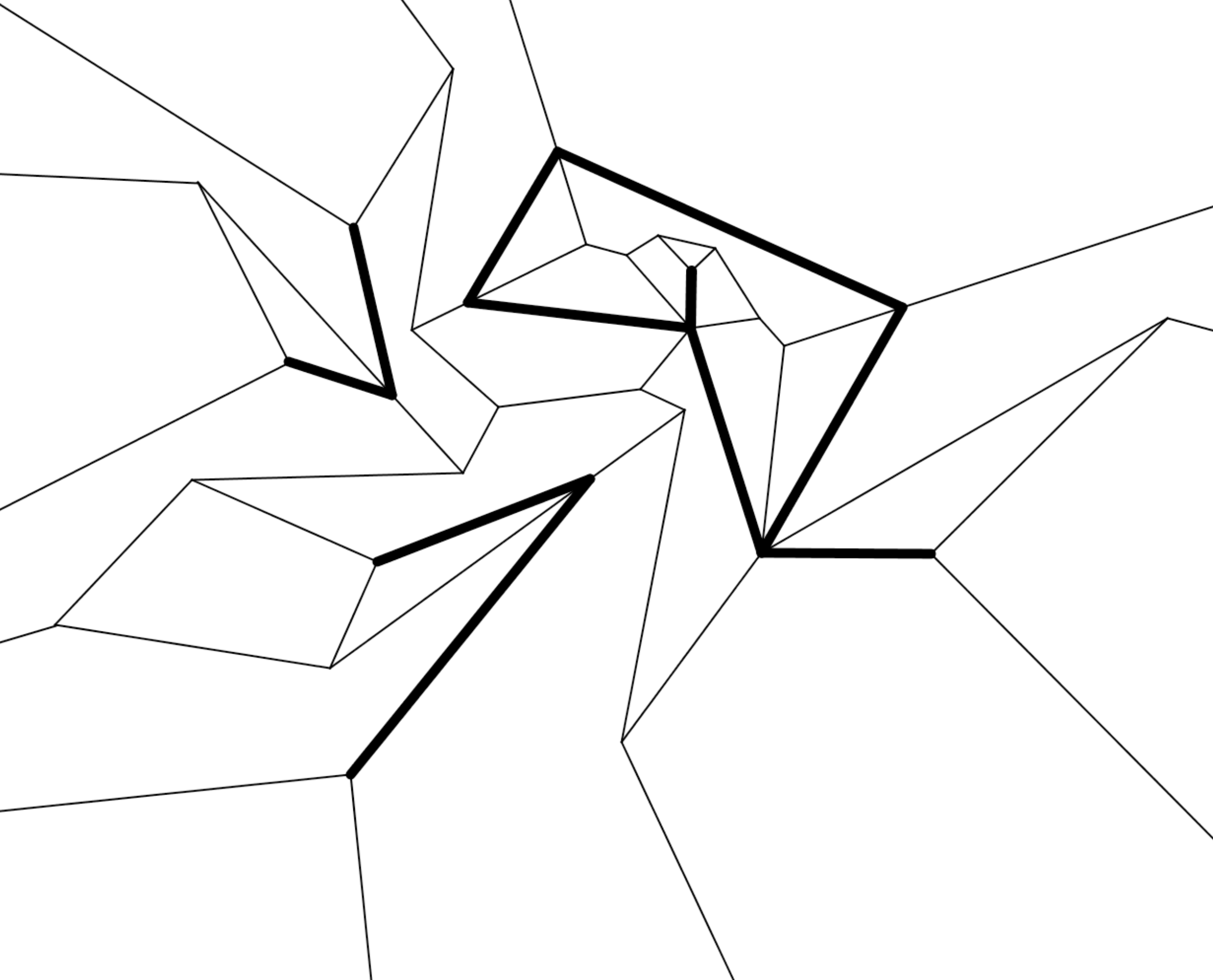}}
\qquad
\subfloat[]{\includegraphics[width=0.45\textwidth]{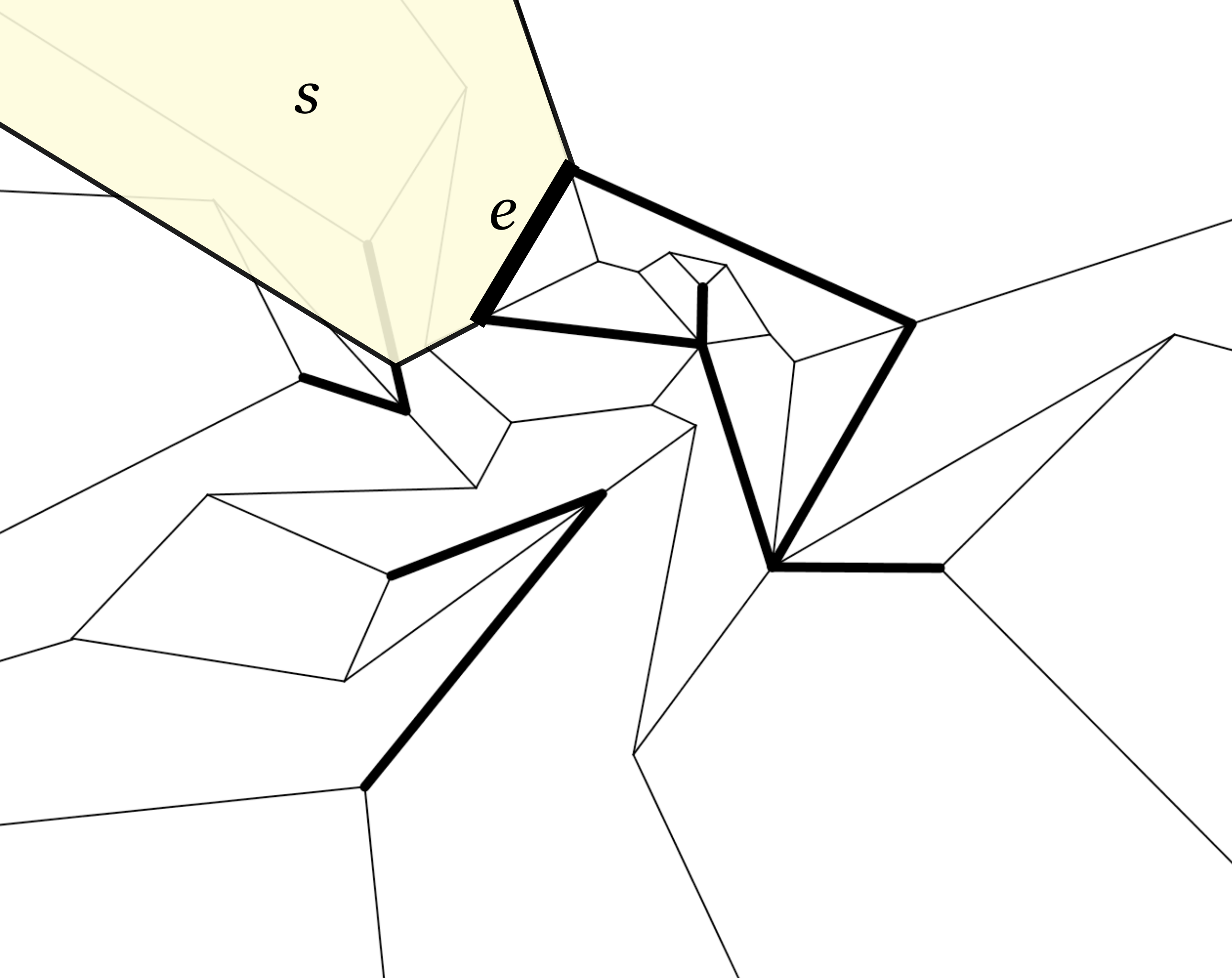}} \\
\subfloat[]{\includegraphics[width=0.45\textwidth]{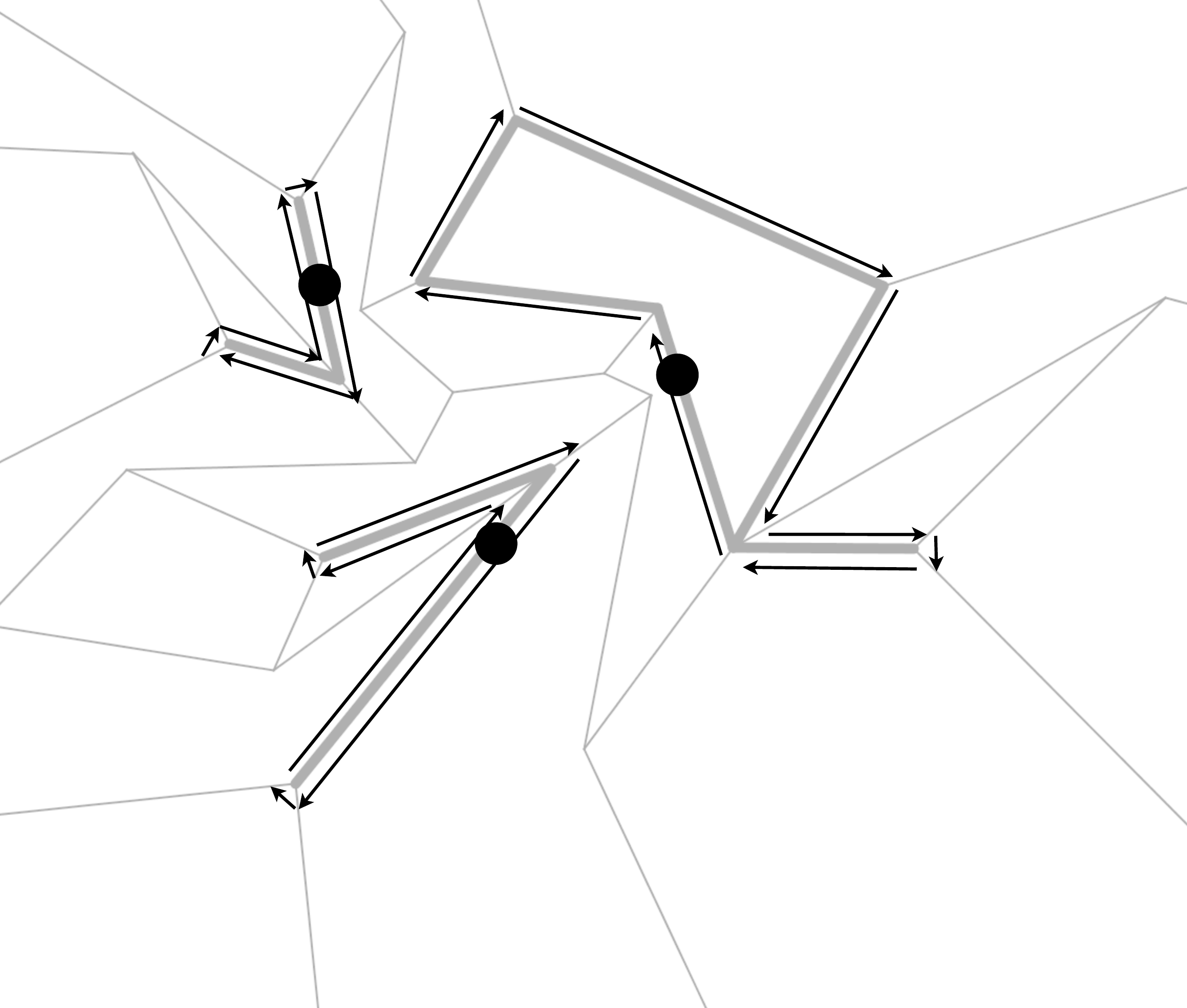}} 
\qquad
\subfloat[]{\includegraphics[width=0.45\textwidth]{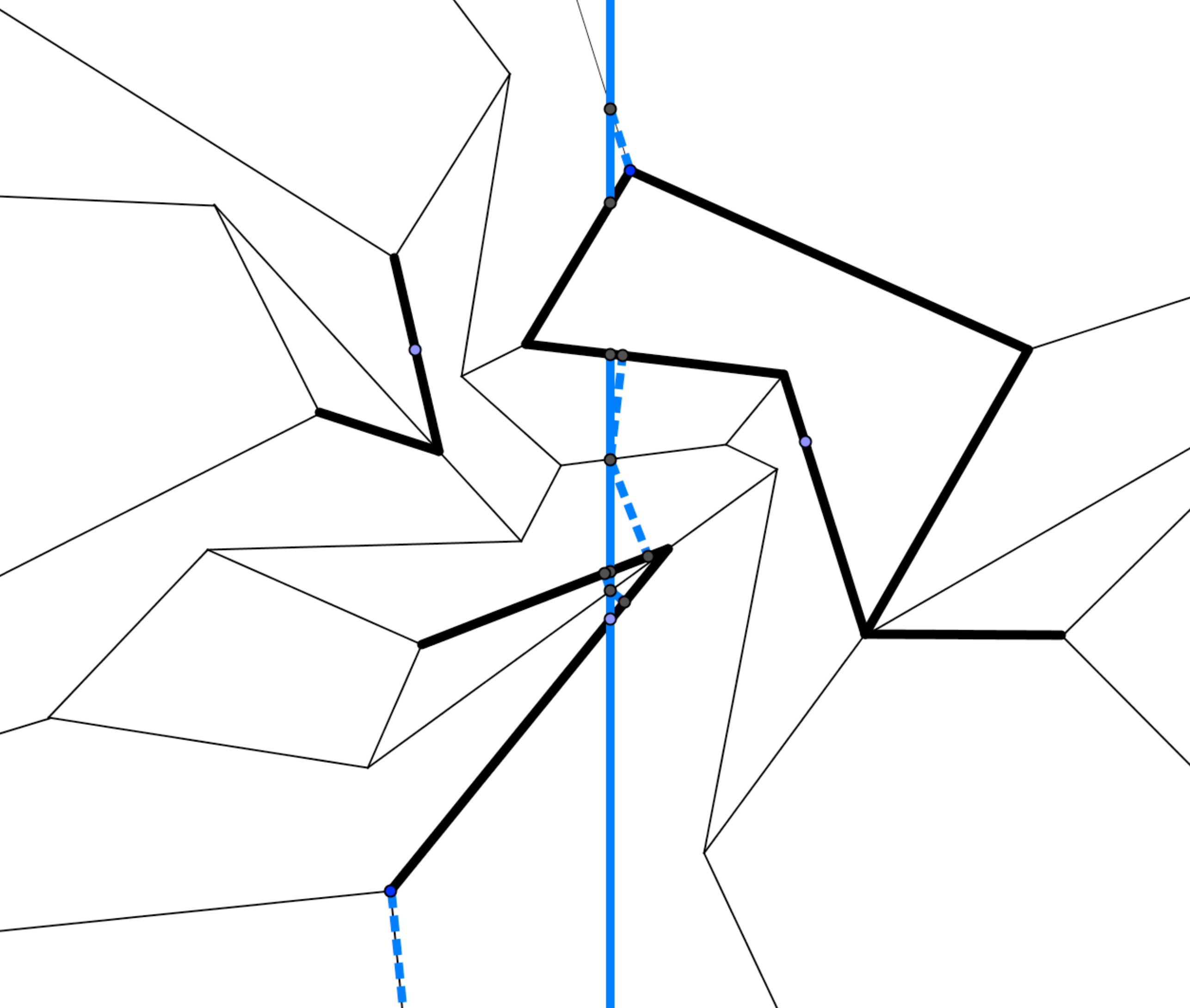}}
\vspace{-4pt}
\caption{\small{The algorithm for PSLGs. (a) A PSLG (thick lines) and its straight skeleton (thin lines). (b) The slab $s$ for an edge $e$ incident to the outer face. (c) An illustration of each boundary component as a (combinatorially) simple polygon given by walking around the boundary (denoted by arrows). Each degree 1 vertex is replaced with a small zero-length edge which we think of as supported by the line perpendicular to the vertex's incident edge. Note that we keep only the edges that are incident to the outer face, and not on the interior of a connected component. The black dots denote the subdivision set $V$. (d) The first subdivision line. The cell boundaries are given by the thick black lines, the solid blue lines, and the dotted blue lines. The solid blue line is the subdivision line $l$ through the point of $V$ with median $x$-coordinate. The dotted blue lines are the descent paths. For the next subdivision see Fig.~\ref{fig:pslg58}. }}
\label{fig:pslg14}
\end{figure}

\begin{figure}
\centering
\subfloat[]{\includegraphics[width=0.45\textwidth]{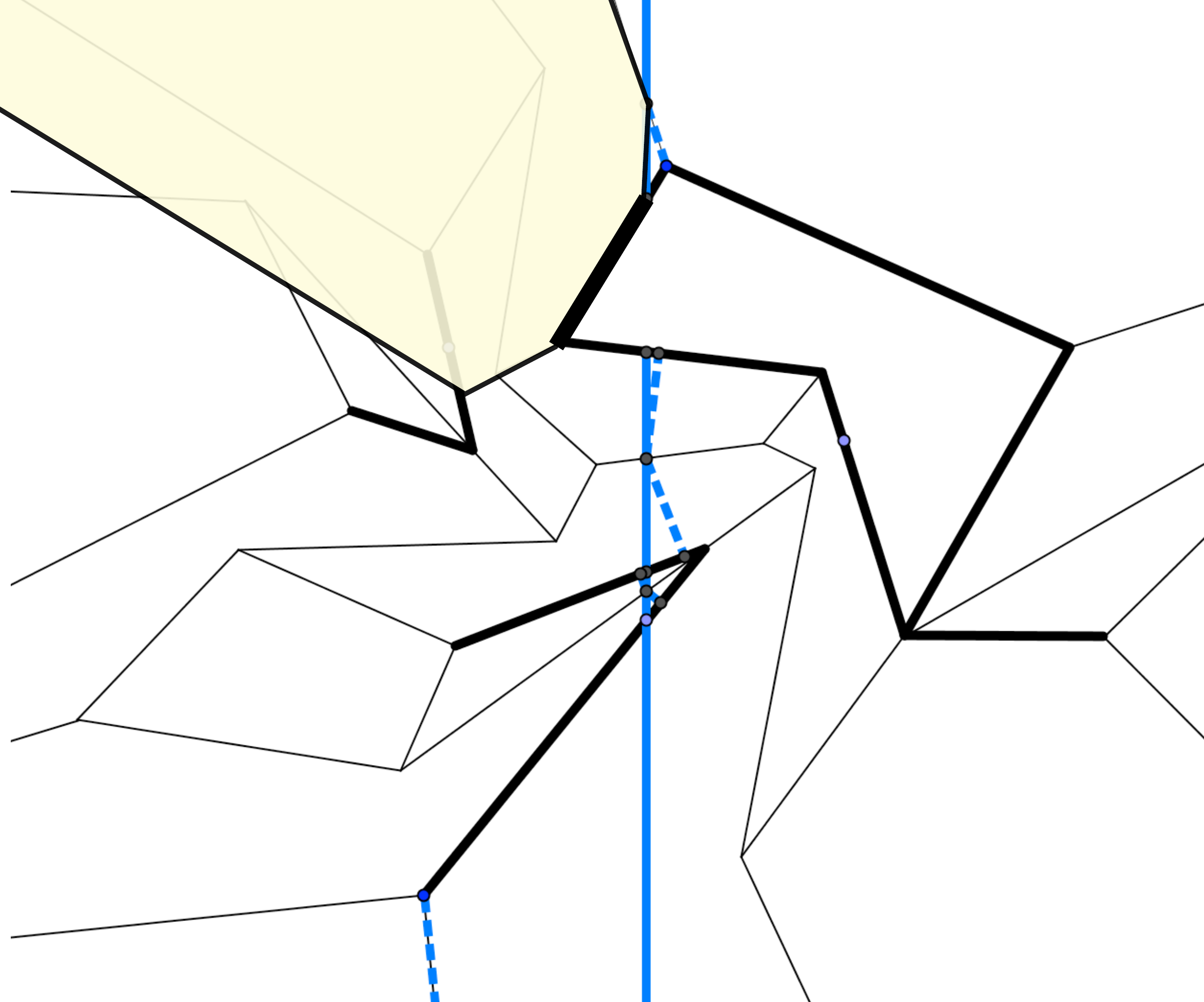}}
\qquad
\subfloat[]{\includegraphics[width=0.45\textwidth]{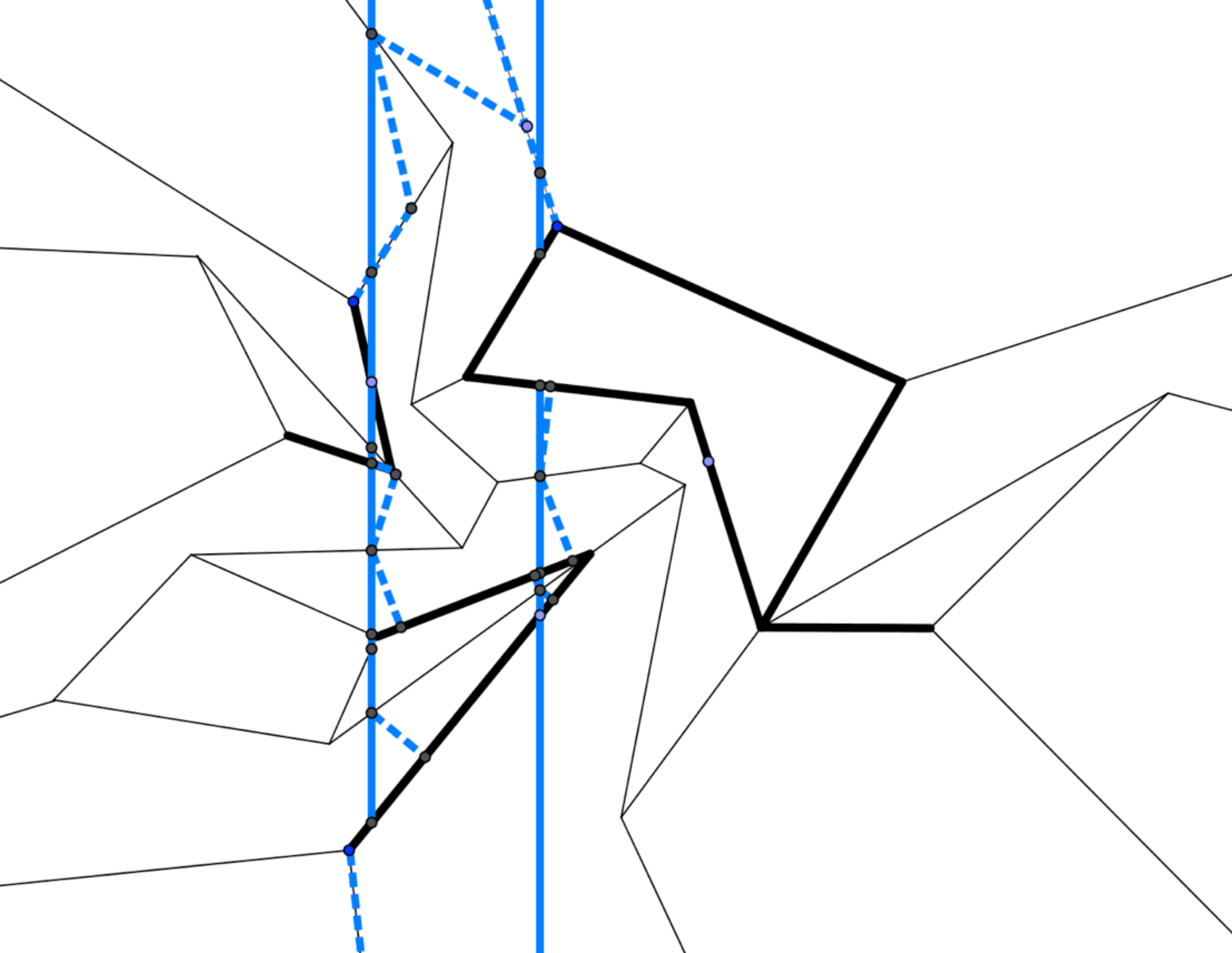}} \\
\subfloat[]{\includegraphics[width=0.45\textwidth]{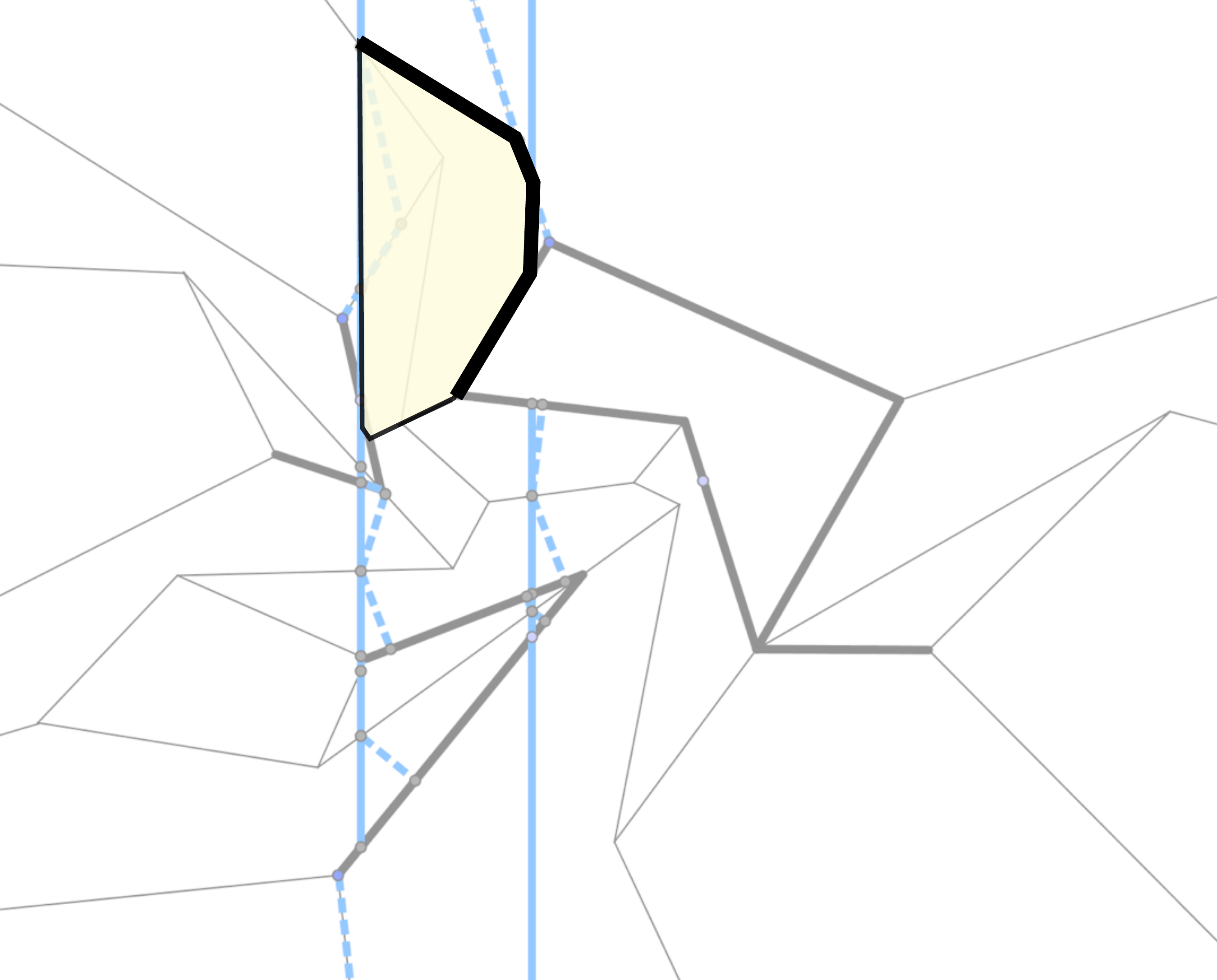}}
\qquad
\subfloat[]{\includegraphics[width=0.45\textwidth]{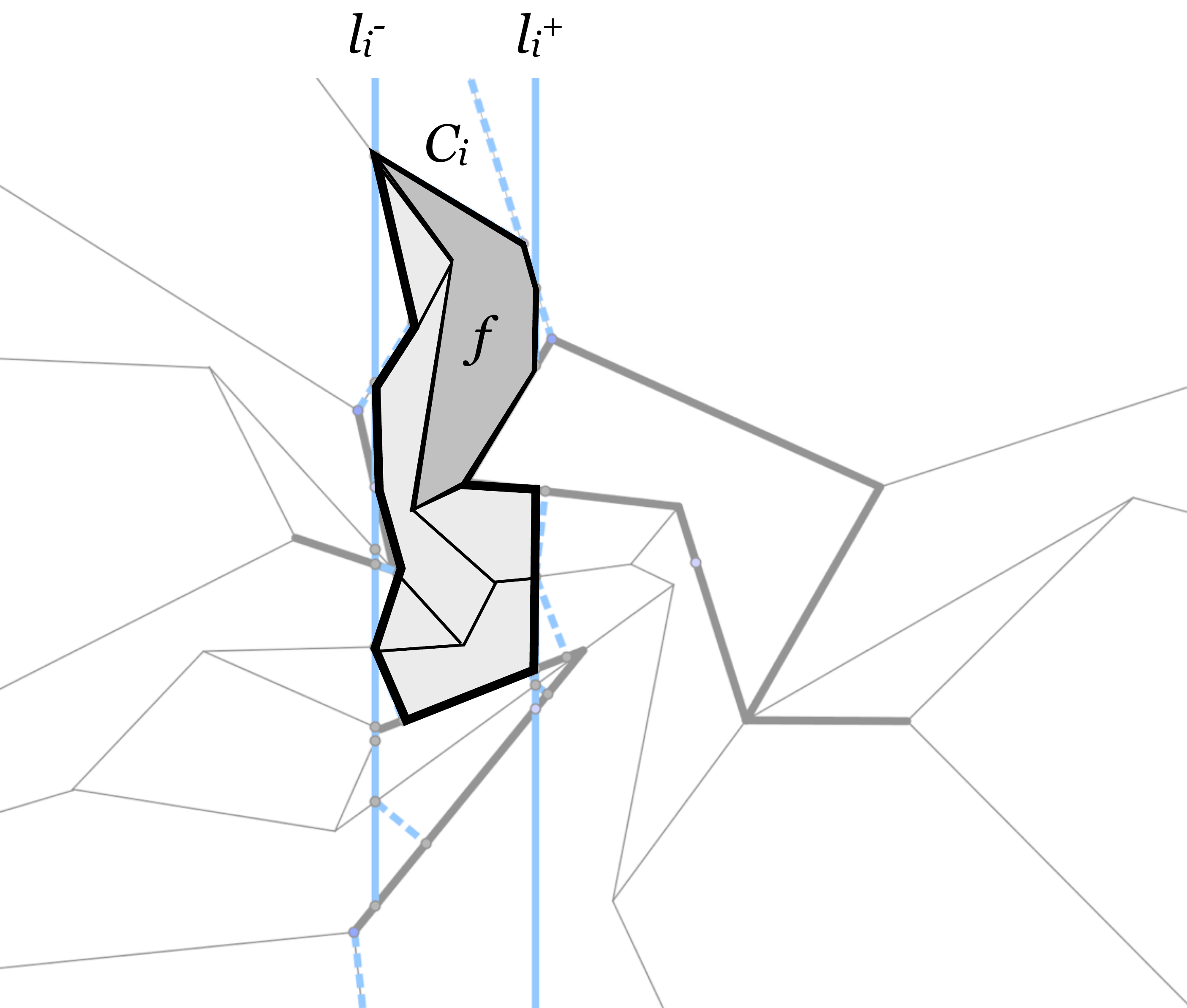}}
\vspace{-4pt}
\caption{\small{Continued from Fig.~\ref{fig:pslg14}. (a) The slab $s$ from the previous figure is subdivided by the subdivision line $l$. Depicted is the part of the slab to the left of the line. (b) The second subdivision line. This is the final subdivision for this figure, because the third subdivision point is not on an interior component of any cell. (c) A further subdivision of the slab $s$ from (a) induced by the new subdivision line. (d) A cell $C_i$ and the roof $R(C_i)$ (as viewed from above). The darkly shaded face $f$ is the face supported by the subdivided slab from (c). The slab is incident $C_i$ along a connected chain of edges (shown as thick lines in (c)). We find $R(C_i)$ by employing a modified version of our algorithm for polygons. The final subdivision, without the overlay of the straight skeleton edges is shown in Fig.~\ref{fig:celldiv}.}}
\label{fig:pslg58}
\end{figure}

\paraskip{Preliminaries.} To understand the vertical subdivision procedure of \cite{chengArxiv2014}, we need several concepts. Given a point $p$ on the final roof $R(F)$, a {\em descent path} is the path of steepest descent from that point $p$ down until it hits an edge of $F$. If $p$ is on the interior of a face $f$ of $R(F)$, there is exactly one descent path from $p$. Otherwise there is one descent path for each face incident to $p$. Each descent path first follows a segment parallel to the slope vector of $f$ downwards until it hits either the base edge of $f$, or a motorcycle edge of $f$. We call this segment the {\em descent edge} of $p$ in $f$. Each motorcycle edge forms a valley in $R(F)$, and so the descent path then travels the rest of the way down the motorcycle edge until it hits the base. If $p$ lies on an edge or vertex of $R(F)$, then it will have a descent path for each face it is incident to. Each descent path lies entirely on a single face, and given a point $p$ which is known to reside on a face $f$ of $R(F)$, the descent path from $p$ can be found knowing only the slab $s$ supporting $f$. In other words, to compute a descent path in a face it suffices to know a single point on the face and the slab containing the face. This property is useful in that it allows us to compute descent paths without first computing the entire roof. See \cite{Cheng:2007go} for more details.  

The subdivision procedure also makes use of {\em vertical lines} and {\em vertical planes}. A vertical line is a line $l$ in the $xy$-plane parallel to the $y$-axis and a vertical plane is plane through a vertical line that is orthogonal to the $xy$-plane. Each point $v$ in the $xy$-plane has a unique vertical line and vertical plane through it. 

\paraskip{Intersecting a vertical plane with $R(C_i)$.} The subdivision procedure makes use of the following subroutine: given a vertical plane $X$ and a cell $C_i$, the intersection of $X$ with $R(C_i)$ can be found in $O(k\log k)$ time {\em without first computing the roof $R(C_i)$} (where $k = |\slabs(C_i)|$). This is done by intersecting each slab $s\in \slabs(C_i)$ with $X$. The intersection of each slab with $X$ is a line segment, and the intersection of $R(C_i)$ with $X$ is the lower envelope (in $X$) of these line segments. The lower envelope of the segments can be found in $O(n\log n)$ using the algorithm from \cite{Hershberger:1989:FUE:79765.79766}. For more information we refer the reader to \cite{Cheng:2007go}. Each of these line segments lies along the roof in $\R^3$, but also has a projection onto the $xy$-plane. For each such edge we will refer to its {\em lifting} into $\R^3$ and its {\em projection} onto the $xy$-plane. The projected edges are used to subdivide $F$ into cells, but we simultaneously keep track of the lifting of the boundary of each cell into $\R^3$.


\begin{wrapfigure}{r}{0.25\textwidth}
\vspace{-14pt}
\centering
\includegraphics[width=0.24\textwidth]{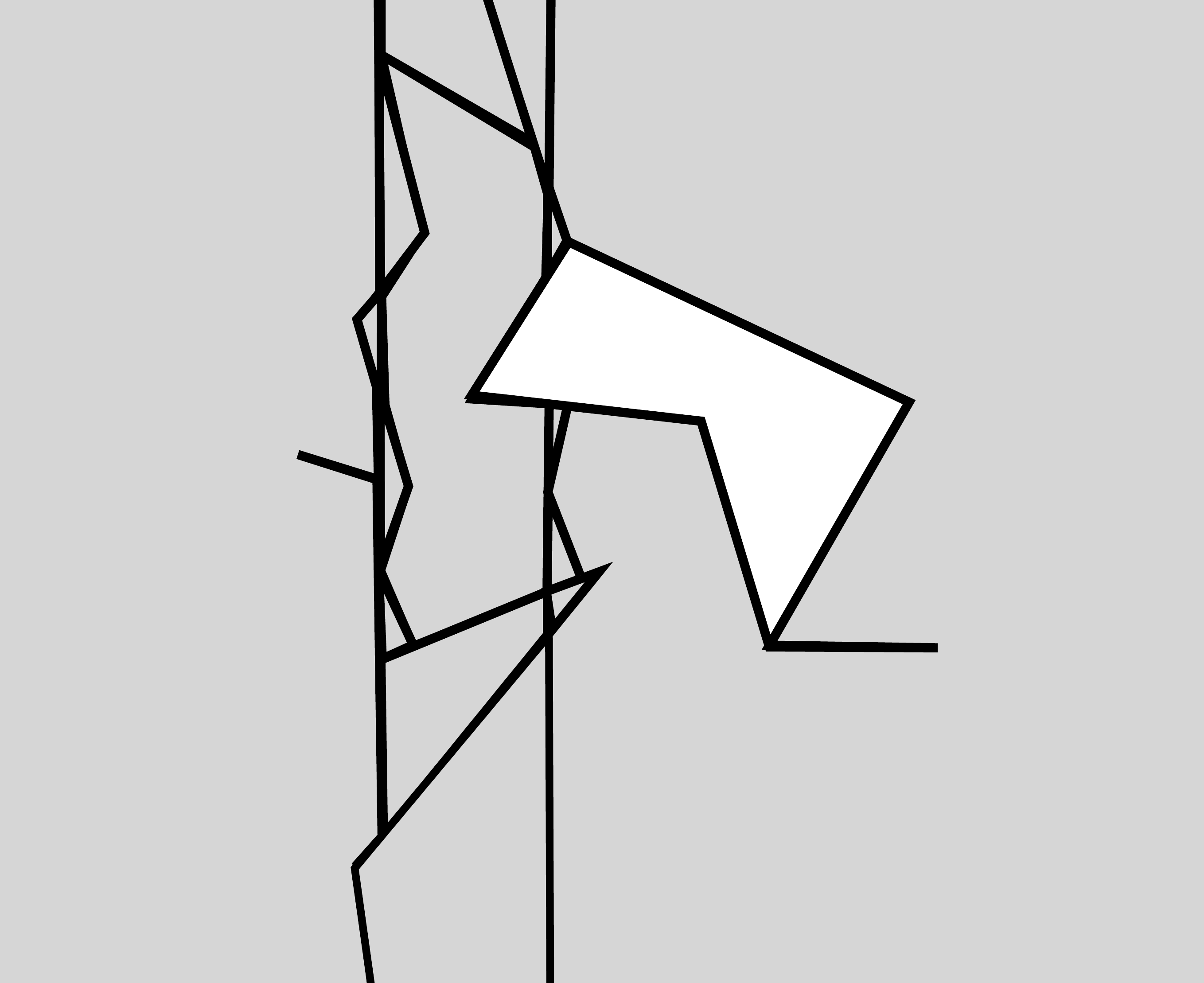}
\vspace{-6pt}
\caption{\small{The subdivision from Fig.~\ref{fig:pslg58} without the final straight skeleton edges depicted. Each connected shaded region is a cell. }}
\vspace{-12pt}
\label{fig:celldiv}
\end{wrapfigure}

\paraskip{Subdividing $F$ into cells.} The subdivision is divide and conquer. At the beginning, we select a set of {\em subdivision points} $V$. At each point in the algorithm we have a division of $F$ into some number of cells $C_1, C_2, \dots$. Each cell $C_i$ is such that the restriction of the lower envelope of its slab set, $\slabs(C_i)$, to the region above $C_i$ is equal to $R(C_i)$. Each cell also maintains a {\em conflict list} $V_i$ of the points of $V$ on its interior.

\begin{wrapfigure}{r}{0.63\textwidth}
\vspace{-14pt}
\centering
\subfloat[]{\includegraphics[width=0.13\textwidth]{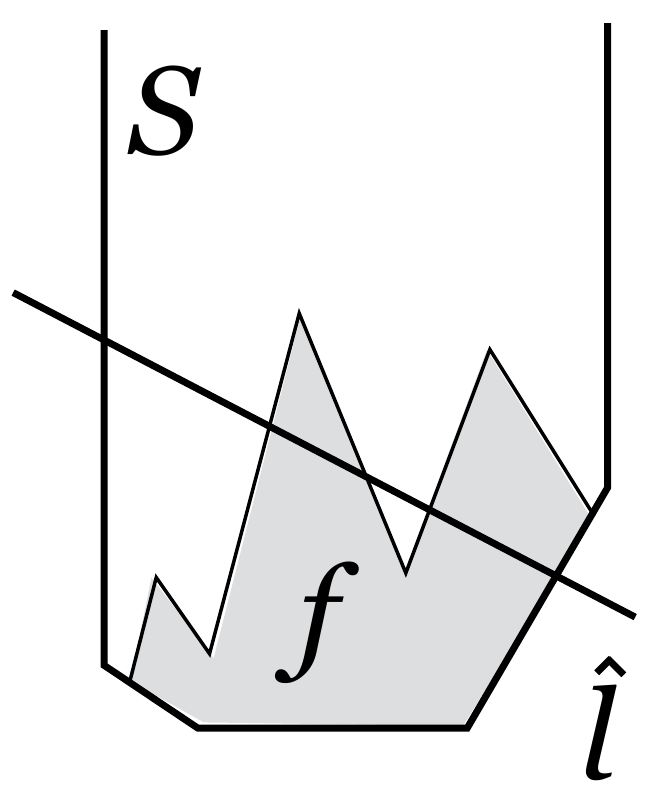}}
\subfloat[]{\includegraphics[width=0.117\textwidth]{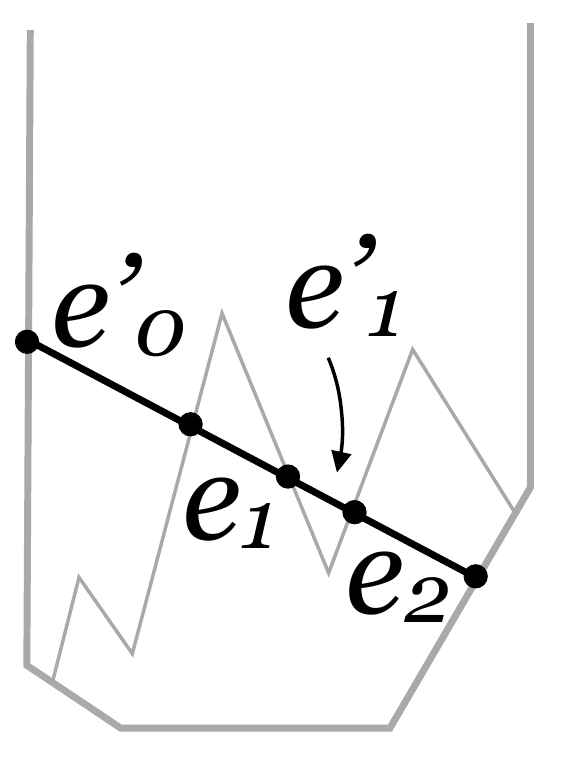}}
\subfloat[]{\includegraphics[width=0.115\textwidth]{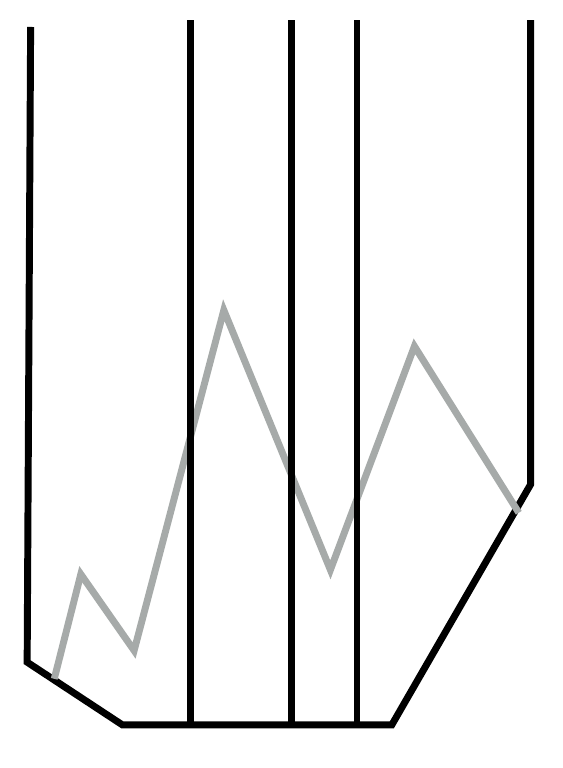}}
\subfloat[]{\includegraphics[width=0.12\textwidth]{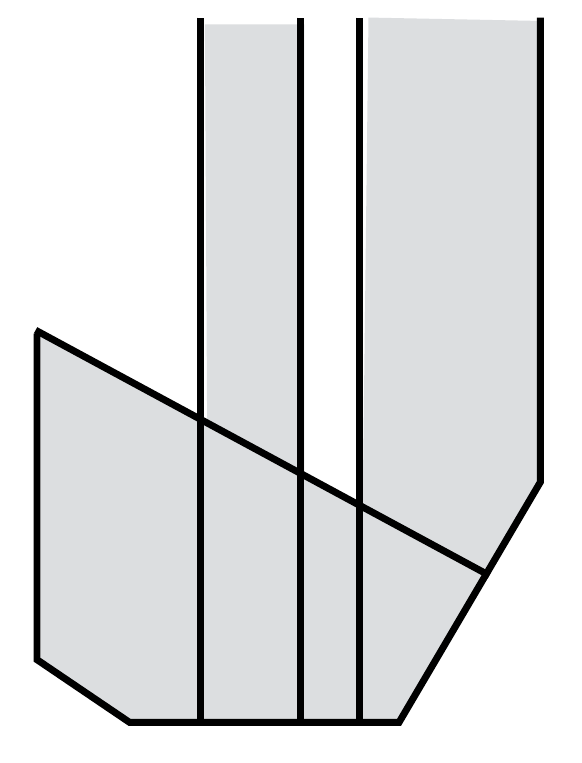}}
\subfloat[]{\includegraphics[width=0.1\textwidth]{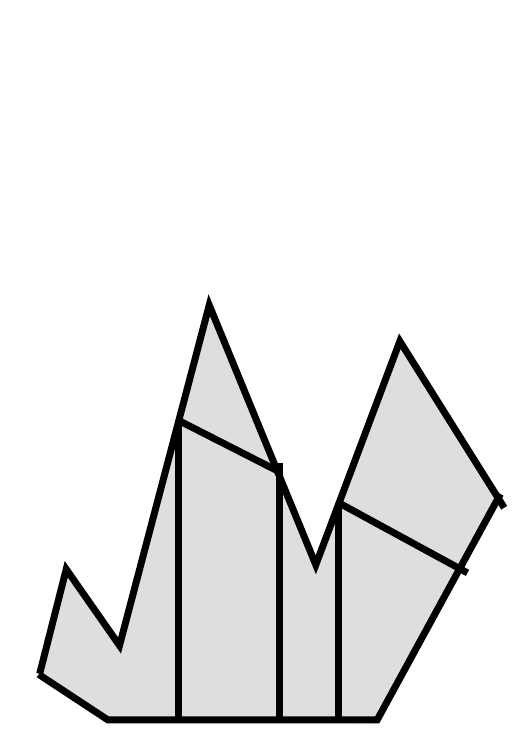}}
\vspace{-6pt}
\caption{\small{The subdivision induced by a line $l$ on a face of $R(F)$ and its supporting slab. (a) The face $f$, its slab $s$, and the lifted subdivision line $\hat{l}$. (b) the edges induced by $\hat{l}$, $e_1$ and $e_2$ lie on $f$, $e'_0$ and $e'_1$ lie on $s$ but not on $f$. (c) The slab-columns and face-columns induced by $\hat{l}$. (d) The final subdivision slabs. (e) The final subdivision faces.}}
\vspace{-8pt}
\label{fig:facesubdiv}
\end{wrapfigure}

The initial cell $C_1=F$. The boundary of $C_1$ is given by the connected boundary components of $F$, $\slabs(C_1) = \slabs(F)$, and $V_1 = V$. Each recursive step takes a cell $C_i$ for which $|V_i|\neq 0$ and $|\slabs(C_i)|>2$. It selects the point $v\in V_i$ with median $x$-coordinate, and then finds the intersection of the vertical plane through $v$ with $R(C_i)$ using the subroutine outlined above. We call the segments along this intersection {\em vertical edges}. The endpoints of each vertical edge lie on edges of $R(F)$ and represent intersection points between slabs on the final roof $R(F)$. From each intersection point we trace the descent paths in each of the incident faces of $R(F)$. The projection of the vertical edges and descent paths onto the $xy$-plane further subdivides the cell $C_i$. See Figs.~\ref{fig:pslg14} and \ref{fig:pslg58}. In \cite{chengArxiv2014}, they distinguish between three types of cells: {\em empty cells}, which have a single slab in $\slabs(C_i)$, {\em wedge cells}, which have two slabs in $\slabs(C_i)$ and the part of the straight skeleton on their interior is just the part of the projection of the intersection line of the two slabs that lies on the interior of $C_i$ (a single edge), and the remaining {\em general cells}, which have more than two slabs in $\slabs(C_i)$. The empty cells and wedge cells are base cases for the divide and conquer and are not further subdivided. 

\paraskip{Subdividing faces and slabs.} The subdivision procedure above induces a subdivision on the faces of $R(F)$. When we subdivide $F$ into cells, we do so along vertical edges and descent paths. In both cases the subdivision edges lie on a particular face of $R(F)$ and the subdivision of $R(F)$ induced by the lifting of these edges is the subdivided roof $\hat{R}(F)$. But, a single slab of $\slabs(F)$, which supports only one face of $R(F)$, may support multiple faces of $\hat{R}(F)$. For this reason we also subdivide each slab in order to maintain that each slab supports exactly one face. 

Let $f$ be a face of $R(C_i)$ which is subdivided during the subdivision of $C_i$. Let $X$ denote the vertical subdivision plane, $l$ denote the vertical line contained in $X$, $s$ be the supporting slab of $f$ in $\slabs(C_i)$, and $\hat{l}$ be the line supporting the intersection of $X$ with $s$. Note that because we assume no edge of $F$ is parallel to the $x$ axis (and thus perpendicular to $l$), $\hat{l}$ is not parallel to the slab's slope vector. Let $e_1, \dots, e_p$ be the vertical edges on $f$ (i.e. the connected components of $f\cap X$). These subdivide $\hat{l}$ into a series of segments: $e'_0, e_1, e'_1, e_2, e'_2, \dots, e'_{p-1}, e_p, e'_p$ where each $e'_i$ lies on $s$ but not on $f$ (Fig.~\ref{fig:facesubdiv}(b)). Let $l_p$ denote the line through a point $p$ on $s$ parallel to the slope vector of $s$. The induced subdivision on both $f$ and $s$ is defined by first subdividing $f$ and $s$ along $l_p$ for all points $p$ that are an endpoint of one of the edges $e_i$ not on the boundary of $s$ (Fig.~\ref{fig:facesubdiv}(c)). This divides $f$ into a series of {\em face-columns}, each of which is bounded below by some part of the boundary of $s$, and above some part of the upper monotone chain of $f$. This divides $s$ into a series of unbounded {\em slab-columns}, each of which contains exactly one face-column. Each slab-column either contains a vertical edge $e_i$ or contains one of the edges $e'_i$. In the first case, we further subdivide both the slab-column and face-column along $e_i$. In the second case we subdivide only the slab-column by $e'_i$ and keep only the part containing the corresponding face-column. We call these the {\em subdivided slabs and faces} (Fig.~\ref{fig:facesubdiv}(d,e)). We show an example of a face split by three lines in Fig.~\ref{fig:threesplit}. This procedure inductively maintains the following properties:

\begin{figure}
\vspace{-10pt}
\centering
\includegraphics[width=0.75\textwidth]{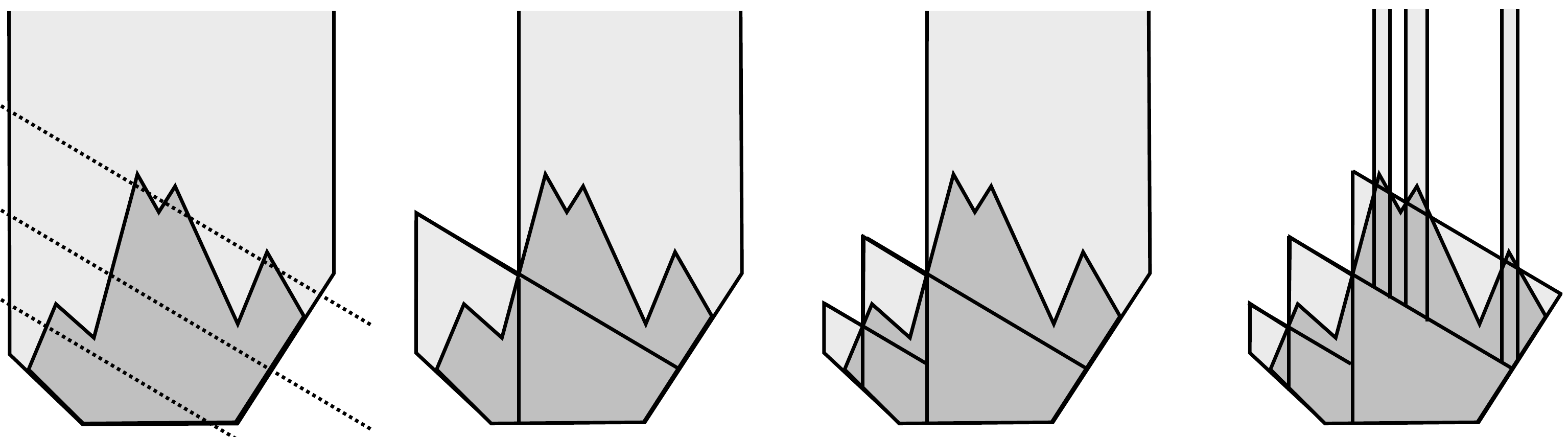}
\vspace{-6pt}
\caption{\small{The recursive subdivision of a slab $s$ and the face $f$of the final straight skeleton roof it supports. The ``local view'' in the slab's coordinate system is shown with the base edge along the bottom. In the left figure, the dotted lines denote lifted versions of three vertical subdivision lines that intersect $f$. The first subdivision results in three subdivided slabs and three subdivided faces. One of the slabs (the bottom right) is itself an entire {\em empty cell} and is not further subdivided.}}
\vspace{-14pt}
\label{fig:threesplit}
\end{figure}

\begin{lemma}[Properties of subdivided slabs.]\label{lem:subdivprops} (1) Each subdivided slab supports a single subdivided face. (2) The edges of a subdivided face that lie along boundary edges of the slab are also edges of one of the cells produced by the subdivision, and these form a connected chain of edges on the boundary of the cell. (3) The combinatorial complexity of each slab is $O(1)$. (4) each subdivided slab and face is monotone with respect to the base edge of the original slab. (5) Letting $\slabs(C_i)$ denote the subdivided slabs incident to the edges in $\partial C_i$, the restriction of the lower envelope of $\slabs(C_i)$ to the region above $C_i$ is the roof $R(C_i)$. (6) Each vertex of $\partial C_i$ is incident to at most two slabs of $\slabs(C_i)$. \end{lemma}
\pf (1) follows by induction from the definition. (2) follows from induction and the fact that the face is subdivided by the the vertical edges and descent paths, which are also the edges used to subdivide the cell that originally contained the face. (3) The original slabs have $O(1)$ complexity. A subdivided slab is given by ``clipping'' the slab between at most two lifted vertical subdivision lines and at most two descent edges parallel to the slab's slope vector. Thus each subdivided slab contains some connected portion of the original slab plus $O(1)$ additional edges. (4) The subdivision into slab-columns and face-columns maintains this property since the cuts are along the direction of monotonicity. The final cut along a vertical edge cuts clear across a column, by definition, since its endpoints define the sides of the column, which necessarily maintains monotonicity. (5) Suppose not. Since the subdivided slabs in $\slabs(C_i)$ cover the faces of $R(C_i)$, this means that one of the slabs in $\slabs(C_i)$ must appear in the lower envelope below some face $f$ of $R(C_i)$ that it does not support. However, since each subdivided slab is a subset of its original slab and each face of $R(C_i)$ is a subset of its original face, and the subdivided slabs cover the faces, then this means some part of $f$ cannot be part of the final roof $R(F)$, a contradiction. (6) This is true initially, since each vertex of $\partial F$ is incident only to the two slabs for its incident edges. Now assume it is true for a cell $C_i$ which is subdivided by a line $l$. Let $p$ be an intersection point along $l$. Then each face of $R(C_i)$ that is incident has a descent path traced from $p$. The descent paths form the new cell boundaries and separate the subdivided faces and slabs so that only two of each end up in any subdivided cell.\qed

\begin{figure}
\centering
\subfloat[]{\includegraphics[width=0.15\textwidth]{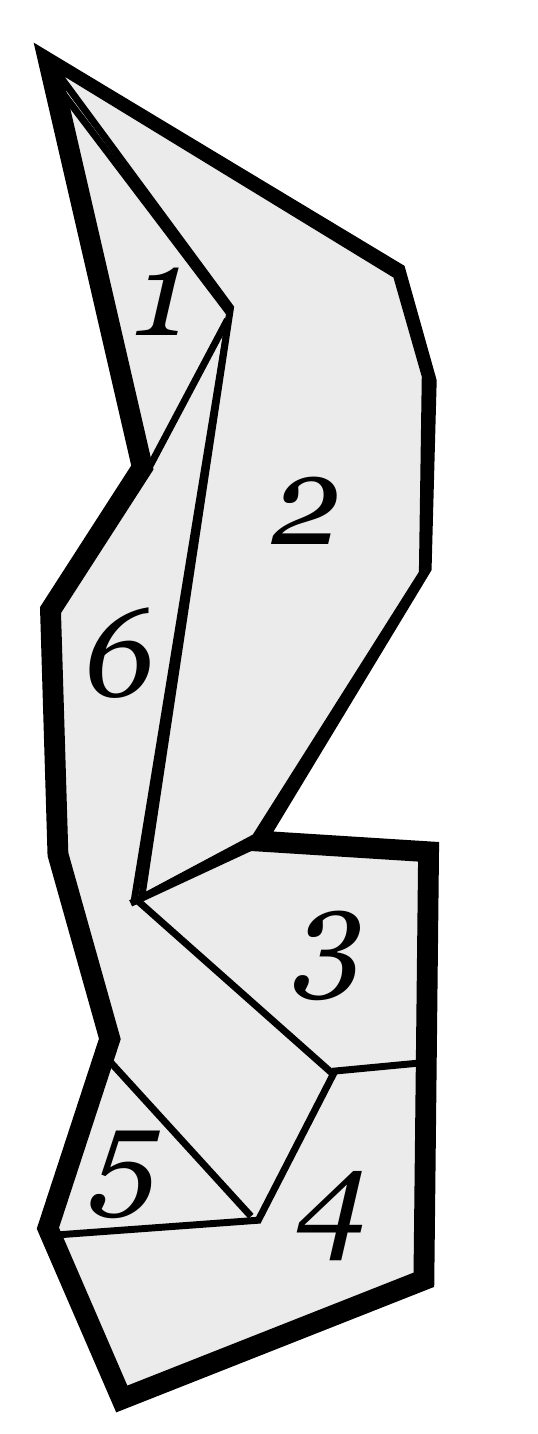}}
\qquad
\qquad
\subfloat[]{\includegraphics[width=0.5\textwidth]{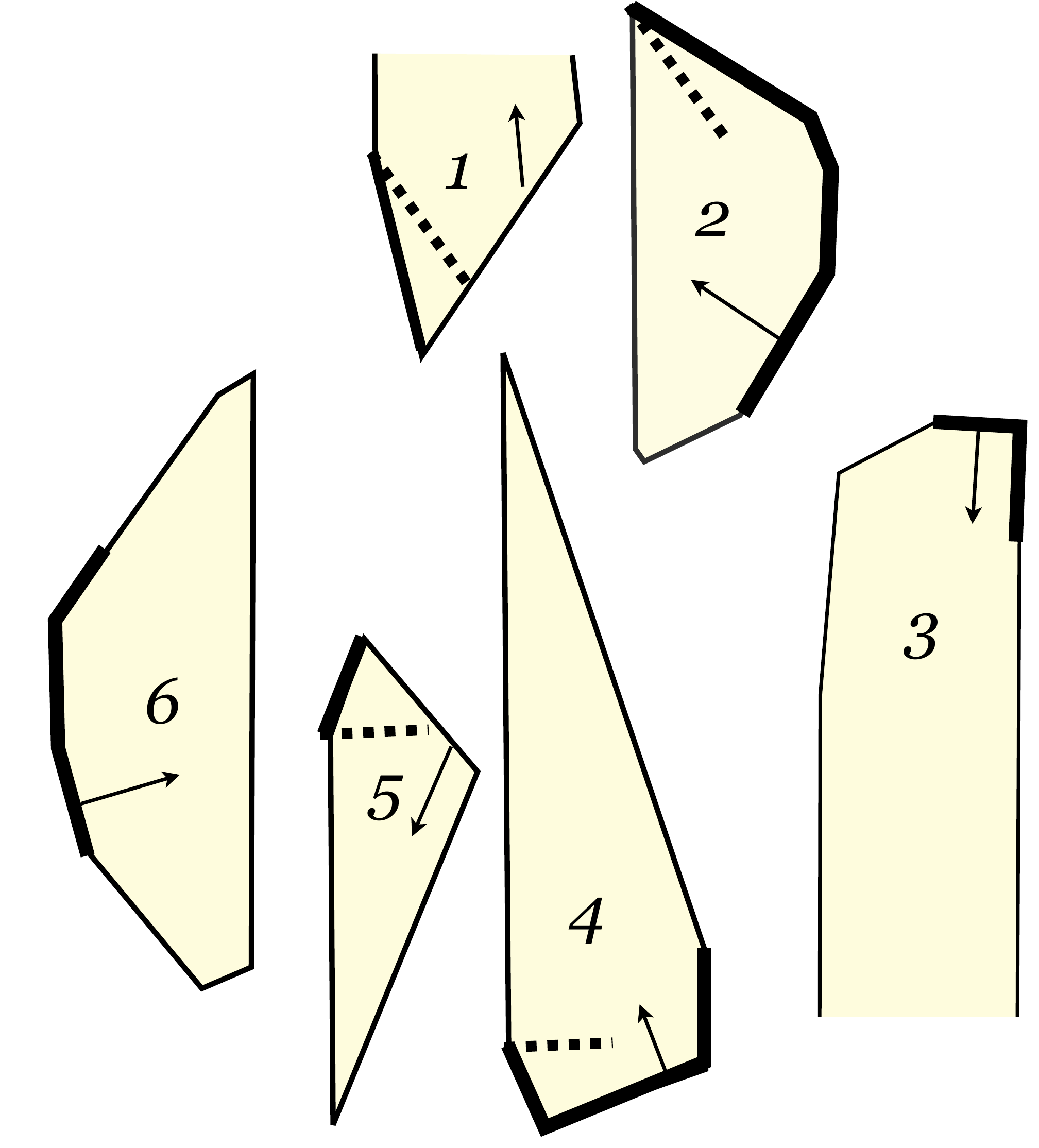}} \\
\subfloat[]{\includegraphics[width=0.45\textwidth]{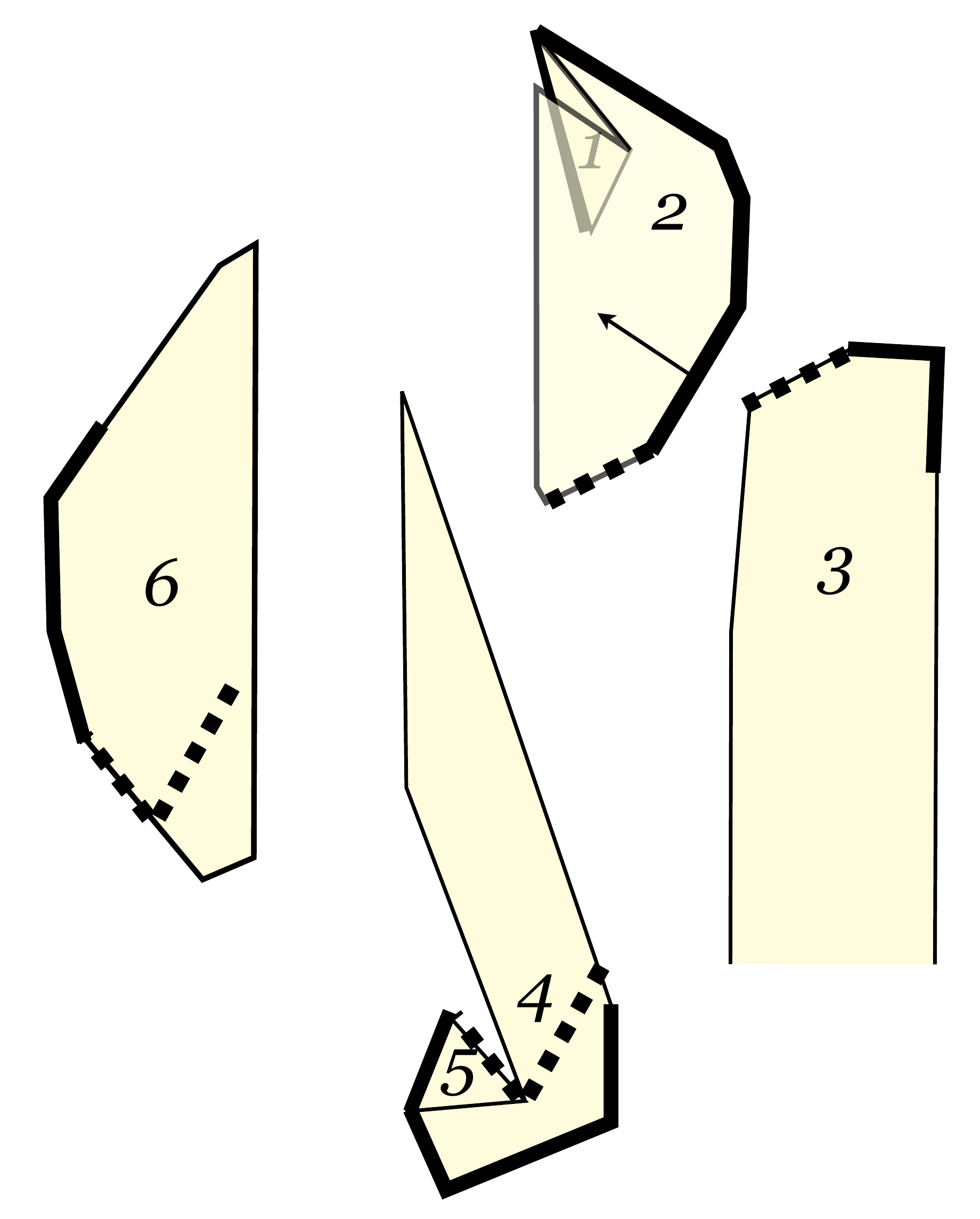}}
\qquad
\qquad
\subfloat[]{\includegraphics[width=0.3\textwidth]{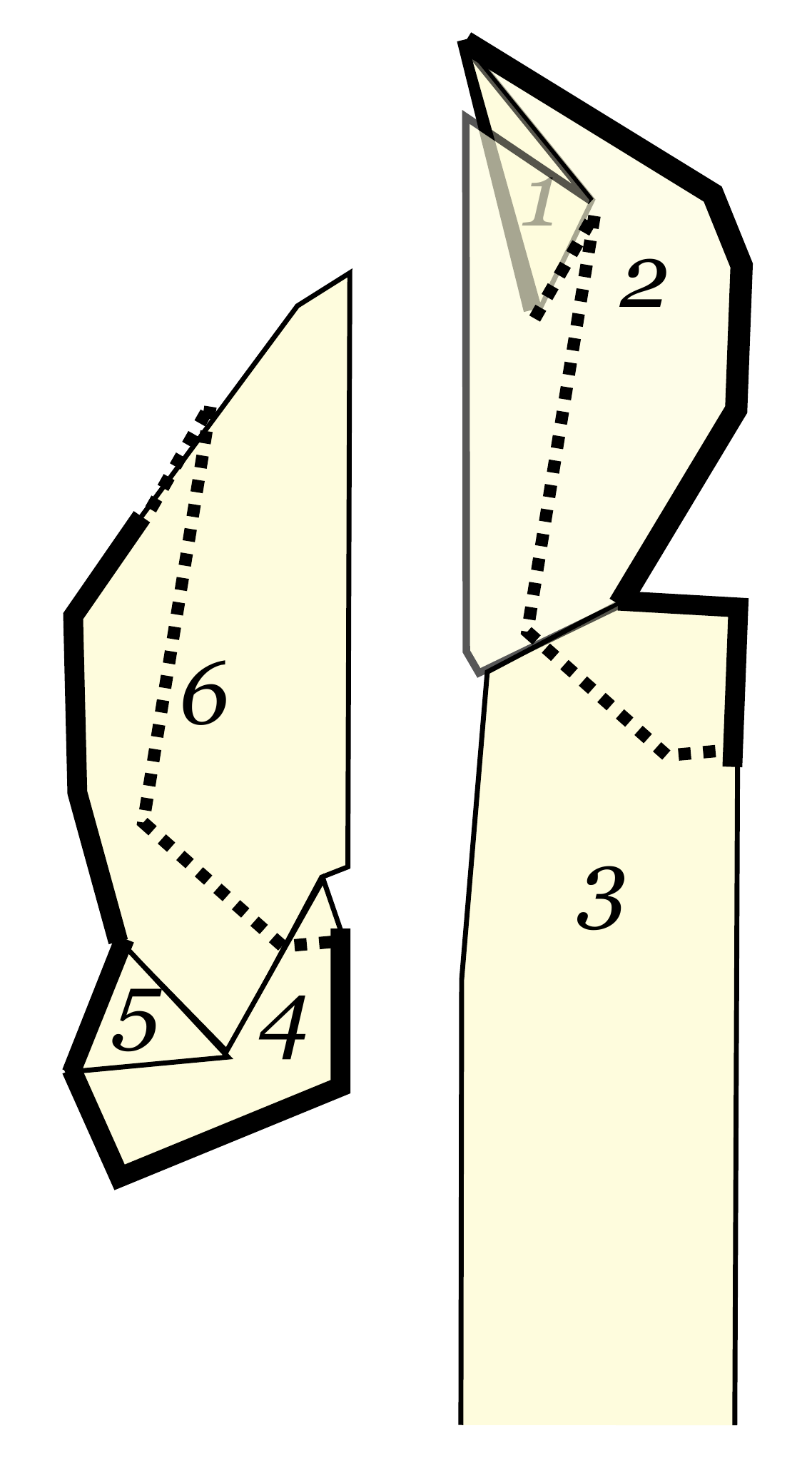}}
\vspace{-4pt}
\caption{\small{Filling in the cell from Fig.~\ref{fig:pslg58}d. (a) The cell with faces numbered. (b) The corresponding subdivided slabs. The small arrows indicate the slope of the face. The thick edges denote the defining edges, i.e. the edges of the slab that are incident to the cell boundary (this takes the place of the base edge from the polygon case). We first apply the merge operation to 1 and 2, and 4 and 5. The dotted lines show the splicing path. (c) After merging 1 with 2 and 4 with 5. We now merge $(1, 2)$ with $3$ and $(4, 5)$ with 6. (d) After the merge operations from (c). The dotted line shows the splicing path for the final merge operation, which produces the cell in (a).}}
\label{fig:cellcompose}
\end{figure}

\paraskip{Properties of the subdivision.} Let $C_1, C_2, \dots$ denote the cells computed during the subdivision (including intermediate cells). We note the following properties, which are proven in \cite{chengArxiv2014}. Let $\kappa_i$ denote the number of edges of $R(F)$ intersecting the interior of $R(C_i)$. Then $\sum_i \kappa_i = O(n\log |V|)$, the number of edges in $\partial C_i$ is $O(\kappa_i)$, and computing the subdivision takes $O(n(\log n)\log |V|)$ time. Note that they use a different (and larger) set of points $V$ than we do, namely their $V$ is the set of vertices of the induced motorcycle graph, but the proofs for the properties above generalize both to unbounded straight skeleton edges, and point sets $V$ in which each point $v\in V$ lies on the boundary of $F$. 
\begin{figure}
\centering
\subfloat[]{\includegraphics[width=0.55\textwidth]{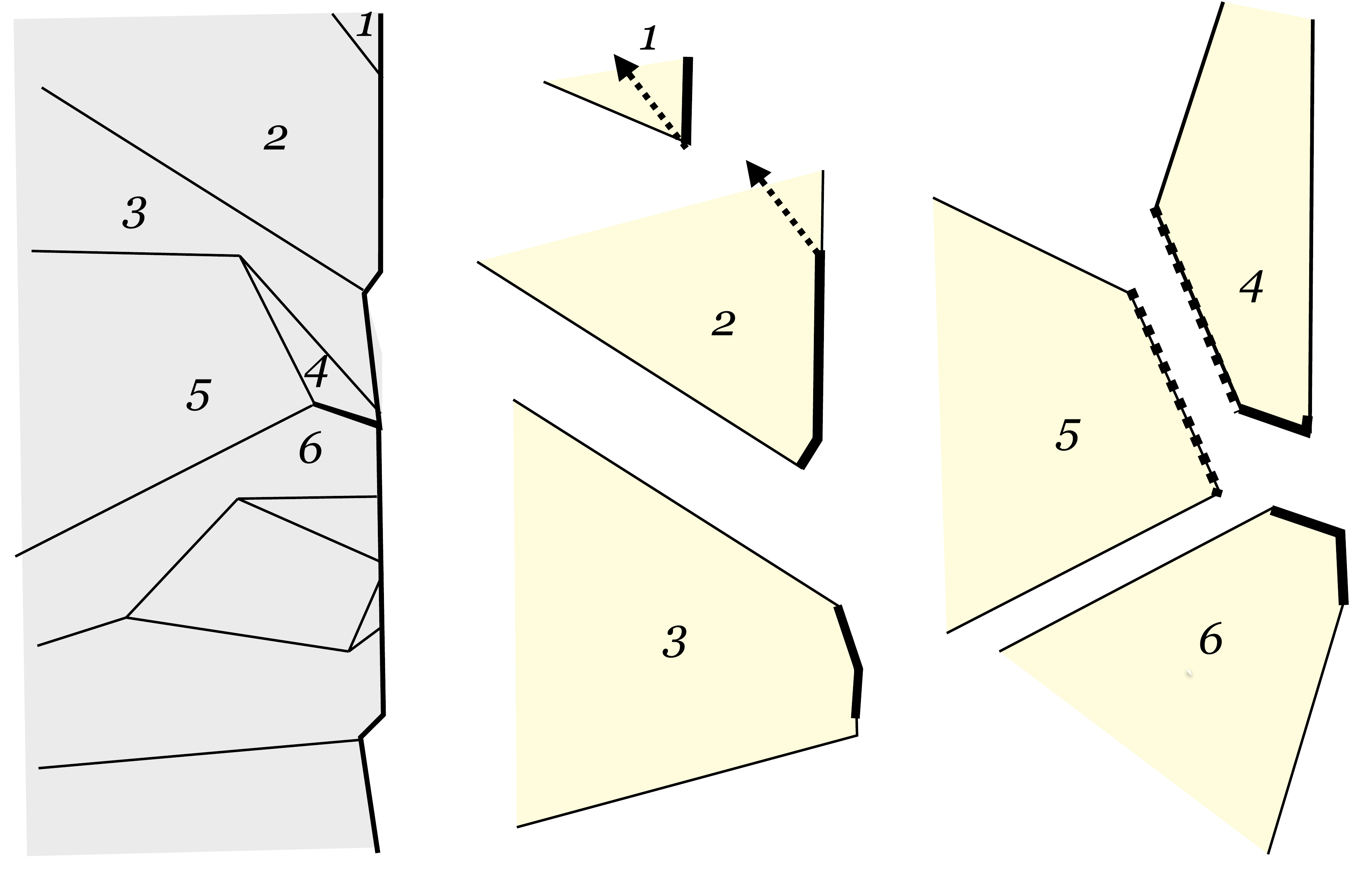}}
\qquad
\subfloat[]{\includegraphics[width=0.38\textwidth]{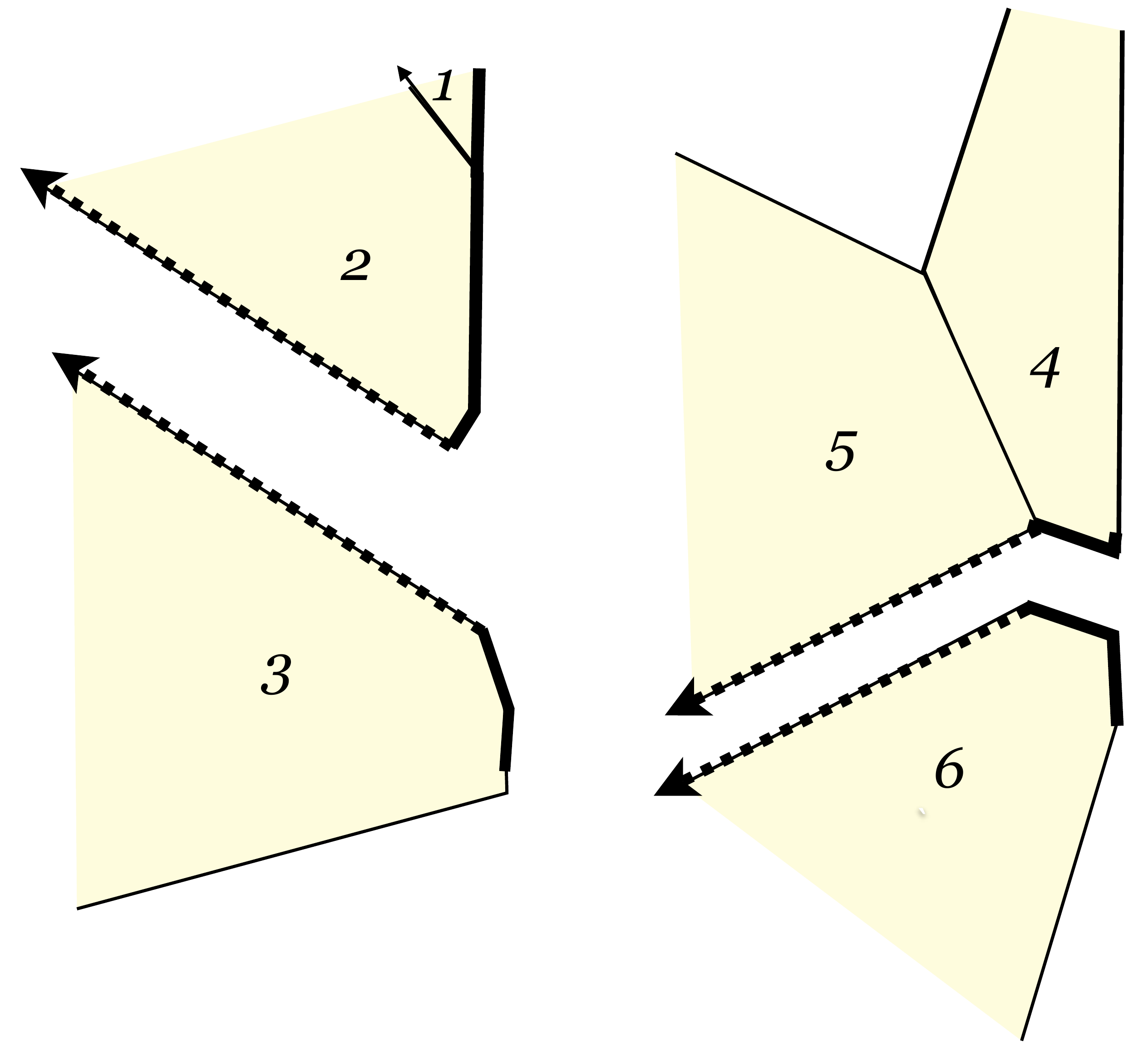}} \\
\subfloat[]{\includegraphics[width=0.4\textwidth]{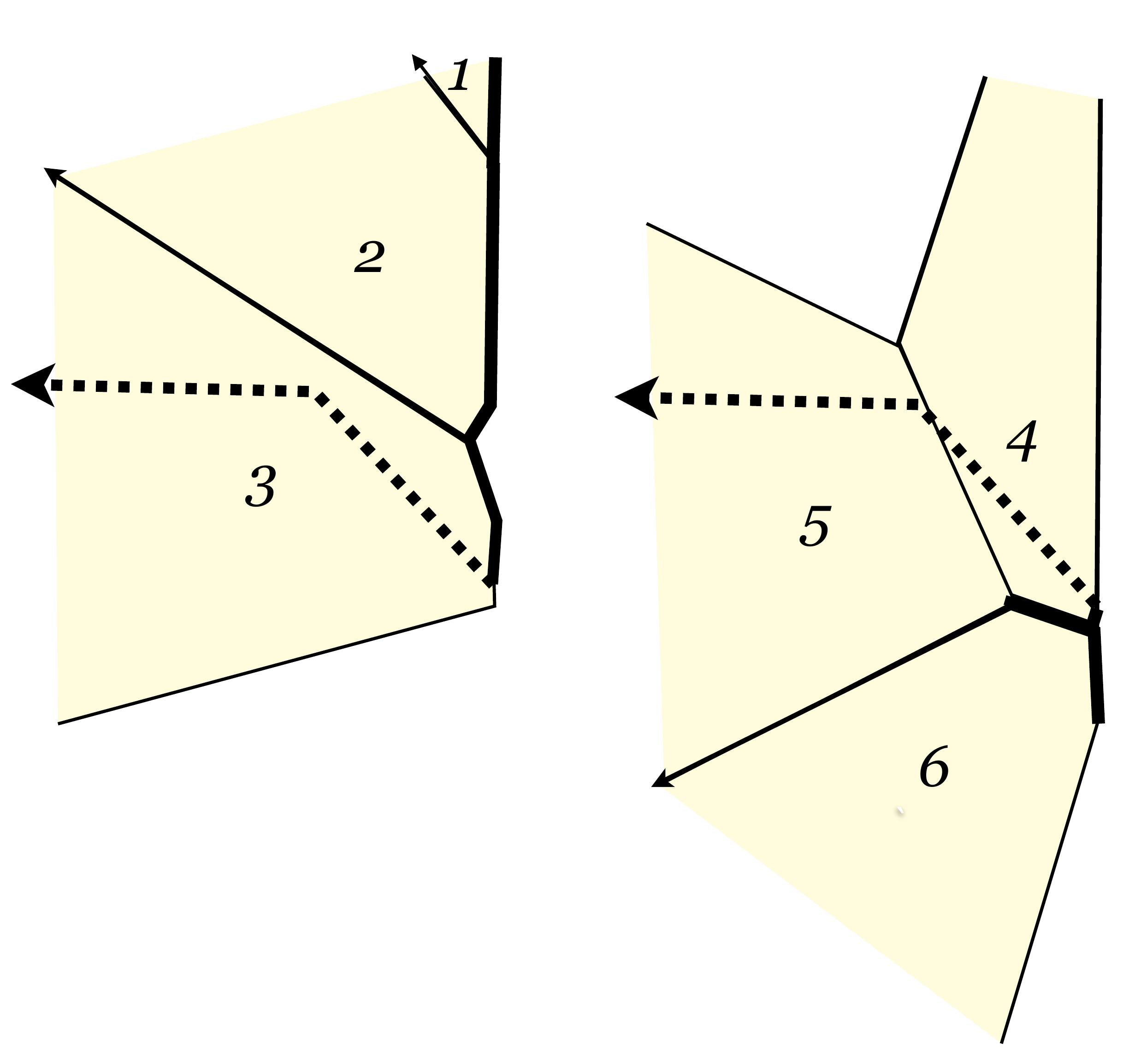}}
\qquad
\qquad
\qquad
\qquad
\subfloat[]{\includegraphics[width=0.2\textwidth]{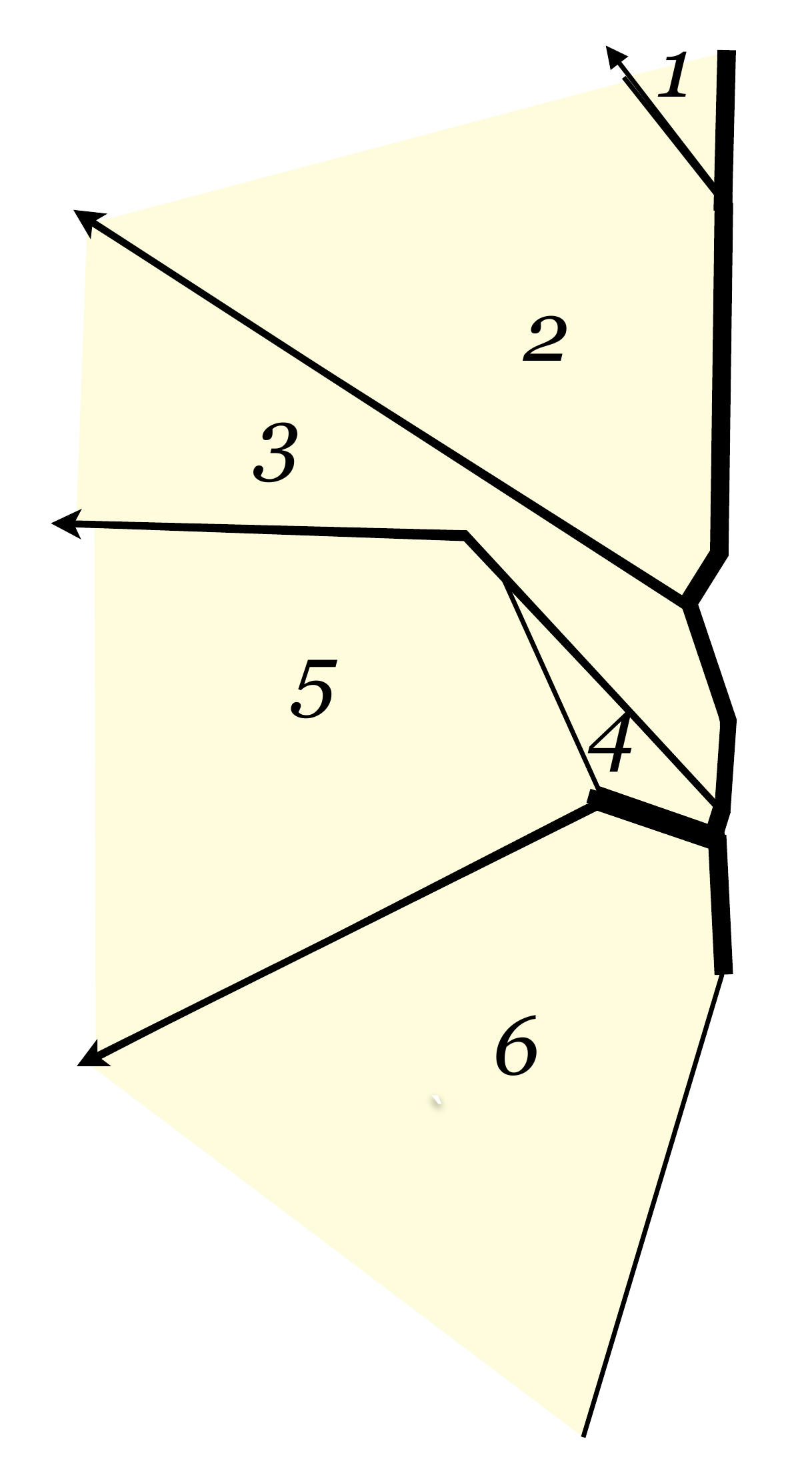}}
\vspace{-4pt}
\caption{\small{Merging the top 6 slabs of the left-most cell of Fig.~\ref{fig:pslg58}. (a) The cell (gray) with top 6 faces numbered, and the slabs for each face (numbered). We first merge 1 with 2 and 4 with 5. Dotted lines show the splicing path. (b)  The next merges are for $(1,2)$ and $3$ and $(4, 5)$ and $6$. Each of these is along an unbounded edge that lies on an unbounded motorcycle edge of the original slabs.  (c) The splicing path for the merge between $(1,2,3)$ and $(4,5,6)$. (d) All six faces merged.}}
\label{fig:cellcompose}
\end{figure}

\paraskip{The subdivision points.} In \cite{chengArxiv2014}, they choose $V$ to be the set of vertices of the induced motorcycle graph, so that $|V| = \Theta(r)$. 
 For our point set $V$ we arbitrarily select a point (which need not be a vertex) from each connected boundary component of $F$. We could, for instance, choose one reflex vertex from each connected component, making our choice of $V$ a strict subset of the one in \cite{chengArxiv2014}. The purpose of our choice of $V$ is to ensure that each cell is (weakly) simple, meaning:\looseness=-1

\begin{lemma}\label{lem:noholes}If $C_i$ is a final cell of the subdivision, then its interior contains no connected component of $F$.\end{lemma}
\pf Assume not. Then there must be a connected component on the interior of $C_i$. We first argue that $C_i$ has at least three slabs. Assume not. Since our connected components are combinatorial polygons, each has at least two edges, and thus $C_i$ has at least two slabs. The boundary of $C_i$ must lie along each of these slabs, but then there exists some intersection point on $C_i$ from which we would have traced a descent path downwards to the base edge of the slab, which is an edge of the connected component, contradicting that the connected component is not part of $\partial C_i$. Now, the connected component on the interior of $C_i$ necessarily has a subdivision point in $V$. Thus $V_i$ is non-empty and $|\slabs(C_i)| \geq 3$, which contradicts that $C_i$ is a final cell of the subdivision.\qed

\begin{lemma}\label{lem:contree} If $C_i$ is a final cell of the subdivision, then the interior edges of $R(C_i)$ are acyclic.\end{lemma}

\pf Assume there is such a cycle. Vertical and descent edges always become cell boundary edges, so all the edges of the cycle are edges of the final $R(F)$. There must be a face on the interior of the cycle, and since it is not sub-divided by any cell boundary it must contain a base edge which is not connected to $C_i$ by cell boundary edges. This implies that $C_i$ contains a connected component of $F$ on its interior, contradicting Lemma~\ref{lem:noholes}. \qed 

We now have:

\begin{lemma}\label{lem:slabcount}The total number of subdivided slabs and the total number of edges in all cells is $O(n\log m)$.\end{lemma}
\pf Each subdivided slab is incident to at least one edge of its cell $C_i$ along its defining chain. The number of edges of $C_i$ is $O(\kappa_i)$ and the total number of subdivided slabs is less than the total number of edges over all cells, which is $O(\sum_i\kappa_i) = O(n\log m)$.\qed

\subsection{The partial roof of a cell}

\paraskip{Filling in each cell.} We now have that each final cell $C_i$ has a set of subdivided slabs, each of which is incident to it along a chain of edges, called the slab's {\em defining edges}. This is slightly different than in the polygon case, where each slab is incident to the polygon only along a single base edge. When we split $C_i$ into subchains, we keep all edges incident to the same slab together (in other words we split $C_i$ into two subchains with equal size slab sets). The definition of partial roofs extends naturally to subchains of $C_i$. 

\paraskip{Partial roof of a subchain of a cell.} The partial roof $R$ of a $k$-length subchain $C$ of a cell $C_i$ is a piecewise linear surface, topologically a disk, which has exactly one face supported by each slab $s\in\slabs(C_i)$. The partial roof satisfies the (modified) edge containment, face containment, face monotonicity, and boundary properties. 

\begin{figure}
\vspace{-10pt}
\centering
\includegraphics[width=0.89\textwidth]{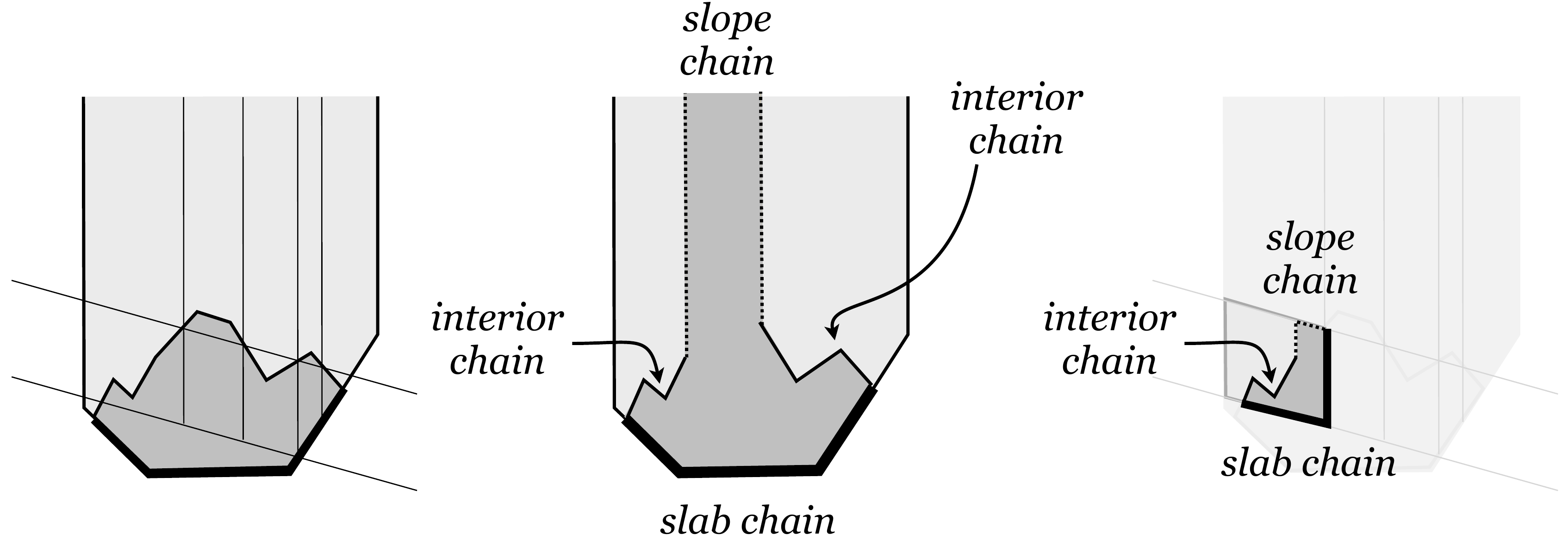}
\vspace{-6pt}
\caption{\small{Left: a slab, a face, and a possible induced subdivision of the slab and face if it is part of a PSLG. Middle: an illustration of the parts of a face of a partial roof that appear in the polygon case. The slab chain lies along the boundary of the slab, which is made up of the base edge and motorcycle edges, has two interior chains which are made up of edges incident to other faces of the partial roof, and a single slope chain, which in this case is given by two unbounded edges. Right: a face of a partial roof (built on the same slab) in the case of a PSLG. In this case, the subdivided slab does not contain the base or motorcycle edges, so the slab chain includes edges along the lifted subdividing line and descent path that define the subdivided slab. The face also has an interior chain of edges incident to other face of the partial roof, and a slope chain, which includes a single edge parallel to the slope vector of the slab, and a cap edge along the upper lifted subdivision line.}}
\vspace{-14pt}
\label{fig:labeledchains}
\end{figure}

\paraskip{(Modified) Properties.} The face and edge containment properties are defined the same way as for simple polygons, except in reference to the faces and edges in $R(C_i)$. The face monotonicity property is that each face $f$ of $R$ is monotonic with respect to the base edge of the original slab from which $f$'s supporting subdivided slab came from. The boundary property is essentially the same. Instead of each face being incident to $\partial R$ along a single base edge, it is now incident to $\partial R$ along the entire defining chain of its slab. The face again has a slab border chain, which includes the defining edges of the slab. Again we allow one or two interior chains of edges that are incident to other faces of $R$. Now the slope chain is defined as before, if the subdivided slab is unbounded. However, the subdivided slab may be bounded above by a lifted subdivision line, in which case we ``cap'' the unbounded edges with an edge along the subdividing line. In other words the slope chain may not either be two unbounded edges along the slope of the slab as before, or two bounded edges and a cap edge. Figure~\ref{fig:labeledchains} illustrates these slight variations. The boundary of $R$, denoted $\partial R$ again is made up of two chains: a defining chain which includes the defining edges of each slab in $\slabs(C)$ and a fringe chain, containing the remaining slab border edges and slope edges of each face. 

\begin{lemma}\label{lem:partialroofcomplexitypslg}The combinatorial complexity of a partial roof $R$ for a $k$-length subchain $C$ of a cell $C_i$ is $O(k)$\end{lemma}
\pf Each face as $O(1)$ vertices on $\partial R$ that are not part of its defining chain. There are $O(k)$ vertices on the defining chain. Each face is incident to $\partial R$ along its defining chain, which contains at least one edge. Thus the same argument as in in Lemma~\ref{lem:partialroofcomplexity} applies directly to prove the result.\qed

And from the face and edge containment properties we have:

\begin{lemma}\label{lem:partialrooftoroofpslg} Let $R$ be a partial roof for the entire cell $C_i$. Then $R = R(C_i)$. \end{lemma}
\pf The proof follows the same reasoning as Lemma~\ref{lem:partialrooftoroof}, which uses only the edge and face containment properties and so applies generally to bounded and unbounded cells $C_i$.\qed

\paraskip{Merging partial roofs.} The merge operation is exactly the same except that vertices of the cell may be at infinity in the positive or negative $y$-direction, and the fringe simplification step has to take into account if the subdivided slab is bounded or unbounded. If a gluing vertex were at infinity, this could present a problem, since merging at this vertex would no longer have the property of starting at a known intersection point. However, since our preprocessing step ensures that no lines of the straight skeleton are vertical in the $xy$-plane, this cannot occur. 

\begin{lemma}\label{lem:finitegluing}No gluing vertex $\hat{v}$ is an infinite vertex.\end{lemma}
\pf Suppose for contradiction that a cell $C_i$ has an infinite vertex $v$ such that the two edges $e_1$ and $e_2$ of $\partial C_i$ incident to $v$ come from different slabs. Recall that the edges of $\partial C_i$ are of three types: edges of the PSLG, vertical edges created by the subdivision step, and descent edges. Of these only the vertical edges can possibly be unbounded, so $e_1$ and $e_2$ are both vertical edges. Furthermore, since they are incident to the same vertex they must appear on different subdivision planes. There must, then, be some unbounded face $f$ incident to $e_1$ that is not incident to $e_2$. Then $f$ must have some unbounded edge projecting onto the interior of $C_i$. This means that the projection of this edge onto the $xy$-plane is parallel to the $y$-axis (otherwise it would have been subdivided by both the subdividing plane through $e_1$ and through $e_2$ and would not be unbounded), a contradiction.\qed

We also note that the cells to the left of the left-most subdivision line and to the right of the rightmost subdivision line are not polygons, but rather polygonal chains unbounded at each end by an infinite vertex. This also cannot be a gluing vertex, since it is an endpoint of the chain, and is therefore not between two edges. By the containment properties, the partial roof for the chain is equivalent to restricted roof for the cell, and so the algorithm works correctly on the cell. When we apply the fringe simplification step to a bounded subdivided slab, instead of extending out to infinity, the two endpoints are extended up to the lifted subdivision line which bounds the slab. 

Finally, letting $c_i = |\slabs(C_i)|$ we have:

\begin{lemma}The roof $R(C_i)$ for a final cell $C_i$ can be computed in $O(c_i\log c_i)$ time and $O(c_i)$ space\end{lemma}
\pf We prove that the merge operation is correct. The lemma then follows from Lemmas~\ref{lem:partialroofcomplexitypslg} and \ref{lem:partialrooftoroofpslg}. Lemma~\ref{lem:finitegluing} ensures that when we start each merge step, the gluing vertex $\hat{v}$ is a known point on $R(C_i)$. This point is, by Lemma~\ref{lem:subdivprops} property (6), incident to two faces, $f_1$ and $f_2$ of $R(C_i)$ that are incident along an edge with $\hat{v}$ as an endpoint. By the face containment property, the two faces intersect at $\hat{v}$, and so the merge operation is able to get started, as in the polygon case.

The face containment property follows the same argument as for the polygon case: since $R(C_i)$ is the lower envelope of $\slabs(C_i)$ and we only cut along intersections between slabs, it is not possible that we cut a face so that it fails to satisfy the property. 

For the edge containment property, the main difference is that before we had a tree and now we have a forest of interior edges on the final $R(C_i)$ (Lemma~\ref{lem:contree}). We need to show that given an edge $e'$ of $R(C_i)$ supported by subdivided slabs $s_1$ and $s_2$, such that $s_1$ is on the defining chain of one of the merging roofs and $s_2$ is on the defining chain of the other, we can get a path of interior edges in $R(C_i)$ back to the gluing vertex $\hat{v}$. Suppose not. Then the tree containing the gluing vertex is disconnected from the tree containing $e'$. Then there must be a face $f$ that separates the tree containing $e'$ from the tree containing $\hat{v}$. But this necessarily means that $f$ must be incident to the defining chain along an edge between the defining chain of $s_1$ and $\hat{v}$ and along an edge between the defining chain of $s_2$ and $\hat{v}$. But this is a contradiction, since each slab is incident to the defining chain along a connected chain of edges. Face monotonicity follows by the monotonicity of the splicing path, which is maintained by the stopping conditions. The boundary condition follows the same argument as before and the definition of the fringe simplification step. \qed

Finally, filling in a cell takes $O(c_i\log c_i)$ time and $O(c_i)$ space, $\sum_i c_i = O(n\log m)$ (by Lemma~\ref{lem:slabcount}), and computing the subdivision requires $O(n(\log n)\log m)$ time. Thus we have:

\begin{theorem}\label{thm:pslg}The straight skeleton of a PSLG with $m$ connected components can be computed from its induced motorcycle graph in $O(n(\log n)\log m)$ time and $O(n\log m)$ space.\end{theorem}

\section{Handling degeneracies} \label{sec:generalposition}

So far we have assumed that the polygon is in general position and non-degenerate. By {\em non-degenerate} we mean that no two motorcycles crash simultaneously. By {\em general position} we mean that no four slabs intersect at a point and no two slabs are coplanar. 
 We now show how to remove these assumptions.

\paraskip{Degenerate polygons.} In a degenerate polygon multiple motorcycles may collide simultaneously. Huber and Held showed how to handle this by launching a new motorcycle in such cases and extended the definition of slabs to include multiple motorcycle edges along the boundary \cite{HuberH12}. The straight skeleton of such polygons is the lower envelope of the (extended) slabs. We follow their approach and extend the definition of slabs in the same way. This requires extending the boundary property for partial roofs. In particular, we allow the {\em slab border chain} of each face to contain an edge on each of the motorcycle edges incident to a face. A single slab (or face in partial roof) may now be incident to $O(r)$ motorcycle edges, rather than just two, and thus a single face may have up to $O(r)$ edges on the boundary of a partial roof. However, the sum of all such edges is still $O(r)$ (cf. \cite{HuberH12}). But each slab appears in $O(\log k)$ merge steps, and thus the amortized cost of a merge operation remains unchanged. We also note that the vertical subdivision algorithm of \cite{chengArxiv2014} works within the same time bound using the extended motorcycle graph and slabs of \cite{HuberH12}. 

\paraskip{Removing general position.} Let us first assume that the intersection of any two slabs is either empty or a line segment. We now show how to deal with the case where more than three slabs intersect at a point. The proof of the linear complexity of partial roofs explicitly allows for vertices of degree higher than 3, so the proof holds without modification. The main difficulty lies in what to do if the splicing path hits a vertex rather than an edge. 

\paraskip{Computing the splicing path.} The splicing path may now cut through a vertex $v$ rather than an edge of the input roofs, and the hit point no longer tells us {\em a priori} which face the splicing path should traverse next as it does when the splicing path traverses an edge. There are two cases we need to deal with, the first is when the splicing path hits a vertex in one of the partial roofs (say $R_1$), but is still on the interior of a face (say $f$) in the other partial roof. The second is when the splicing path simultaneously hits vertices in both partial roofs.

In the first case, let $e$ denote the incoming edge of the splicing path into $v$ and assume without loss of generality that $e$ is oriented so that the downward slope of the supporting slab $s$ of $f$ is to its right. We use the fact that the final face supported by $s$ is the lower envelope of the line segments given by intersecting all other slabs with $s$. Intersect the slabs supporting the other faces incident to $v$ with $s$ to get a list of line segments $s_1, s_2, \dots$. If $e$ is an edge required by the edge containment property and for one of the segments $s_i$, to satisfy the edge containment property we need an edge along $s_i$, then $e$ and $s_i$ will be part of the lower envelope of $e$ and $s_1, s_2, \dots$ in $s$. This is equivalent to saying that $s_i$ will be the segment making the sharpest right hand turn from $e$ at $v$, and thus it can be found in $O(\operatorname{deg}(v))$ time, where $\operatorname{deg}(v)$ denotes the degree of $v$.

In the second case, let $c_r$ denote a cylinder centered at $v$ with radius $r$ with rotational axis parallel to the $z$-axis and let $F_1 = (f_1, f_2, \dots, f_i)$ and $F_2 = (f'_1, f'_2, \dots, f_j)$ denote the fans of faces incident to $v$ in $R_1$ and $R_2$ (resp.). Choose the radius $r$ small enough that no edge of either fan incident to $v$ lies entirely on the interior of $c_r$ (for instance half the length of the shortest edge of either fan incident to $v$. We compute a walk of each fan starting at the intersection of the splicing edge we just computed with $c_r$, and walk along the local intersection between each fan and $c_r$. As with the merge operation, this walk is guided only by the local intersection between the current intrinsic point on which it lies on a fan, and the intersection of a small neighborhood of that point with $c_r$. Each walk traces out a path of curved segments along cylindric sections of $c_r$. Note that two (curved) segments on $c_r$ may intersect at most at two points (rather than just one) and each segment is monotone in $c_r$ with respect to the $z$-axis. Each walk traces a (curved) polygonal chain on $c_r$. We stop the walk if either we hit a boundary edge of the fan (i.e. an edge such that the face on the other side is not incident to $v$), have traveled one complete turn around the cylinder (i.e. the projection of the walk into the $xy$-plane subtends an angle greater than $2\pi$), or the next edge of the walk is non-monotone with the previously computed segment with respect to the $z$ direction.

The intuition behind this is: suppose there is an edge $e$ in the final straight skeleton roof $R(P)$ along an intersection that we must detect in order to satisfy the edge containment property. Intersect $c_r$ with the final straight skeleton roof $R(P)$, then because $R(P)$ is a terrain, we obtain a polygon on the surface of $c_r$ that is monotone with respect to the $z$ direction. Thus in the walk of either fan, if we arrive at a point violating monotonicity, we know that that the rest of the walk cannot possibly be part of the lower envelope. The same general inductive argument as was used in the proof of Lemma~\ref{lem:mergeproducespartialroof} shows that the beginning of the walk of each fan lies along the intersection between $c_r$ and the final roof $R(P)$ and will only diverge at the edge of $R(P)$ that needs to be detected by the splicing path. This is necessarily at the first intersection between the two polygonal chains produced by the two walks. To find this point we compute the lower envelope in $c_r$ of the two chains with respect to the $z$ direction. 

\begin{wrapfigure}{r}{0.5\textwidth}
\vspace{-10pt}
\centering
\includegraphics[width=0.5\textwidth]{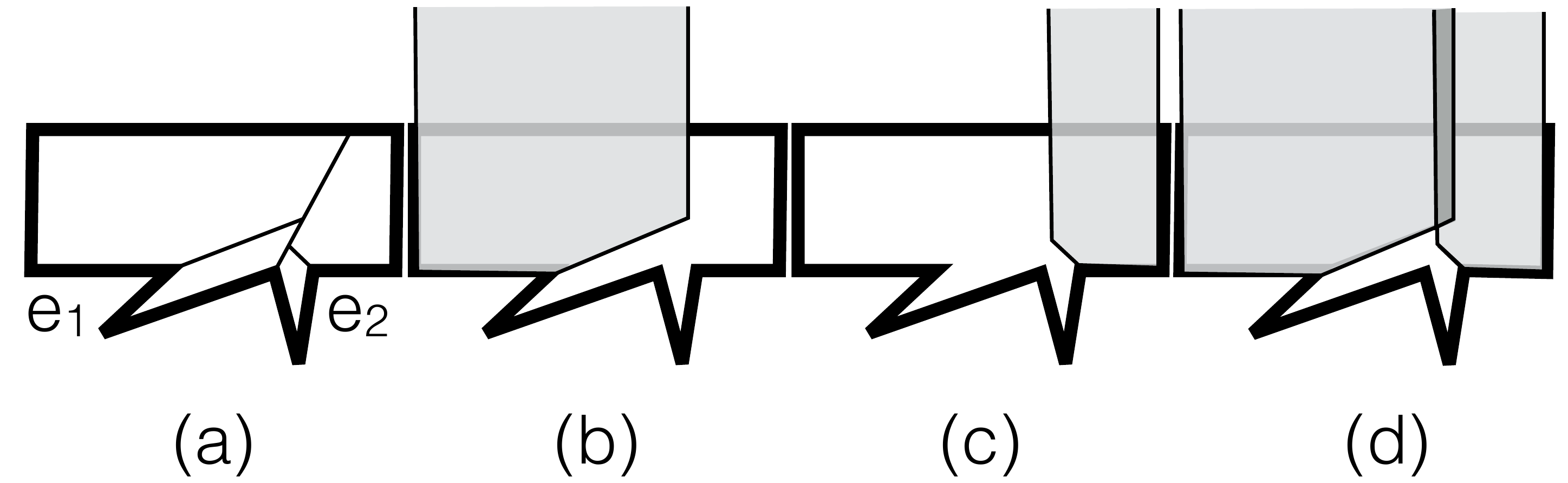}
\vspace{-10pt}
\caption{\small{An example where coplanar slabs overlap in a degenerate manner. (a) A polygon and its motorcycle graph. (b, c) the slabs for parallel edges $e_1$ and $e_2$. (d) the intersection of the two slabs is the darkly shaded region.}}
\vspace{-8pt}
\label{fig:overlap}
\end{wrapfigure}

Choosing the radius takes $O(\operatorname{deg}(v))$ time (where $\operatorname{deg}(v)$ denotes the larger degree of $v$ in either fan). On the walks we traverse each face once, and the stopping conditions are checked in constant time, so computing the walks requires $O(\operatorname{deg}(v))$ time. We then compute the lower envelope of the two monotone chains using standard sweep techniques in $O(\operatorname{deg}(v))$ time. Therefore computing the next edge of the splicing path takes $O(\operatorname{deg}(v))$ time instead of $O(1)$ time, and $\operatorname{deg}(v) = k$. However, we bound the amortized cost of a single merge operation by observing that (by the handshaking lemma), the sum of the degrees over all vertices is twice the number of edges in the partial roof, which is $O(k)$. This similarly bounds the number of trapezoids created by the ray-shooting sub-routine: the number of trapezoids incident to any vertex is at most twice the degree of that vertex, so the total number of trapezoids required for a merge step is $O(k)$. Thus the merge operation takes amortized $O(k)$ time.

\paraskip{Coplanar slabs.} The remaining ambiguity that arises when we remove the general position assumption is that when two base edges have coplanar slabs the intersection of the two slabs may be a region of the plane supporting the two, rather than a simple line segment.  See Fig.~\ref{fig:overlap}. However, if the faces of the final straight skeleton roof supported by these two slabs are incident along an edge, then using the wavefront definition of the straight skeleton it is easy to show that their motorcycles crash simultaneously. (Their base edges are parallel, and the wavefront moves outwards at unit speed in parallel, so they can only crash if their endpoints are reflex and these reflex vertices collide during the wavefront propagation. It was shown in \cite{HuberH12} that the motorcycle edges cover the traces of the reflex vertices and so this implies the motorcycles crash simultaneously.) However, following \cite{HuberH12}, such a simultaneous crash necessitates the creation of a new motorcycle which becomes a boundary edge of both the slabs, negating that the slabs intersect. Therefore if the two slabs intersect in a non-degenerate way, their faces in the final roof are not co-incident along an edge. For this reason, we can add one more stopping condition to the splicing path computation: if we get to a point where the splicing path should traverse across two faces which are coplanar and do not share a common motorcycle edge (i.e. the ambiguous situation above), we stop. This constitutes a proof that the splicing path has already computed all of the edges necessary to satisfy the edge containment property. 

\section{Conclusion.} The main theorem is proven by Theorems~\ref{thm:polygon} and \ref{thm:pslg}. This gives us faster algorithms for computing the straight skeleton of polygons and PSLGs by first computing the motorcycle graph and then using our reductions. However, there still exists a theoretical gap between the known lower bounds of $\Omega(n)$ for polygons and $\Omega(n\log n)$ for PSLGs. This remains an intriguing open problem.

  \bibliographystyle{elsarticle-num} 
  \bibliography{biblio.bib}

\end{document}